\documentclass[english]{iopart}
\usepackage[T1]{fontenc}
\usepackage[latin9]{inputenc}
\usepackage{geometry}
\usepackage{color}
\usepackage{array}
\usepackage{rotating}
\usepackage{booktabs}
\usepackage{multirow}
\usepackage{graphicx}
\usepackage{setspace}
\usepackage{esint}
\usepackage[numbers,numbers,sort&compress]{natbib}
\makeatletter

\providecommand{\tabularnewline}{\\}

\usepackage{iopams}
\usepackage{setstack}

\usepackage{graphicx}
\usepackage[numbers,sort&compress]{natbib}

\newcommand{\eqref}[1]{(\ref{#1})}

\makeatother

\usepackage{babel}
\begin{document}

\title{Climate Change and Grain Production Fluctuations}

\author{Giuliano Vitali, Sergei Rogosin \& Guido Baldoni}
\begin{abstract}
\textcolor{black}{50 year-long time series from a Long Term Agronomic
Experiment have been used to to investigate the effects of climate
change on yields of Wheat and Maize. Trends and fluctuations, useful
to estimate production forecasts and related risks are compared to
national ones, a classical regional climatic index as Western Mediterranean
Oscillation Index, and a global one given by Sun Spot Number. Data,
denoised by EMD and SSA, show how SSN oscillations slowing down in
the last decades, affects regional scale dynamics, where in the last
two decades a range of fluctuations (7-16 years) are also evident.
Both signals reflects on yield fluctuations of Wheat and Maize both
a national and local level.}
\end{abstract}


\pacs{89,92}

\ams{60,62,86}
\maketitle

\section{\textcolor{black}{Introduction}}

\subsubsection*{\textcolor{black}{a) formulation of the problem (in brief)}}

\textcolor{black}{Climate change is considered a matter of fact, and
}food security is probably one of more worrying connected problem
(\citep{Godfray-2010},\citep{Ray-2019}) as global and local trends
and oscillations of temperatures and rainfalls are related to the
variability of crop yields. To assess the dimension of the problem
it is fundamental analyzing time series of yields in relation to other
climate related variables. 

\subsubsection*{\textcolor{black}{c) which are the main methods applied in the area}}

\textcolor{black}{In the last decades, a growing collection of methods
have been developed to analyze time series. Most of them are aimed
at finding global or local trend}s, and identifying noise component
from trend is a main issue\textcolor{black}{{} (\citep{Vilar-2009}).
On the other hand signals characterizing natural phenomena or artificial
apparatus often have regular fluctuations, which are classically approached
by the Fourier Analysis (FA), assuming an underlying system linear}
and \textcolor{black}{quasi-s}tationary.

\textcolor{black}{For non-stationary signals other approaches have
been developed }(\citep{Robinson-1980}) as local FA approaches, including
Windowed Fourier transform (WFT), a forerunner of Wavelet Analysis
(WA), and Hilbert Transform (\citep{Bruns-2004}).

Jet, w\textcolor{black}{hen signal is characterized by asymmetric
or skewed behavior, non-parametric methods can be used, as the Empirical
Mode Decomposition (EMD, \citep{Huang-1998}), and the Singular Spectrum
Analysis (SSA, \citep{Vautard-1989}). Both EMD and SSA generalize
the Principal Component Analysis (PCA) in a Functional way (\citep{Ramsay-2005}),
decomposing the original time series into a complete set of components,
which in the second case are also orthonormal\citep{Golyandina-2013}.}

Both parametric and non-parametric approaches are used to study signals
separately or to find relations, basing on finding of similar signals
more or less dumped or delayed.

\subsubsection*{b) what people are doing in the area}

Harrison (\citep{Harrison-1974}) is one of first authors who looked
for a relation between sunspot cycles and crop yields, and Kozlowsky
(\citep{Kozlowsky-2007}) analyzed the impact of weather factors on
crop yields.

To date most of analysis have been oriented to relate yields to observed
climatic variables with linear regression methods (e.g.\citep{Lobell-2010}),
aimed at develop previsions for different climate scenario (e.g.\citep{Wiebe-2015}).

WFT has been used to investigate yield fluctuations in important crops
(\citep{Vitali-1999}), WA have been used by \citep{Kozlowsky-2007}
to study the impact of weather factors on yields, and by in Crowley
\citep{Crowley-2005} to investigate the impact of climate on yields
from a market perspective.

\textcolor{black}{\citep{Usowicz-2004}used a selection of non-parametric
methods included EMD to analyse spatial \& temporal variability of
environmental determinant of crop performance.}

SSA has been used to characterize ecosystem-atmosphere interactions
from short to inter-annual time scales (\citep{Mahecha-2007}).

\subsubsection*{d) what is the specific of the question which is discussed in the
paper}

What is still needed is a deeper investigation of relation of yields
with variables (indicors) related to climate to find confirmation
of possible teleconnections, which could give a better basis of perceived
effects of climate change on yields.

\subsubsection*{e) main questions addressed in the paper}

The purpose of the present analysis is investigating oscillations
of crop yields grown in a single site to see if then can be recognized
in those of country average data, if both can be related to a regional
climatic indicator, and to finally find if some components of fluctuations
can be identified in solar activity.

\subsubsection*{f) structure of the paper}
\begin{enumerate}
\item 50 years records of yield of Wheat and Maize from both from a local
(Bologna,IT) LTAE and at country-level (FAO assessed values) have
been collected together with yearly averages of two climate indicators
at regional and worldwide level, namely\textcolor{black}{{} Western
Mediterranean Oscillation index (WeMO) and Sun Spot Number (SSN). }
\item The 4 scale data-sets are formerly detrended with classical non-parametric
methodologies. \textcolor{black}{Stationarity test is performed on
detrended signal, successively EMD and SSA are used to identify noise
components and finally a wavelet analysis is performed on main signal.}
\end{enumerate}

\section{Problem Formulation and data Source}

\subsection*{Problem Formulation}

It is possible to see a crop yield as the result of an integration
process affected from solar radiation, temperature, relative humidity,
wind velocity, water and nutrient availability: simulation models
as DSSAT, APSIM and STICS already implement most of processes occurring
in a cropping system.

Therefore a crop yield may be considered a candidate climatic indicator
useful to detect the effects of climate-change \citep{Uusitalo-2015}
and linked to important issues as food security. To test such hypothesis
yield record has to be compared to other climatic indicators. 

\subsection*{Data Source}

Bologna LTAE No.64 has been established in 1967, and is still in progress.
Situated in the Southeast Po valley (Italy, $44{^\circ}33$ N, $11{^\circ}21$
E; 32 m a.s.l., silty-loam soil with a OM 1,3\%w.w. in the layer $0-40$
cm, climate sub-humid, with average annual temperature of $13{^\circ}$C
and $700$ mm rainfall, water table $0.5$ to $2.5$ m depth), the
experiment compares 5 rotations: a $9$-year rotation (corn (Zea mays
L.)-wheat (Triticum aestivum, L.)-corn-wheat-corn-wheat-alfalfa(Medicago
sativa L.)-alfalfa, two $2$-year successions (corn-wheat and sugarbeet(Beta
vulgaris L.)-wheat, continuous corn and continuous wheat. Crops, all
rainfed, are grown every year under three mineral fertilizing doses
(the maximum being the optimum-different for each crop) combined to
three rates of cattle manure. For the present paper we have used corn
and wheat yield relative to the maximum level of mineral fertilization
without manure \citep{Triberti-2016}.

The described local yields values have been compared to an higher
scale time series, that of average yields for Italy collected from
FAO's (from \textit{http://www.fao.org/faostat/en/\#data/QC}), and
to two classical climatic indicators, the Western Mediterranean Oscillation
index (WeMO, from \textit{http://www.ub.edu/gc/en/wemo/}), and the
Sun Spot Number (SSN, also known as Wolf number, collected from Royal
Observatory of Belgium in Brussels since 1749, available at\textbf{
}\textit{http://www.sidc.be/silso/home}, data available at \textit{https://crudata.uea.ac.uk/cru/data/moi/}).
SSN affects on Total Solar Irradiance (TSI) seems to call out sun
from the causes of climate change (e.g.\citep{Yang-2010}) even if
the complexity of the Sun-Earth system make conclusions still hard
to be done (e.g. \citep{Jager-2013}).

\begin{figure}[h]
\begin{centering}
\begin{tabular}{cc}
\includegraphics[width=6cm]{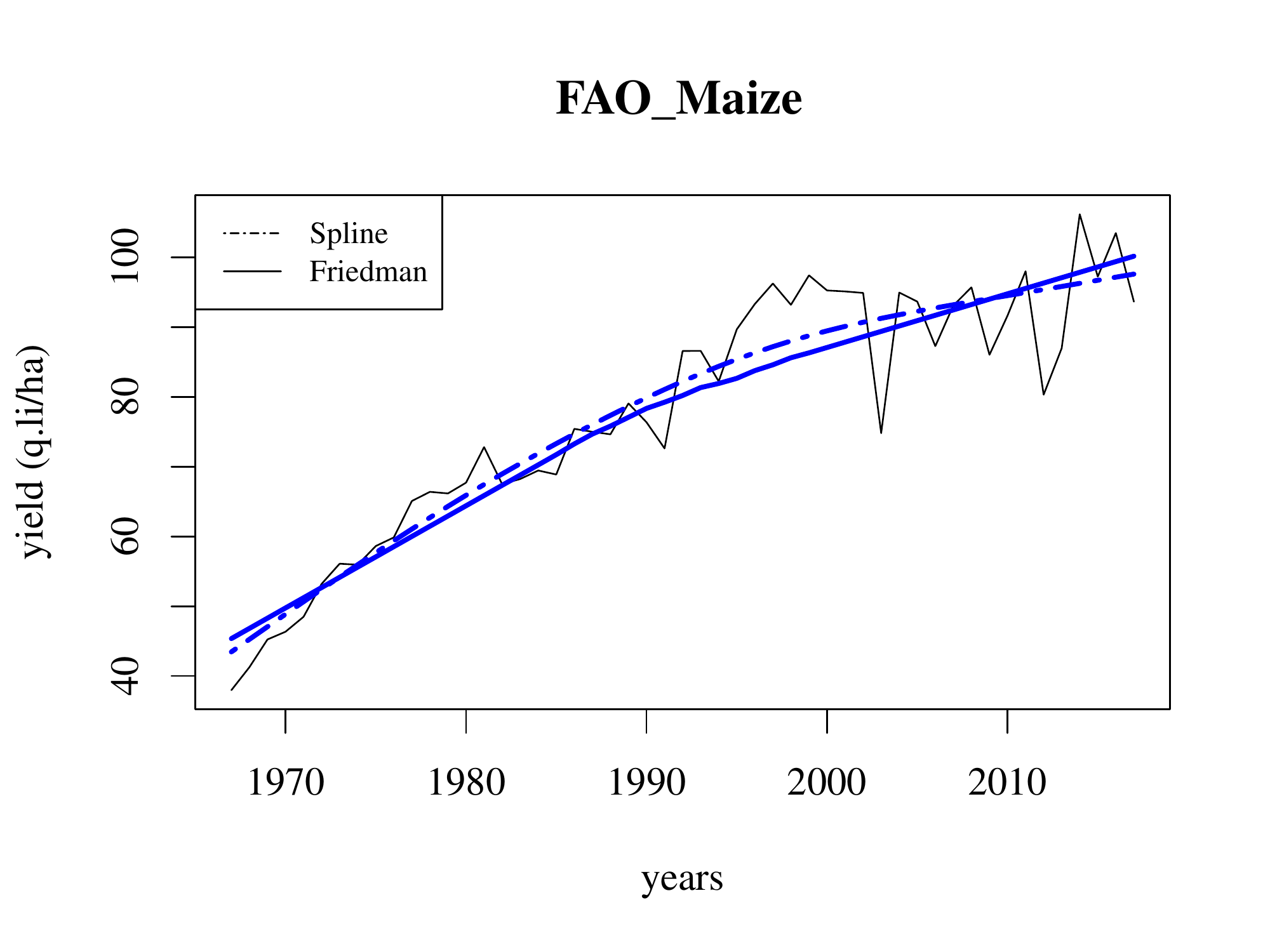} & \includegraphics[width=6cm]{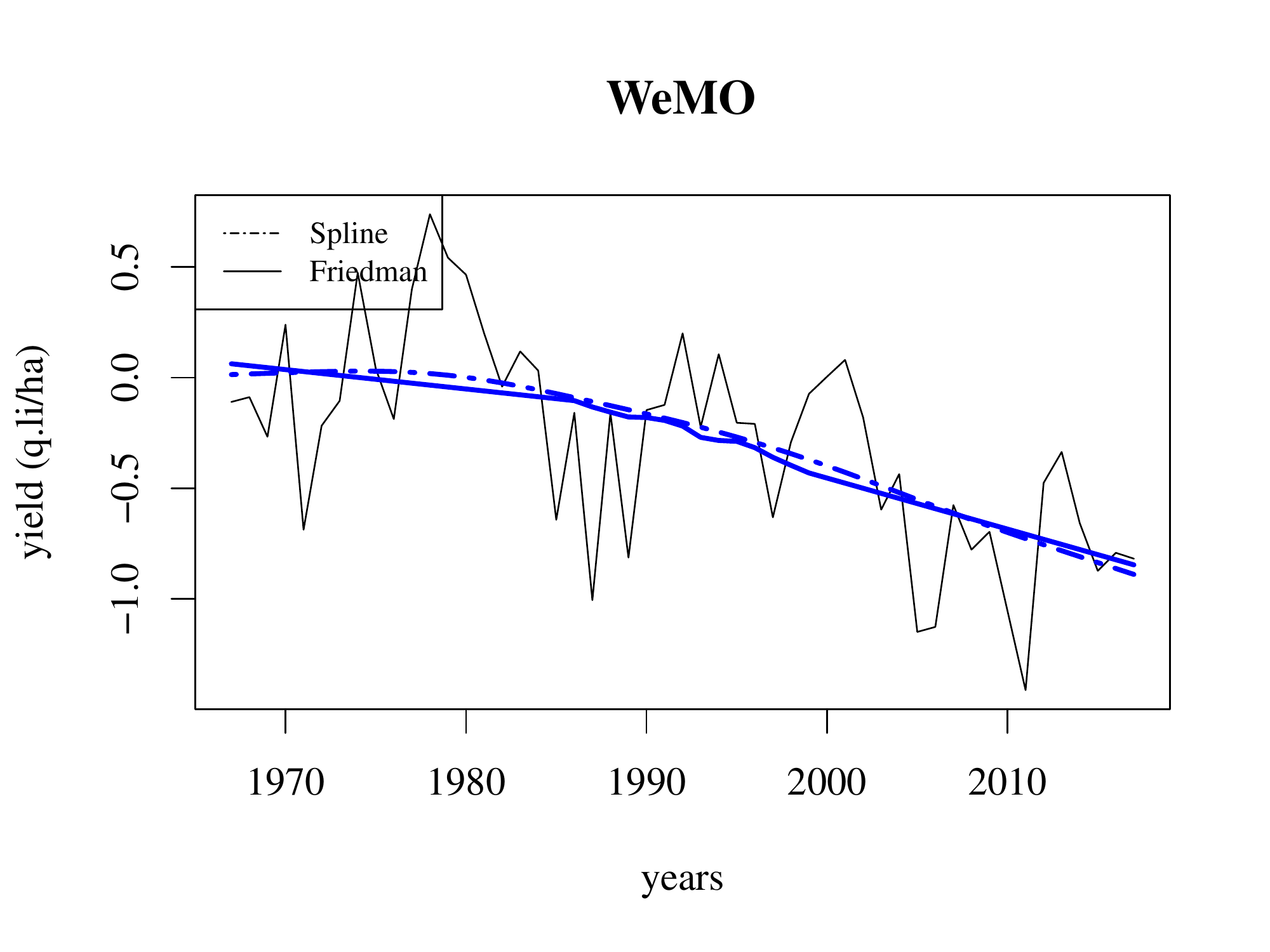}\tabularnewline
\includegraphics[width=6cm]{01_TREND_m} & \includegraphics[width=6cm]{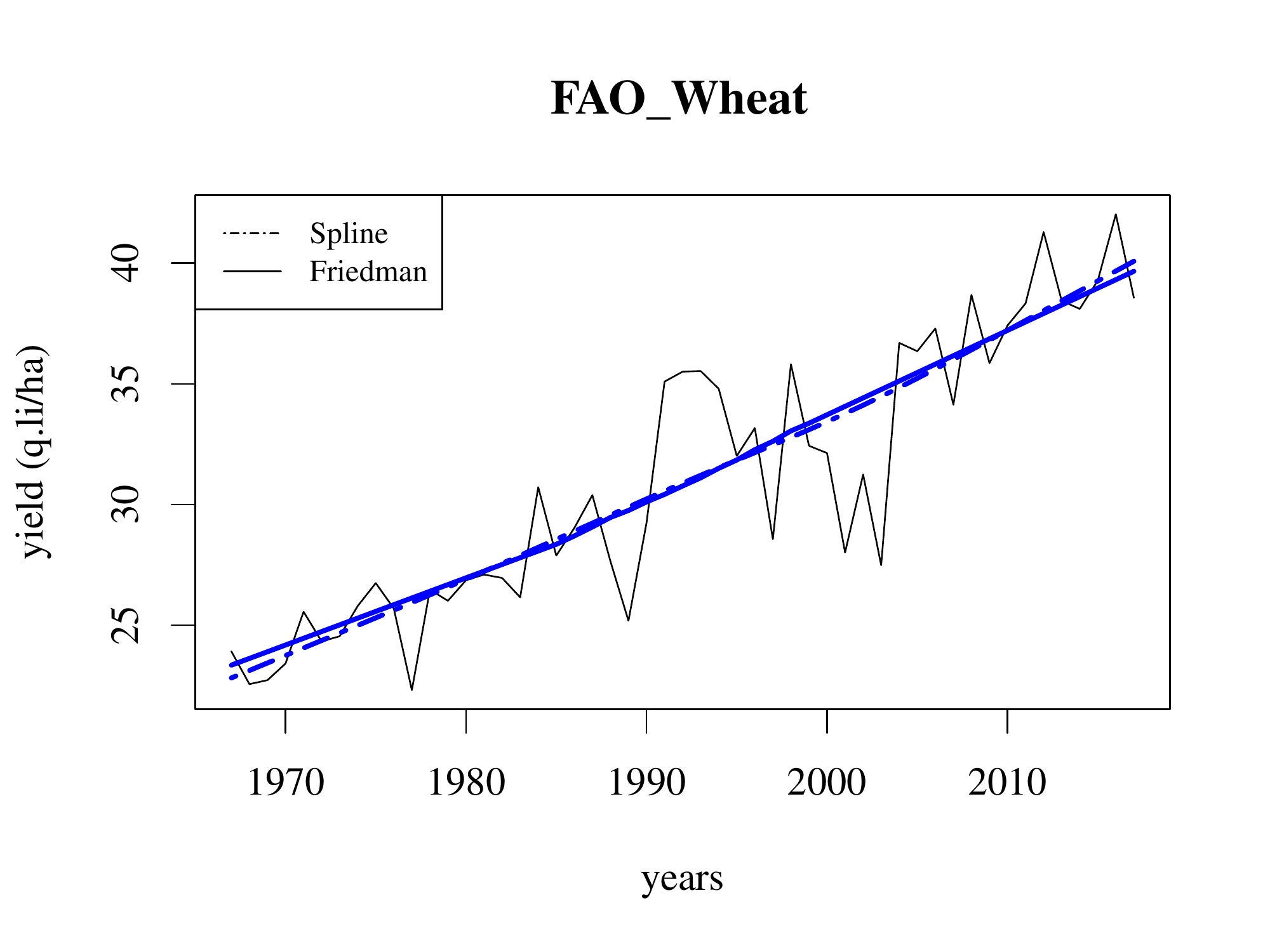}\tabularnewline
\includegraphics[width=6cm]{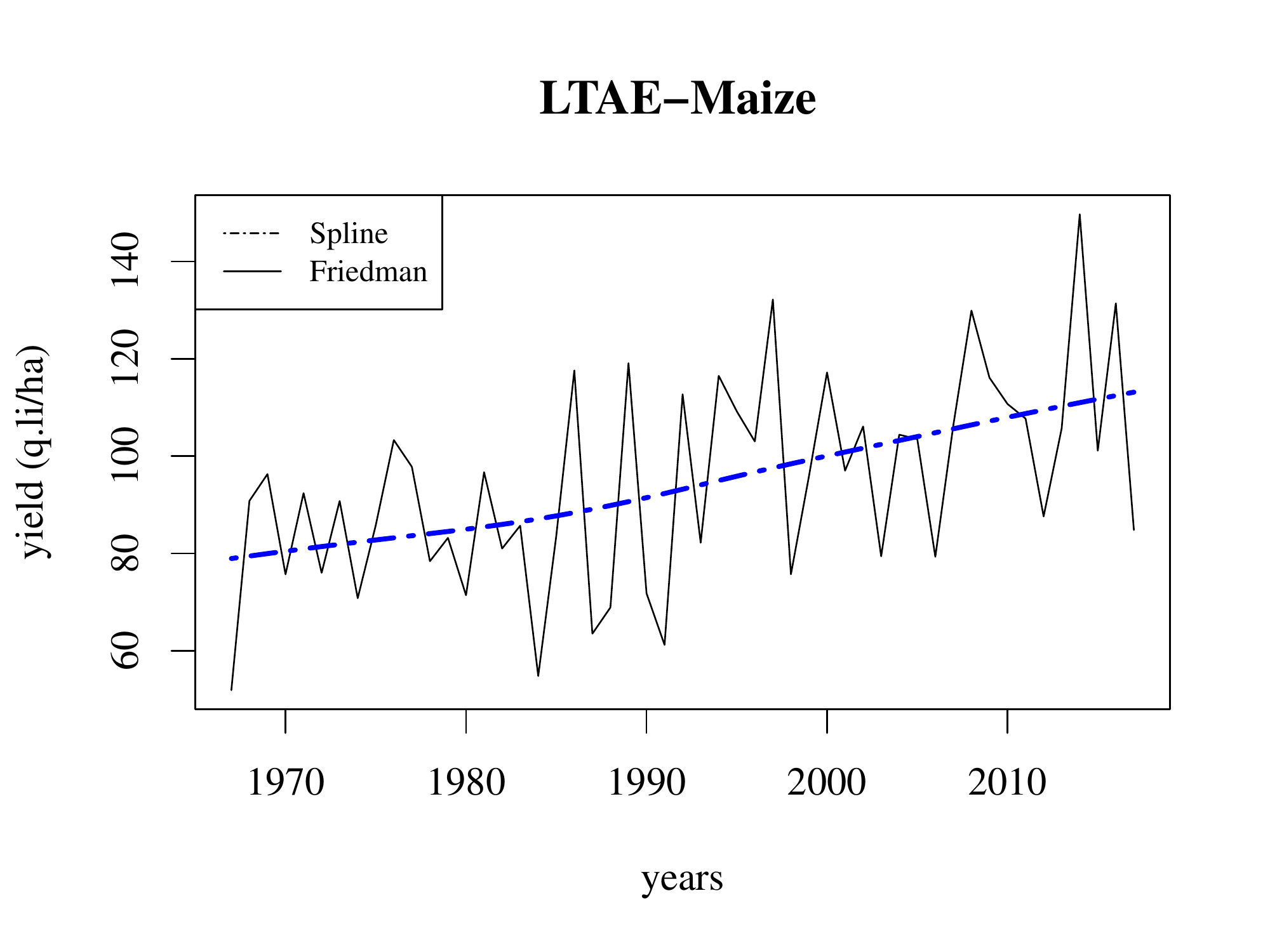} & \includegraphics[width=6cm]{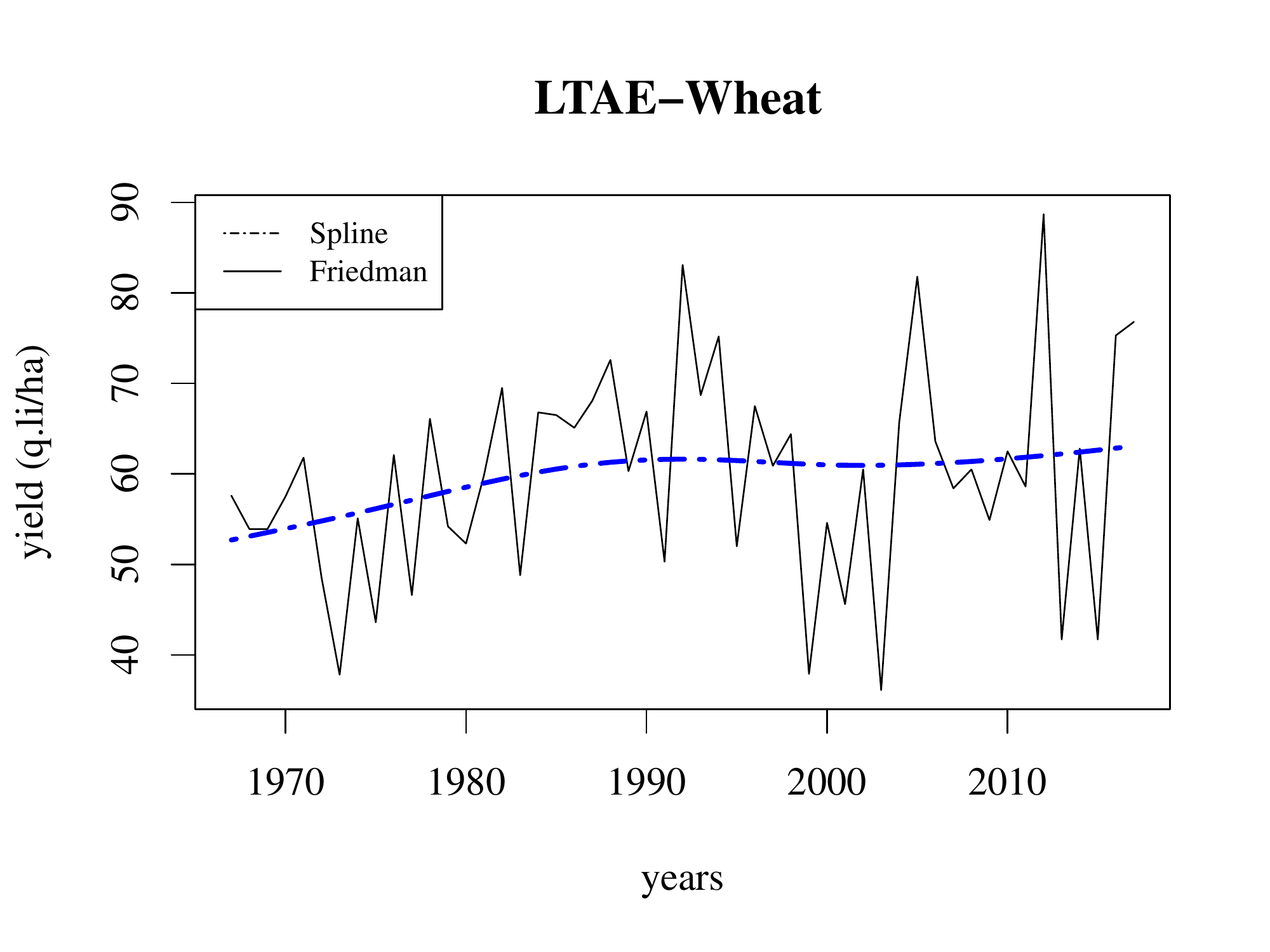}\tabularnewline
\end{tabular}
\par\end{centering}
\caption{\label{fig:TimeSeries}Time series considered in the analysis during
the period 1967-2017 with computed trends.}
\end{figure}

As SSN and WeMO series have a time resolution finer than those of
yields \citep{Clette-2014}, their yearly average has been considered
for the sake of homogeneity. Series analyzed are shown in figure \ref{fig:TimeSeries}
together with their trends (described below).

\begin{figure}[h]
\centering{}\includegraphics[scale=0.6]{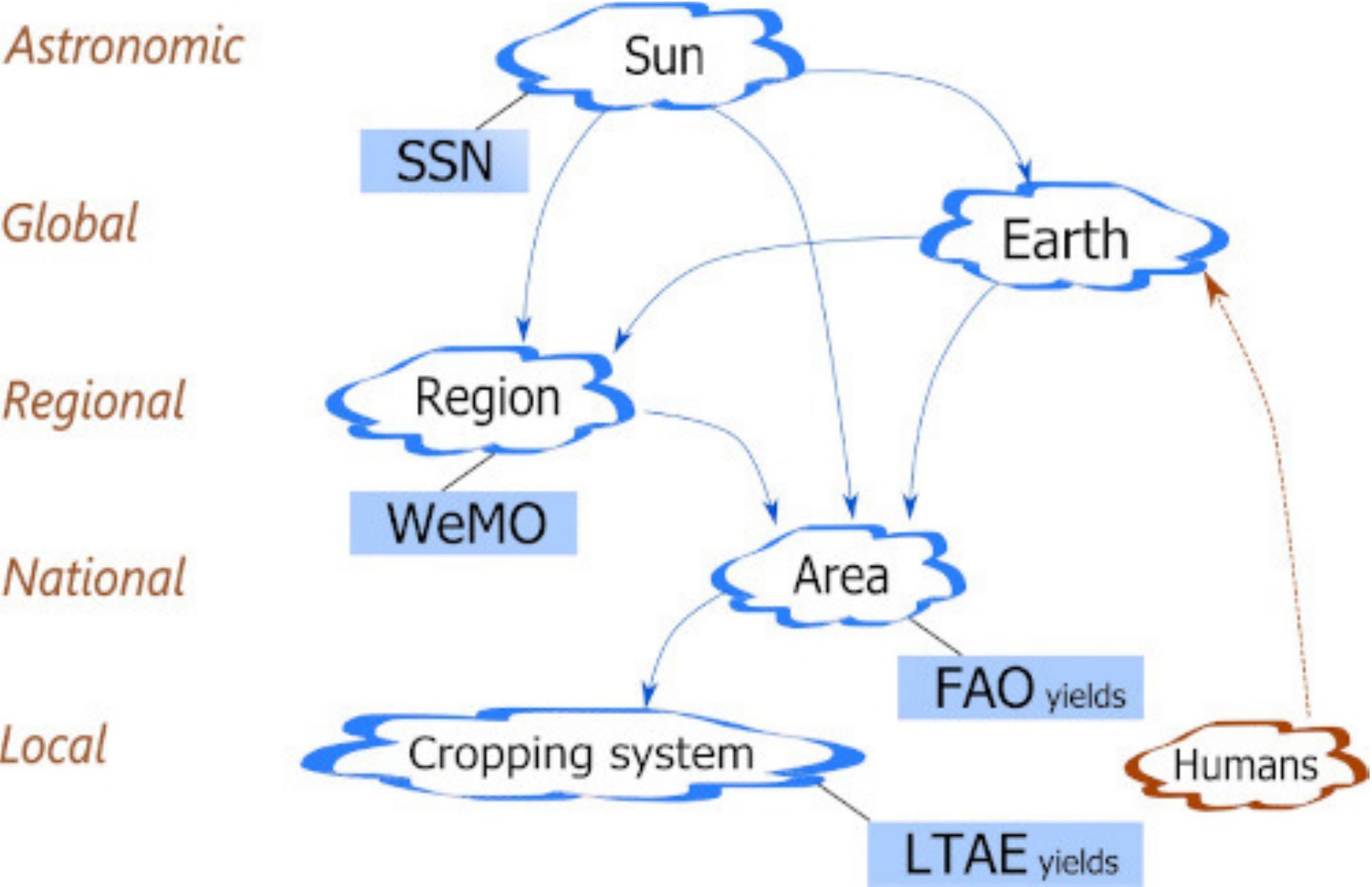}\caption{\label{fig:causality}Relation between observed indicators and dynamical
systems involved (clouds) with related scales on the left.}
\end{figure}

\textbf{Causality} - Any causality issue has been disregarded in present
analysis because of the yearly time interval of data, which make series
to be assumed as synchronous \citep{Davies-1990}. SSN is an indicator
of solar activity, which affects Global Circulation, that in turn
affects regional climates, whose trends can be revealed by indicators
as WeMO. Regional weather conditions certainly affects those of a
country, whose yields aggregate those attained locally. The diagram
shown in figure \ref{fig:causality}, the causality chain, displayed
in figure is therefore embedded in spatial aggregation.

The diagram also shows that causality could derive from relationships
between observed indicators meant as state variables of ``black box''
systems as Sun and General Circulation System still are. Such teleconnections
are therefore embedded in observation scales.

\section{Applied Method}

\subsection*{Time Series Representation}

Each time series ${x_{i}}$ is commonly represented in terms of three
main components, a \textbf{trend} $\mu_{i}$ (sometimes decomposed
into an offset-intercept and a more or less linear trend), a (more
or less regular) \textbf{cyclical} component $y_{i}$ , and an \textbf{noise}
$\varepsilon_{i}$ \textbf{}.\textcolor{black}{{} In fact recognizing
these components is strongly dependent on the method used and the
hypothesis assumed. }The components can be assumed to appear in an
additive model, where they are independent of each other $x_{i}=\mu_{i}+y{}_{i}+\varepsilon_{i}$\textcolor{black}{{}
(further discussed in Appendix 1).}

\subsection*{Trend \& Stationarity}

Trend is in most discipline considered the most important component
of time series, therefore a lot of statistical methods are oriented
to model it (e.g. regression, fitting). On the other hand when signal
analysis focuses on cyclical components, most of methods requires
the time series be stationary (see \textbf{Appendix 2}), and a preliminary
detrend\textcolor{black}{ing is required. This is a common practice
when having to do with time series too short to identify trend as
long-time oscillation. Such a detrending is often performed by smoothing
(acting like an high-pass filter) or local linear regression, as spline
functions driven by a stiffness parameter, or with self-adaptation
of span value \citep{Friedman-1984}. }First components from parametric
and non-parametric methods can be also used to the purpose.

\subsection*{Parametric analysis}

Parametric approach assumes that the signal can be represented by
a family of cyclic signal are represented by a family of functions.
For a continuous time signal $f(t)$, and a 1-parameter family of
functions, a parameter distribution, called transform $F(\omega)$,
can be obtained by the integral:

\begin{equation}
F(\omega)=\int_{\tau=0}^{t}K(\tau,\omega)\cdot f(\tau)\:d\tau\label{eq:1}
\end{equation}

$F(\omega)$ represents the distribution of weights / amplitudes (and
phases) of $K(\tau,\omega)$, the kernel of transformation. For a
uniformly sampled signal $X={x_{i};i=1..N}$ \textcolor{black}{transformation
can be can be discretized to:}

\begin{equation}
F_{k}=\sum_{i=1}^{N}K_{k,i}\cdot x_{i}\Longrightarrow F=K\cdot X
\end{equation}

\textcolor{black}{Such a linear form is the same encountered both
in parametric and non-parametric approaches.}

Periodical components in natural signals (astronomy, earth, living
organisms) have long been studied with Fourier Analysis (FA), based
on a simple harmonic $K(t,\omega)=e^{i\omega t}$, $\omega$ being
the radian frequency and $t$ the observation time. Fourier Transform
is applied on sampled signals by its discrete version (DFT):

\begin{equation}
F_{k}=\sum_{i=1}^{N}e^{i2\pi kt_{i}}\cdot x_{i}
\end{equation}

where the summation is extended to a number of frequencies $k$ multiple
of the sampling one and in number equal to that of the observations,
$N$. To optimize computation times DFT is usually performed via the
Fast Fourier Transform (FFT) requiring $N'=2^{n}$, where $n$ are
the possible frequencies.

\subsection*{Non-parametric analysis}

Non parametric analysis is used when the signal analyst does not want
or can not use pre-defined functions, and may be regarded as an extension
of Functional Principal Component Analysis. To date the more diffused
methods are represented by EMD and SSA. They are often considered
as non-linear analysis tool, as they focus on signals hard to be generated
by linear system (plane harmonics). 

\textbf{EMD} is based on an iterative methodology called \textbf{sifting}:
on the original (detrended) signal is identified an envelope described
by a couple of curves obtained by maximum and minimum values, whose
average identify the points of the so-called Intrinsic Mode Function
(IMF). IMF points are used to detrend the signal, while the residual
undergoes the procedure repetitively till a tolerance ($\epsilon$)
is reached:

\begin{equation}
\sum_{t}\left(\frac{h_{j}(t)-h_{j-1}(t)}{h_{j-1}(t)}\right)^{2}<\epsilon
\end{equation}
where $h_{i}(t)$ is the IMF at step $i$. Since when it has been
proposed \citep{Huang-1998}, IMFs have been recognized to need to
be interpreted as amplitude and frequency variable functions, and
Hilbert-transform (\textbf{Appendix 3}) has followed, allowing for
obtaining a generalized Fourier expansion:

\begin{equation}
Y(t)=\sum_{j=1}^{n}h_{j}(t)\:e\:^{i\intop\omega_{j}(t)\:dt}
\end{equation}
where $Y(t)$ is the reconstructed signal, and amplitude distribution
$h_{j}(t)$ is known as Hilbert-Huang transform.

\textbf{SSA} allows for detecting amplitude and phase modulated oscillations
\citep{Allen-1996} in terms of Empirical Orthogonal Components (EOCs).
The method is based on building lagged series from the original one
(of length $N$) taking $L$ records only (window size) obtaining
$K\quad(=N-L-1)$ rows of the so-called trajectory matrix $T$ (embedding
step). The successive $T$ is diagonalization completes the Singular
Value Decomposition (SVD):

\begin{equation}
T=U\;D\;V'
\end{equation}

allowing to obtain the eigenvectors (diagonals of $D$), and $K$
eigenvectors of length $L$ (rows of $U$), representing the components
of the signal. Components may also be collected into groups to facilitate
recognition of trend, periodicity and separation of noise \citep{Golyandina-2014}.

\textbf{\textcolor{black}{Non stationarity}} - Both parametric and
non/parametric methods required the signal be stationarity together
with the possibility to reconstruct the signal by a linear superposition
of components.

\textcolor{black}{To assess stationarity of a (detrended) signal,
Kwiatkowski--Phillips--Schmidt--Shin (KPSS) test can be used to
verify (HP-0) that the time series is stationary around a deterministic
trend, or the (complementary) Augmented Dickey Fuller (ADF) tests
(HP-0) that there is a unit root, that the characteristic equation
represents an AR process (see e.g. \citep{Kwiatkowski-1992}).}

When the detrended signal is non-stationary, because of non-homeostaticity
of the underlying system, changing frequency, phase or amplitude of
signal components, the methods described above have been extended
by means of \textbf{local} operators. 

SSA has been extended for non-stationary series adding circularity
in Circulant SSA (CSSA, \citep{Bogalo-2017}) and Sliding SSA (SSSA,
\citep{Harmouche-2018}), which are ways to analyze locally the series
by means of windows. The windowing strategy has been introduced by
\citep{Gabor-1946} in WFT by superimposing to the traditional kernel
a windowing function $w$ :

\begin{equation}
K(t,to,\tau,\omega)=w(t;to,\tau)\cdot e^{i\omega t}
\end{equation}

WFT allows to maintain the orthogonality of the transformation within
the same interval, that is the invertibility of the Fourier transform
\citep{Luis-1997}. Such an approach is also known as Short Time Fourier
Transform and available by algorithms as the sliding DFT \citep{Heinzel-2002}
well fitting the issue of having $N>N'$, that make possible to perform
the analysis on several data sets ($L=N-N'$)\footnote{Integral transform generalizes the statistical concept of correlation
between two sets of data G and F, $C=\sum_{j}G_{j}\cdot Y_{j}$ ,which
can be generalized to continuous realm, to get the correlation function:
$c(y)=\int\:g(x+y)\:f(x)\:dx$ where $y$ is the delay of the signal
$g$ with respect to the \textbf{$f$} (the correlation parameter). }. Such approaches, though, do not guarantee ortho-normality\footnote{the importance of the feature has been realized by Vercelli in its
1940s' \textit{cimanalysis}, where DFT weights have been applied to
cancel out components in the linear system of signal components \citep{Vercelli-1954}\textbf{\textcolor{blue}{{}
}}\textcolor{blue}{(historical value note).}} . Wavelet Transform (WT) \citep{Addison-2002,Mallat-2009} is identified
by a two parameter convolution kernel $K(x,y)$ made of a mother wavelet
$\psi(t)$, which can be interpreted as an high-pass filter, and a
scaling function $\phi(t,s)$ (also called father function) which
can be regarded as a low-pass filter, $s$ being the scale\textbf{.
}Its discrete version can be written as a weighted average:

\begin{equation}
W_{k}(s)=\sum_{i=0}^{N-1}Q_{k,i}(s)\cdot x_{i}
\end{equation}

where $Q_{i,j}=c(s)\;\psi(t_{i})\circ\phi\left(t_{i}-t_{j},s\right)=c(s)\;\psi\left((t_{i}-t_{j})/s\right)$: 

Wavelet approach can adopt several mother functions, each with peculiar
properties (e.g. Haar wavelet, deriving from the box function, has
a real transform) that make it appropriate to certain data sets. In
geophysical data it is quite used the Morlet wavelet, inheriting features
of ``tapering'' method\textbf{ }\citep{Thomson-1982}:

\begin{equation}
\psi(\omega,t)=\pi^{-1/4}e^{i\omega t}e^{-t^{2}/2}
\end{equation}

Wavelet technique can be used to evidence pattern superposition of
two series by means of the Cross Wavelet Transform (CWT). Power spectrum
$P_{k}$ definition may be extended to obtain the cross power spectrum:

\begin{equation}
P_{k}^{XY}=\left\langle W_{k}^{X}\;W_{k}^{Y*}\right\rangle 
\end{equation}

where $\left\langle \right\rangle $ stands for expected value and
($^{*}$) for complex conjugate. 

Following the scheme described above, the cumulated probability distribution
of cross power spectrum for a given $p$ is :

\begin{equation}
D\left(\frac{\left|W_{n}^{X}(s)\;W_{n}^{Y^{*}}(s)\right|}{\sigma_{X}\;\sigma_{Y}}<p\right)=\frac{Z_{m}(p)}{m}\sqrt{P_{k}^{Y}\;P{}_{k}^{Y}}
\end{equation}

where $Z(p)=\sqrt{\chi_{m}^{2}(p)\;\chi_{m}^{2}(p)}$ . As $P_{k}^{XY}$
is complex, analyzing coherency is also important, the latter being
defined by:.

\begin{equation}
R_{k}^{2}(s)=\frac{\left|S(W_{k}^{XY}(s)/s)\right|^{2}}{S\left(\left|(W_{k}^{X}(s))\right|^{2}/s\right)\cdot S\left(\left|(W_{k}^{Y}(s))\right|^{2}/s\right)}
\end{equation}

where smoothing both in scale and time is applied: $S_{k}(W_{k})=S_{time}[S_{scale}(W_{k})]$
(see \citep{Grinsted-2004} for details).

\textbf{Denoising and Periodogram readability }- In many cases noise
corresponds to high frequency stationary components (2-nd order stationary),
and removed applying high pass filters \citep{AlvarezMeza-2013,Boudraa-2005}.

Smoothing is also applied on periodograms (power spectrum along years),
to increase their readability, and interpolated values are commonly
considered consistent estimator of the true spectrum \citep{Maraun-2004}.
In WT (2D periodograms) things are more complex though, because neighboring
points in time and scale are correlated, and expected power spectrum
$P_{k}$ can be obtained from a Fourier power spectrum extending the
Wiener-Khinchin theorem\textbf{ }\citep{Maraun-2004} :
\begin{equation}
P_{k}=\left\langle W_{k}\cdot W_{k}^{*}\right\rangle 
\end{equation}

where scaling factor $s$ is omitted for sake of simplicity. Applying
it to a Fourier spectrum of an AR(1) $x_{t}=x_{t-1}+\alpha\cdot\varepsilon_{t}$
\citep{Allen-1996} it becomes:

\begin{equation}
P_{k}=\frac{1-\alpha^{2}}{1-\alpha\cdot e^{-i2\pi k}}
\end{equation}

where the error $\varepsilon_{t}$ is modeled as a Gaussian model
(white noise) and $\alpha$ is estimated from data. If $P_{k}$ is
sufficiently smooth, the cumulated probability of a power for a given
significance level $p$ is :

\begin{equation}
D\left(\left|W_{k}\right|^{2}<p\cdot\sigma_{k}^{2}\right)=\frac{1}{2}P_{k}\;\chi_{m}^{2}(p)
\end{equation}

where $m$ is 1 for real WT (e.g.Haar) and 2 for complex ones (e.g.
Morlet) \citep{Grinsted-2004}.

\textbf{\textcolor{black}{Analysis steps }}\textcolor{black}{- Starting
from methods described above, many analysis recipes have been developed,
and a variety of code libraries have been produced that already implement
such methods in several platforms. In the present study, analysis
has been carried out with RStudio (ver.1.1.456), following the steps
below:}
\begin{itemize}
\item Original signals have been formerly detrended (by $detrend.series$,
package dplR, by \textbf{Bunn}) by ''Spline'' and ``Friedman''
approaches, and trends are discussed.
\item On detrended signals, stationarity has been successively tested by
\textcolor{black}{Kwiatkowski--Phillips--Schmidt--Shin (KPSS) and
Augmented Dickey Fuller (ADF) test (}procedures $KPSS.test$ and $ADF.test$
from package aTSA, by\textbf{ Debin Qiu}).
\item EMD (procedure\textbf{ $extractimf$ }, package EMD, by \textbf{Kim})
has been used to identify noise of big amplitude which can be found
in the first component (first sifting passage); on such component
FFT (by\textcolor{black}{{} $periodogram$ , package TSA, by }\textbf{\textcolor{black}{Chan
\& Ripley}}\textcolor{black}{) }has been used to identify the related
noise model.
\item SSA has been successively performed on (by $ssa$ , package Rssa,
by \citep{Korobeynikov-2010}) on EMD filtered signal to identify
high frequency components corresponding to lowest eigenvalues.
\item Wavelelet analysis has been performed on reconstructed signals (by
$analyze.wavelet$ , package ``WaveletComp'', by \citep{Roesch-2018}).
\item Cross Wavelelet Analysis to verify signal coherency has been finally
performed by $analyze.coherency$ (package ``WaveletComp'', by \citep{Roesch-2018}).
\end{itemize}

\section{Results}
\begin{enumerate}
\item At a first glance (see figure \ref{fig:TimeSeries}), time series
looks very different to one another. The large fluctuations visible
in SSN (figure \ref{fig:TimeSeries}), corresponding to the well known
$\sim11$ year cycle (e.g. \citep{Yang-2010}), are not readily recognizable
in the other series.
\item Trends by spline and Friedman techniques are very similar to one another
(figure \ref{fig:TimeSeries}) and the former has been chosen for
the following analysis.
\item SSN and WeMO trends looks very similar, being stable in the first
decades and decreasing in the last period. A\textcolor{black}{{} regression
between the two trends}\textbf{\textcolor{black}{{} ($R^{2}=0.99$)}}\textcolor{black}{{}
could easily suggest a strong dependence of regional climate on solar
activity.}
\item Strong quasi-linear trends appears especially in FAO-yield, showing
the effects of the so called ``green revolution'' due, \textcolor{black}{after
the WWII, to genotype improvements and diffusion of mechanization,
chemical fertilizers and pesticides. S}uch a rise in yield is almost
reaching a plateau in Maize, still growing in Wheat. Such behavior
can be explained from genetic research, which is still intensive for
crops grown in dry conditions (lot of congresses and debates on drought)
as wheat, whereas maize is mostly grown in irrigated areas.
\item In experimental plots (LTAE), the more stable yields of wheat with
respect to maize can be ascribed to a conservative approach aimed
at growing the same varieties. This want though turned to be feasible
for wheat, not for maize, where seed availability is bond to crop
variety commercial availability (and related patents). Moreover, despite
a particular care into experimental crop management\textcolor{black}{{}
(manual weed control, fertilization management) }yearly fluctuation
are still large \textcolor{black}{when compared to country values
(FAO) }which are an average of a large number of values with a variability
due to different climates, cropping techniques, crop varieties,etc.
\item Stationarity tests have been performed on detrended signals (using
spline method) for the base model (no drift, no trend) and results,
displayed in table \ref{tab:STATIONARITY-TEST}, show that SSN represents
the only critical issue. Autocovariance functions are reported in
appendix A2.
\end{enumerate}
\begin{table}[h]
\begin{tabular}{cccccccc}
\cline{3-8} 
 &  & SSN & WeMO & FAO-W & FAO-M & LTAE-W & LTAE-M\tabularnewline
\hline 
\multirow{2}{*}{Spline} & KPSS & >0.1 & >0.1 & >0.1 & >0.1 & >0.1 & >0.1\tabularnewline
\cline{2-8} 
 & ADF & 0.022 & <0.01 & <0.01 & <0.01 & <0.01 & <0.01\tabularnewline
\hline 
\multirow{2}{*}{Friedman} & KPSS & >0.1 & >0.1 & >0.1 & >0.1 & >0.1 & >0.1\tabularnewline
\cline{2-8} 
 & ADF & 0.022 & <0.01 & <0.01 & <0.01 & <0.01 & <0.01\tabularnewline
\hline 
\end{tabular}\caption{\label{tab:STATIONARITY-TEST}p values of stationarity test applied
to detrended time series}

\end{table}

As KPSS are almost every time > 0.1 and ADF values < 0.05 detrended
data can be considered stationary. 

\begin{figure}[h]
\begin{centering}
\begin{turn}{90}
\begin{tabular}{cccc}
\includegraphics[width=6cm]{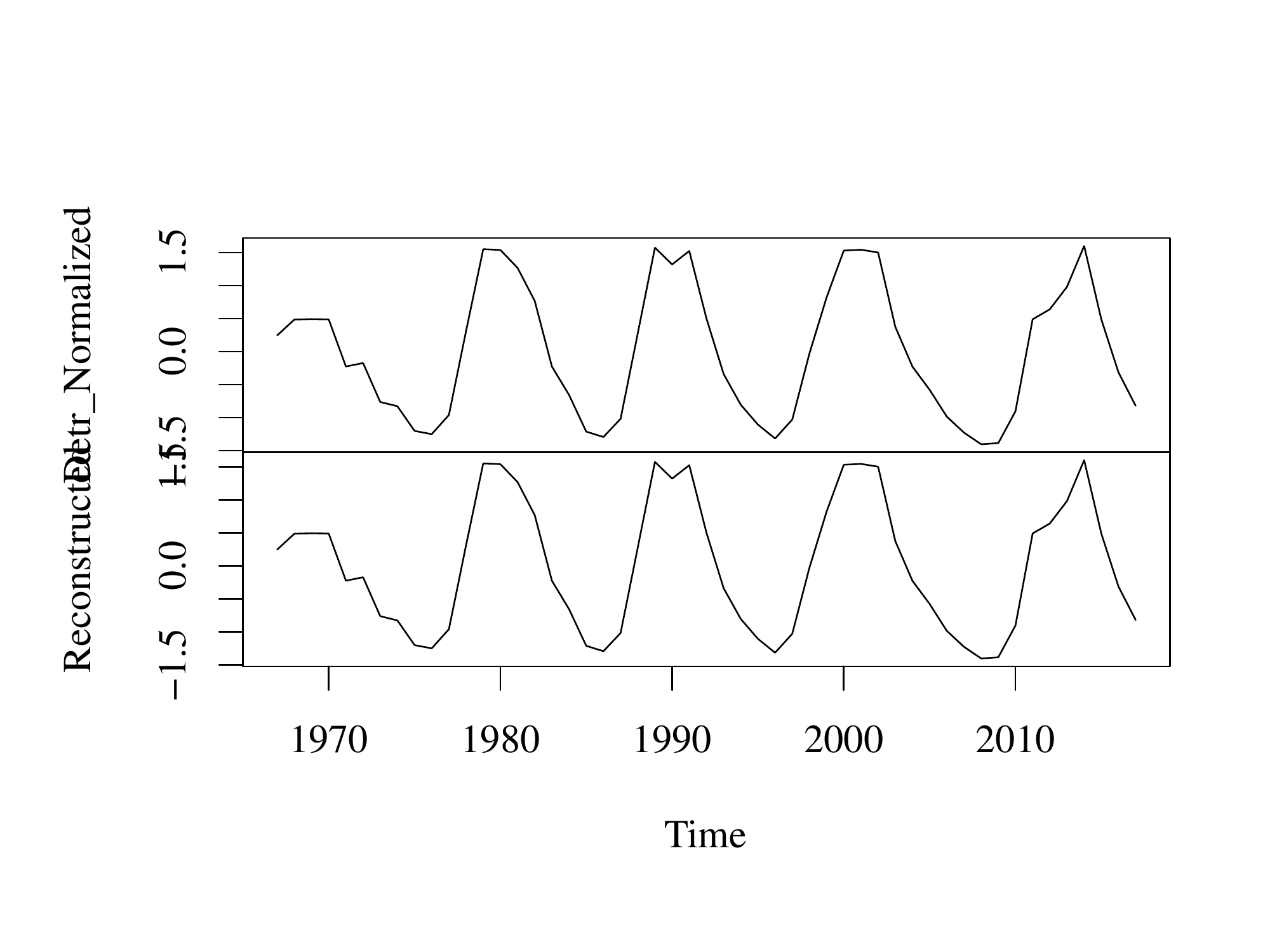} & \includegraphics[width=6cm]{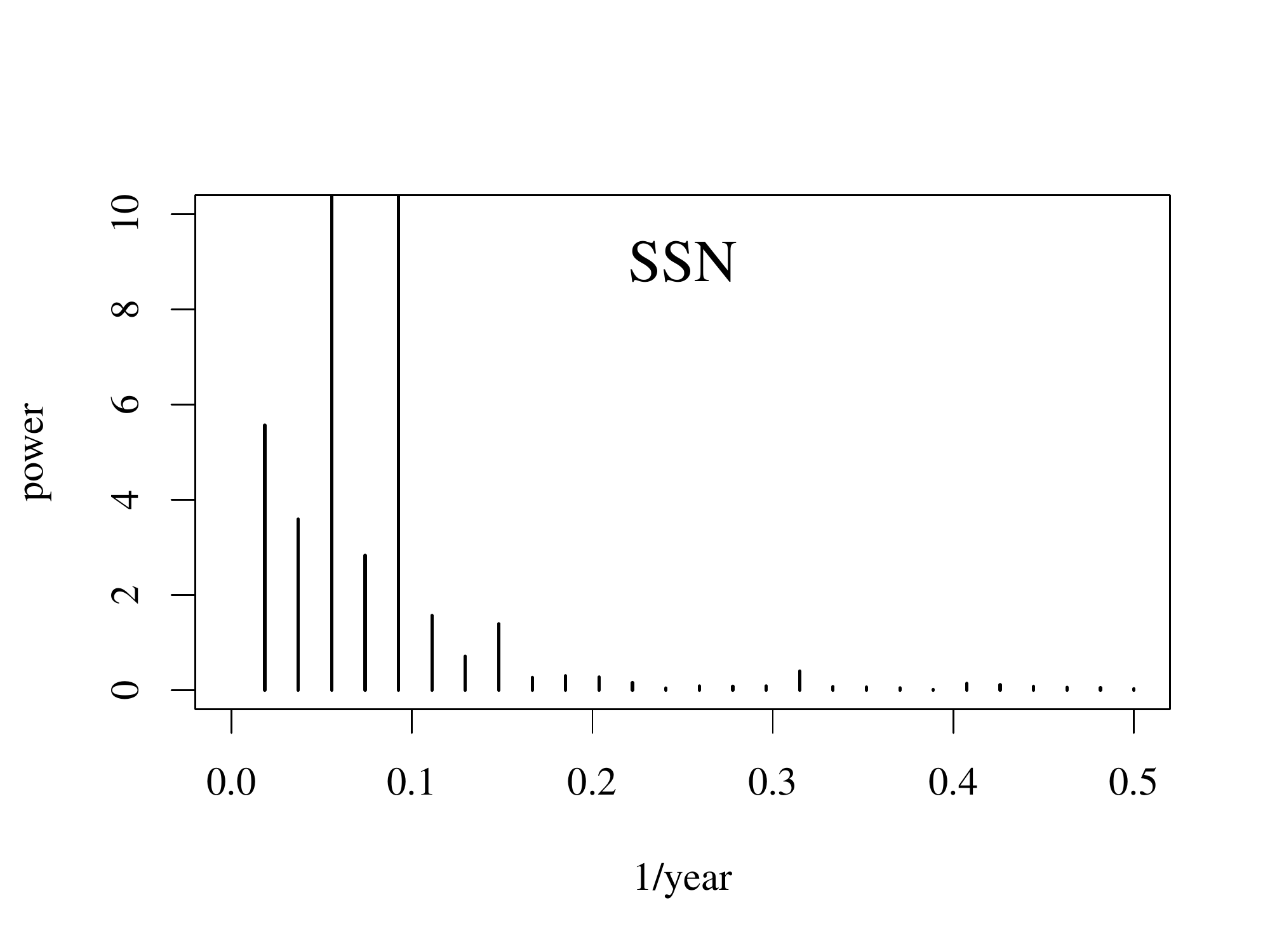} & \includegraphics[width=6cm]{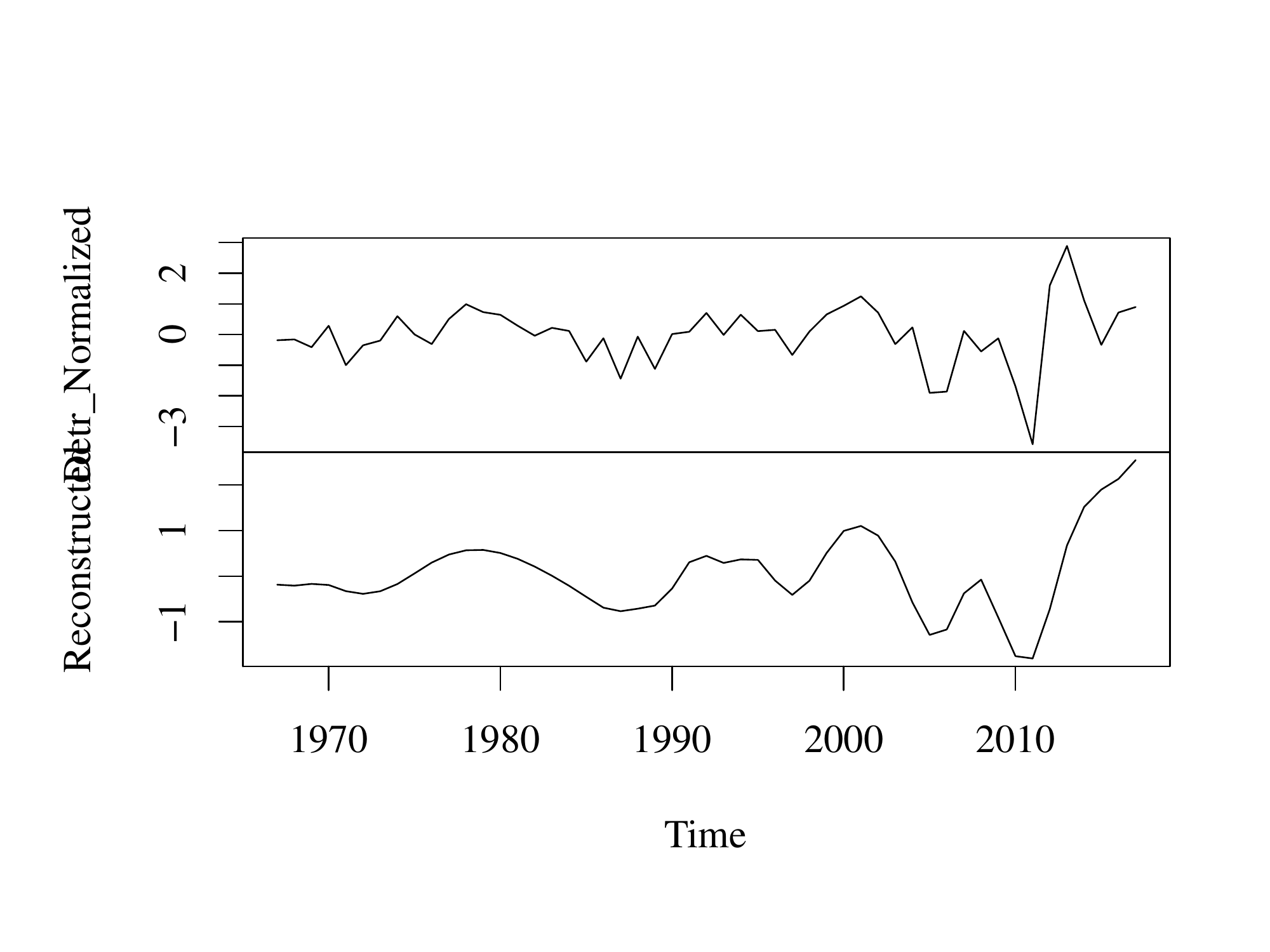} & \includegraphics[width=6cm]{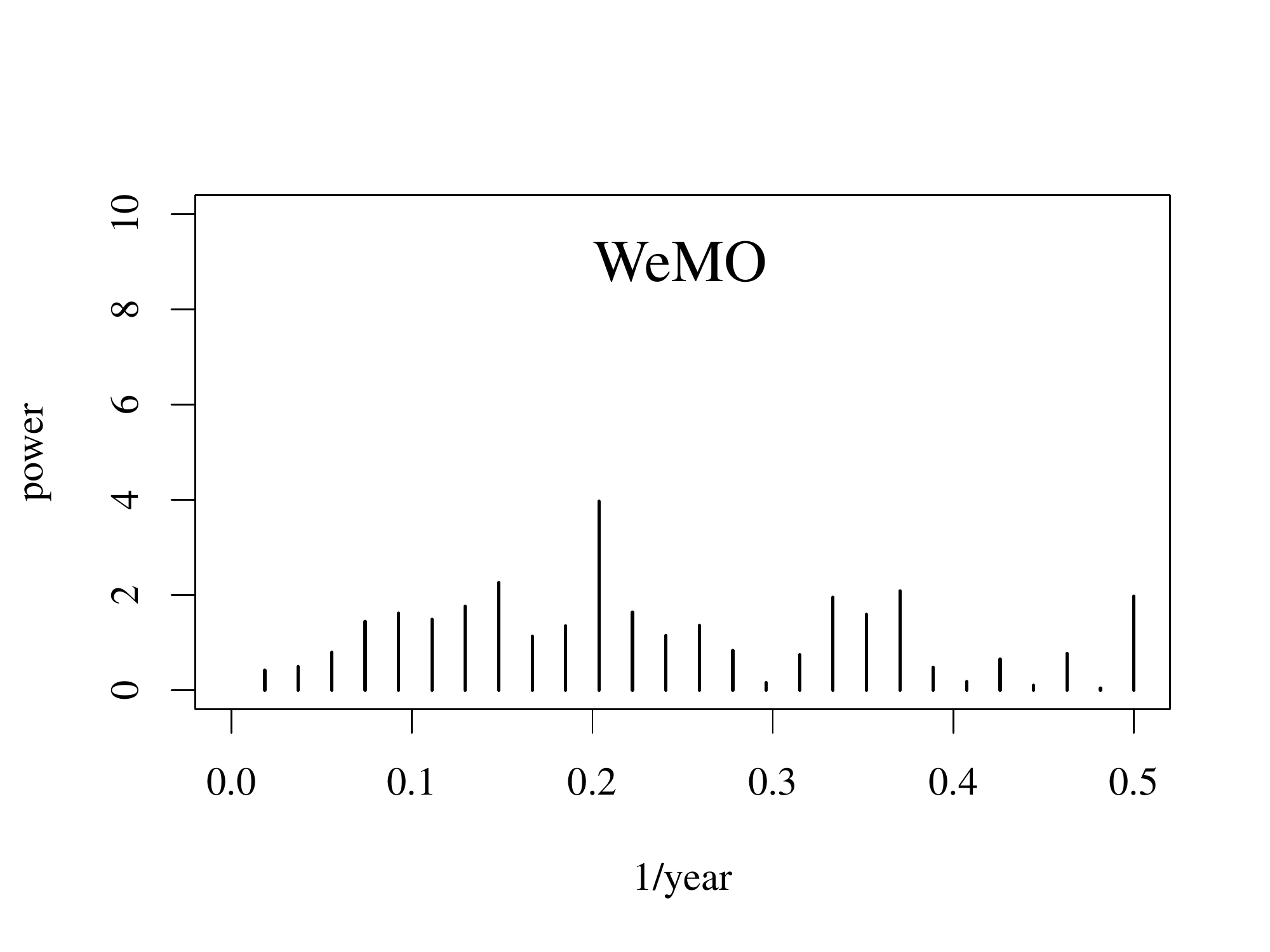}\tabularnewline
\includegraphics[width=6cm]{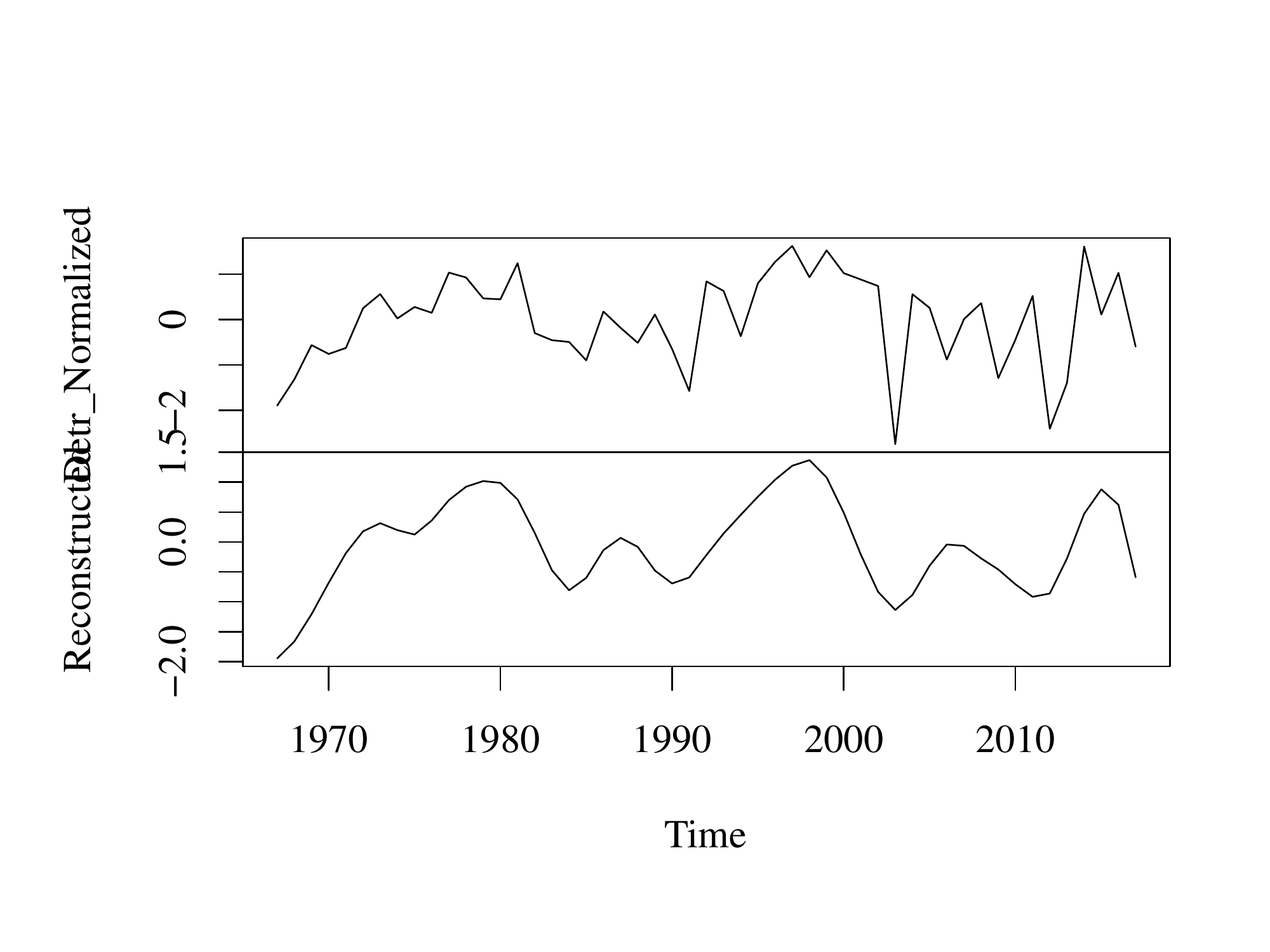} & \includegraphics[width=6cm]{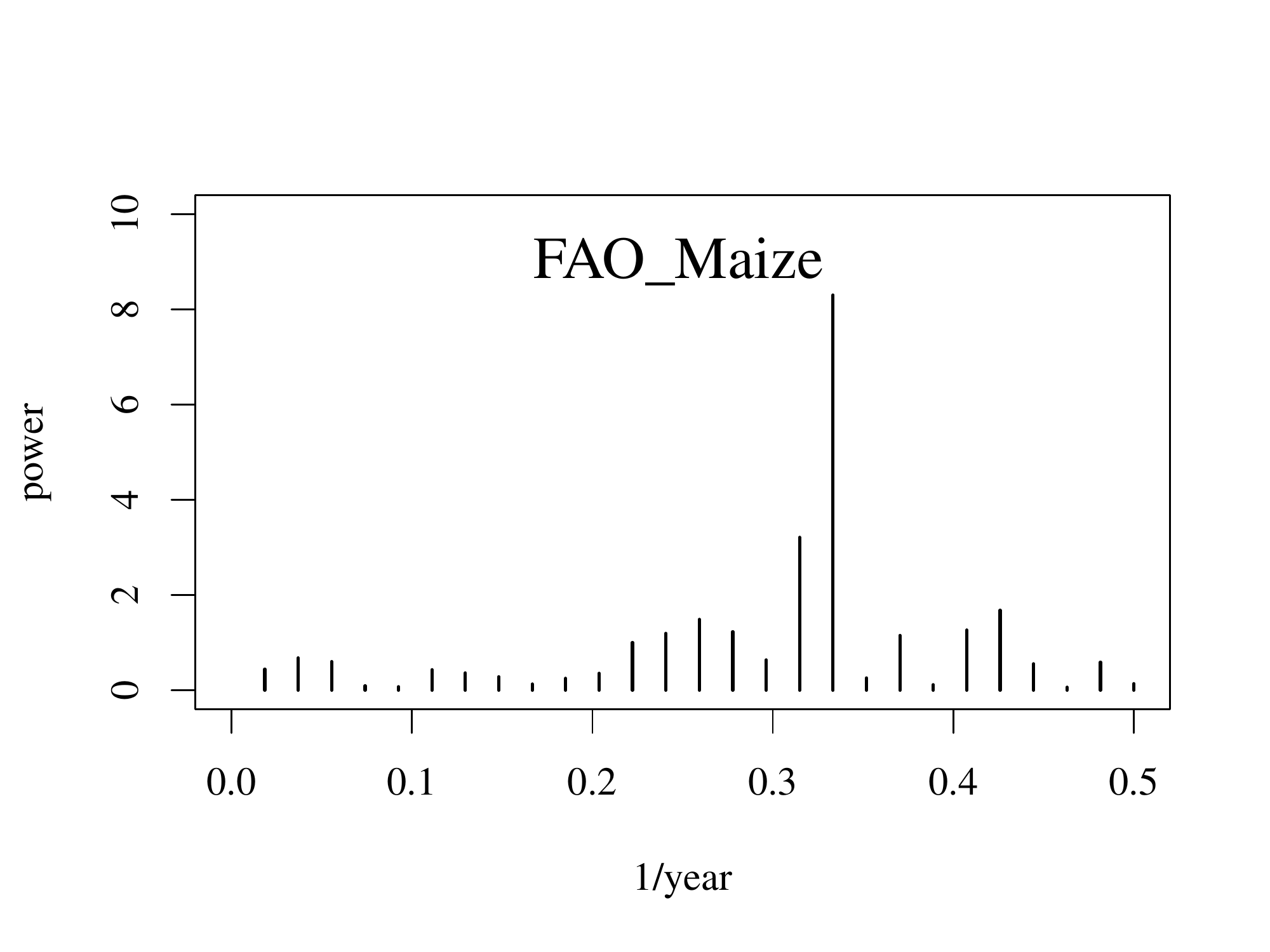} & \includegraphics[width=6cm]{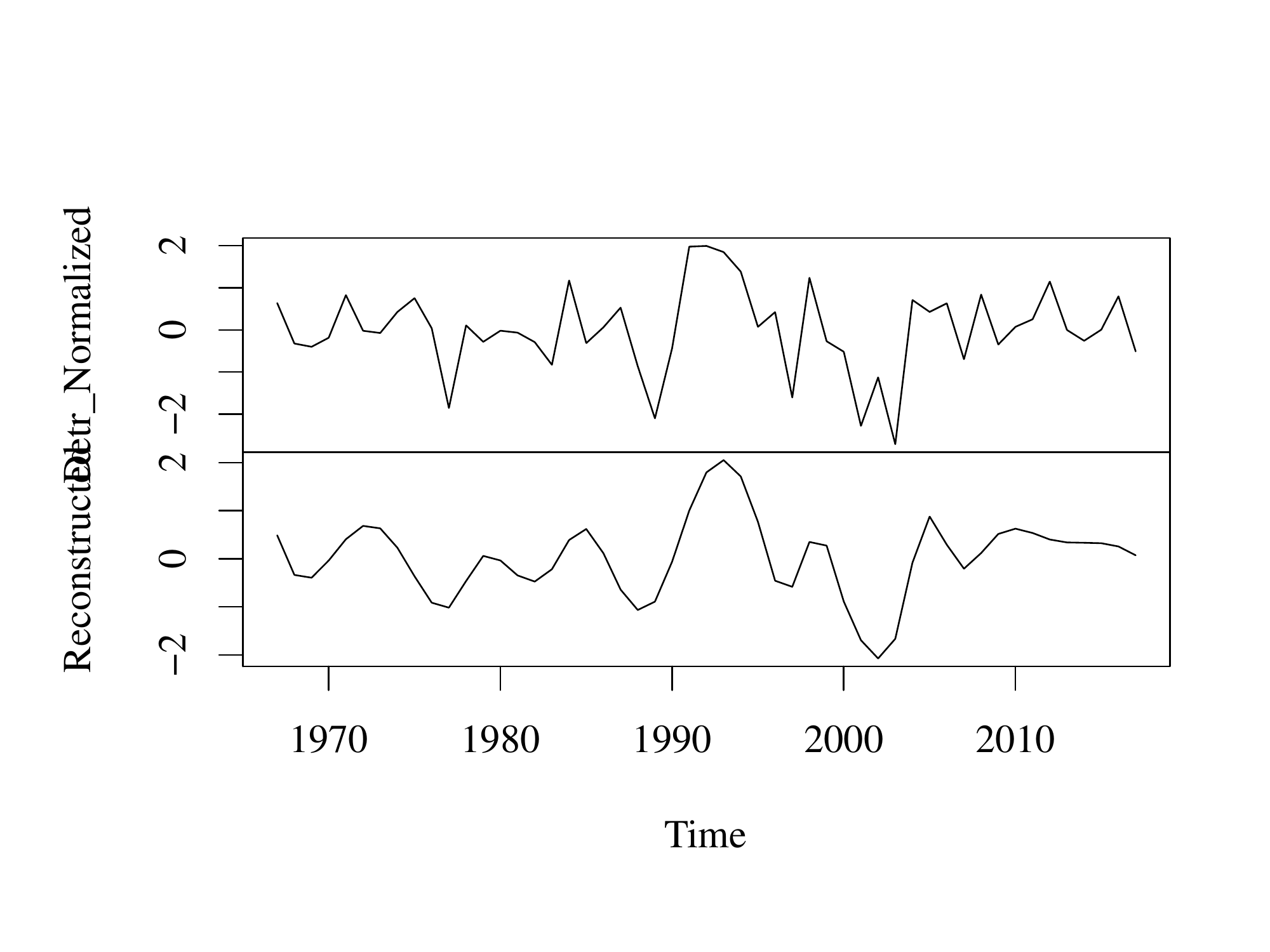} & \includegraphics[width=6cm]{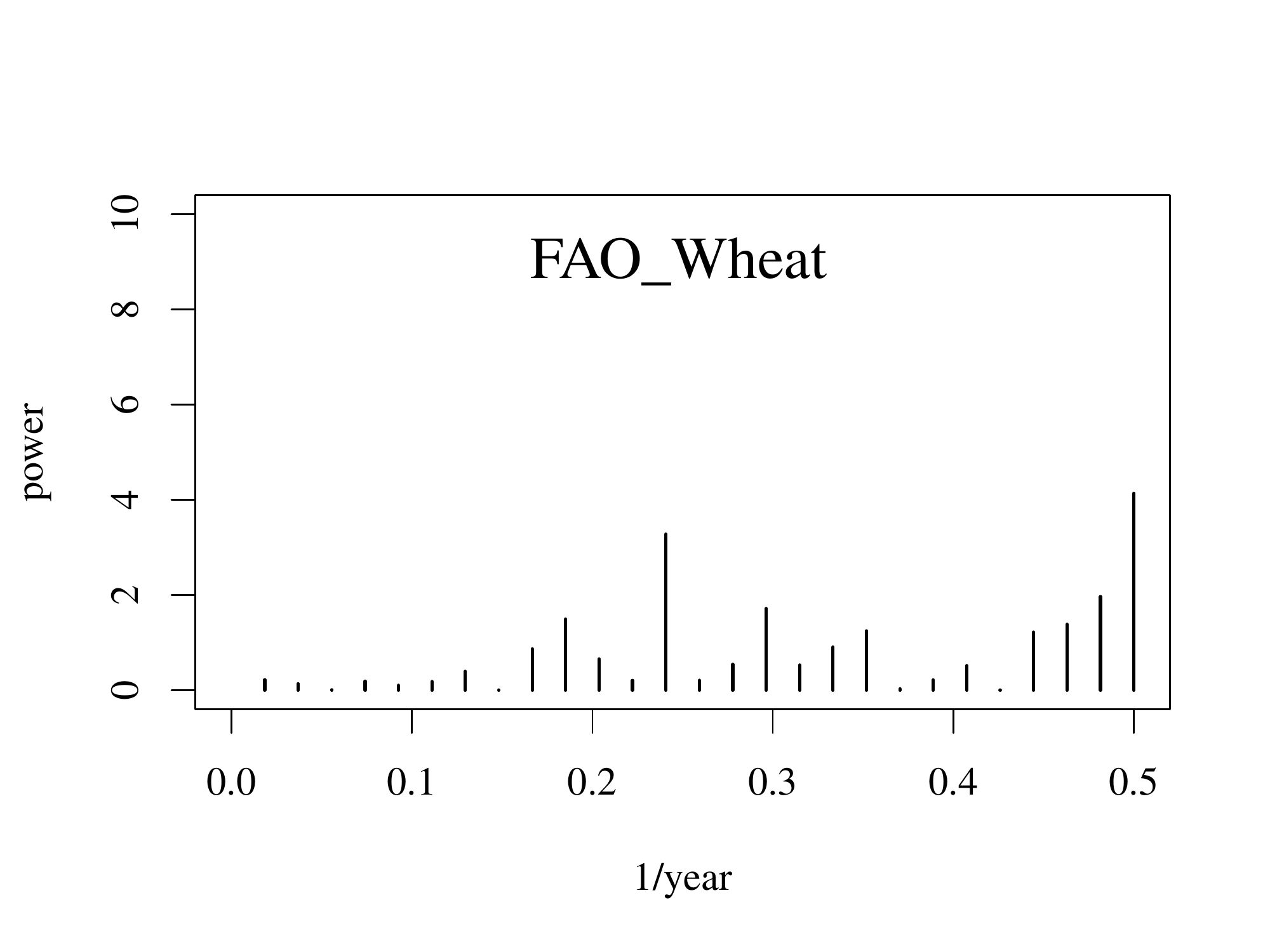}\tabularnewline
\includegraphics[width=6cm]{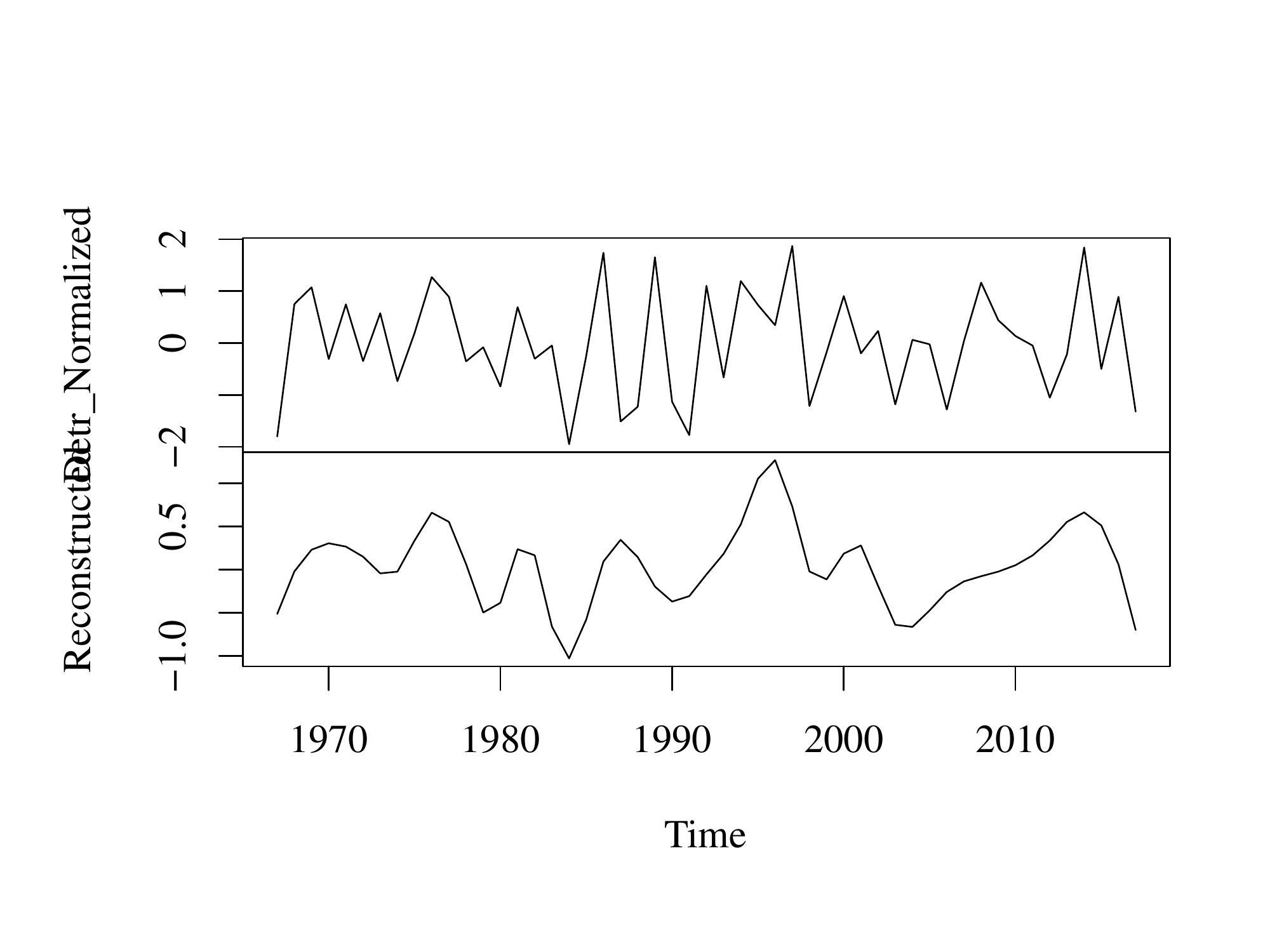} & \includegraphics[width=6cm]{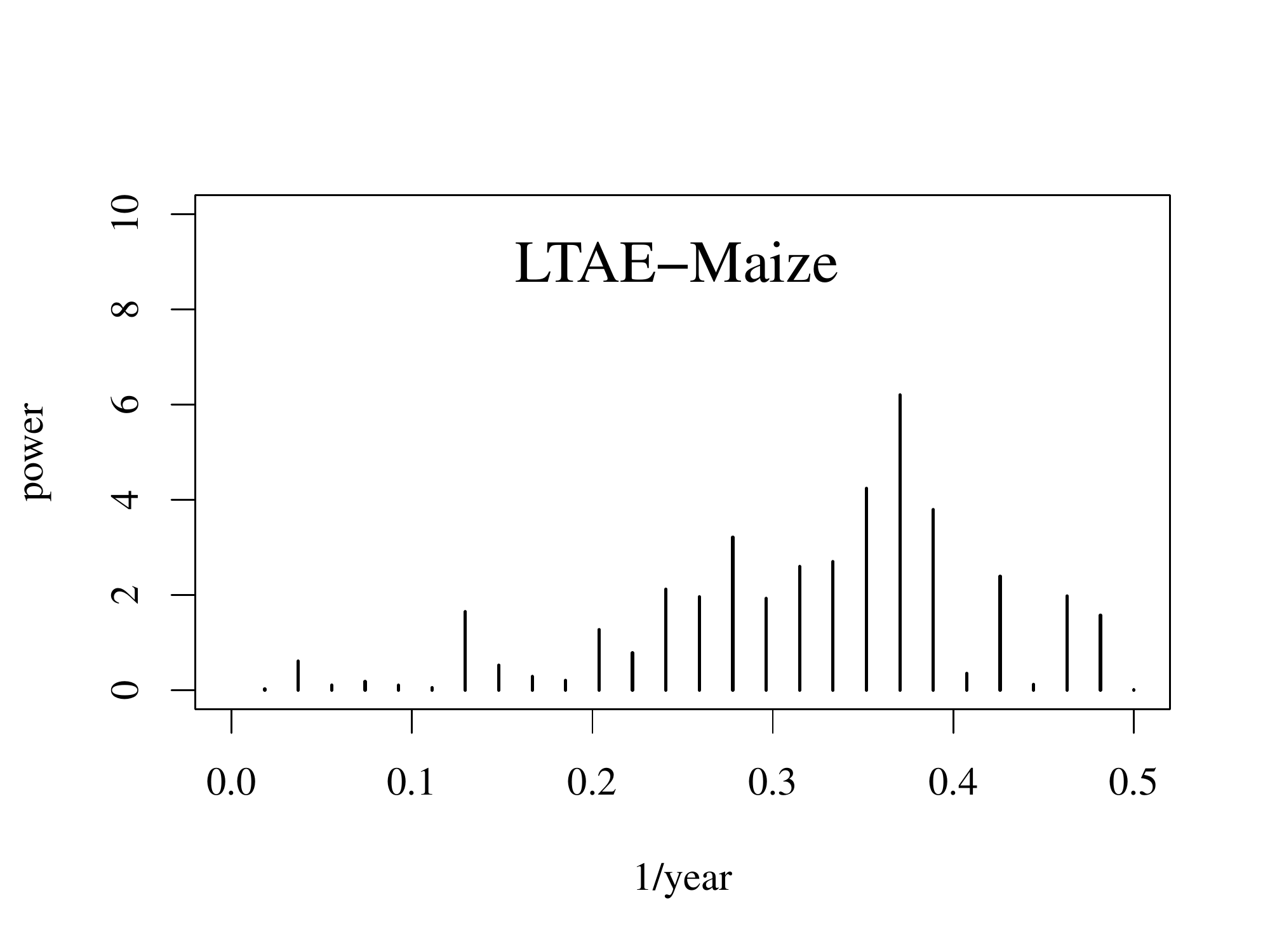} & \includegraphics[width=6cm]{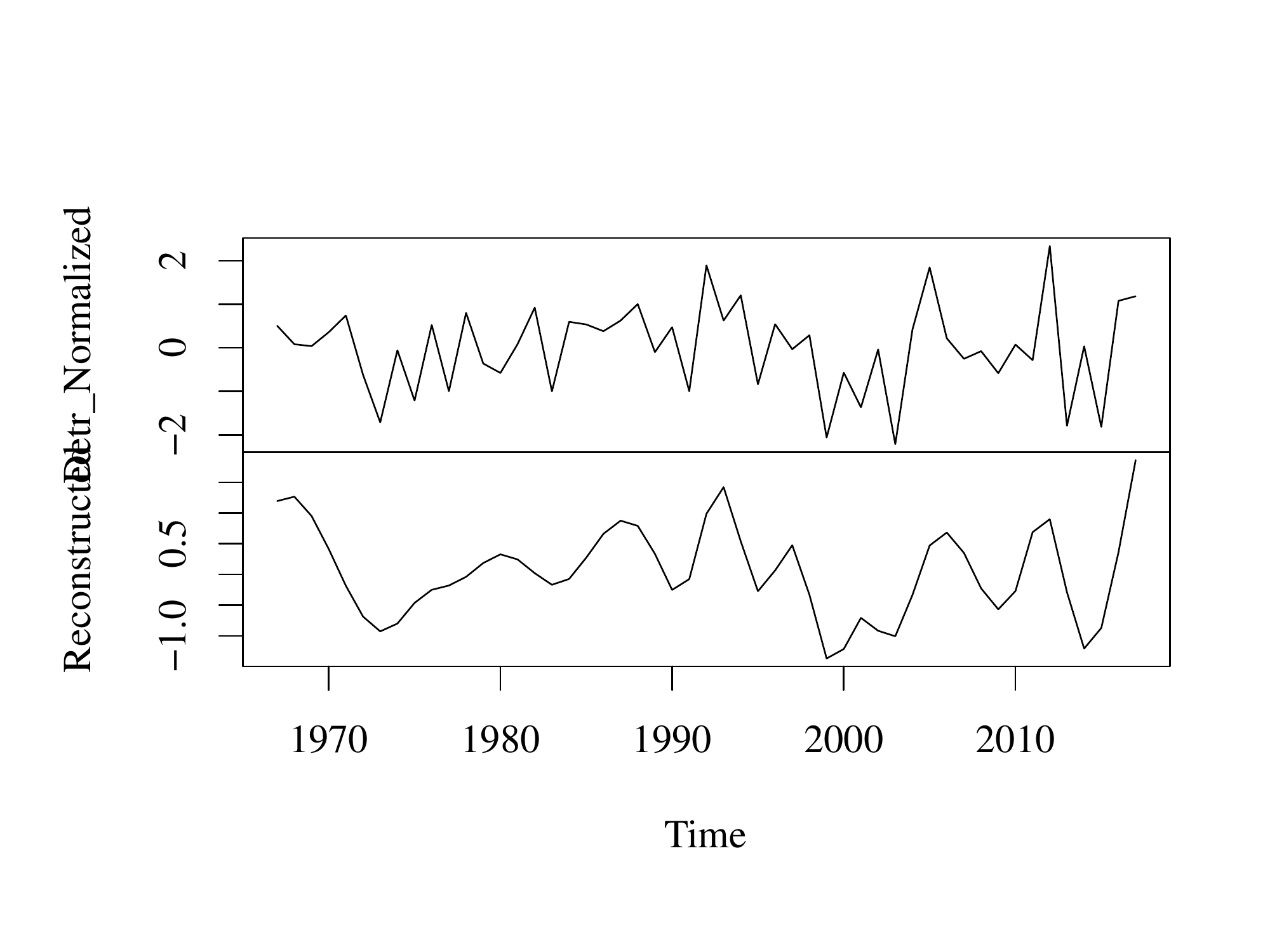} & \includegraphics[width=6cm]{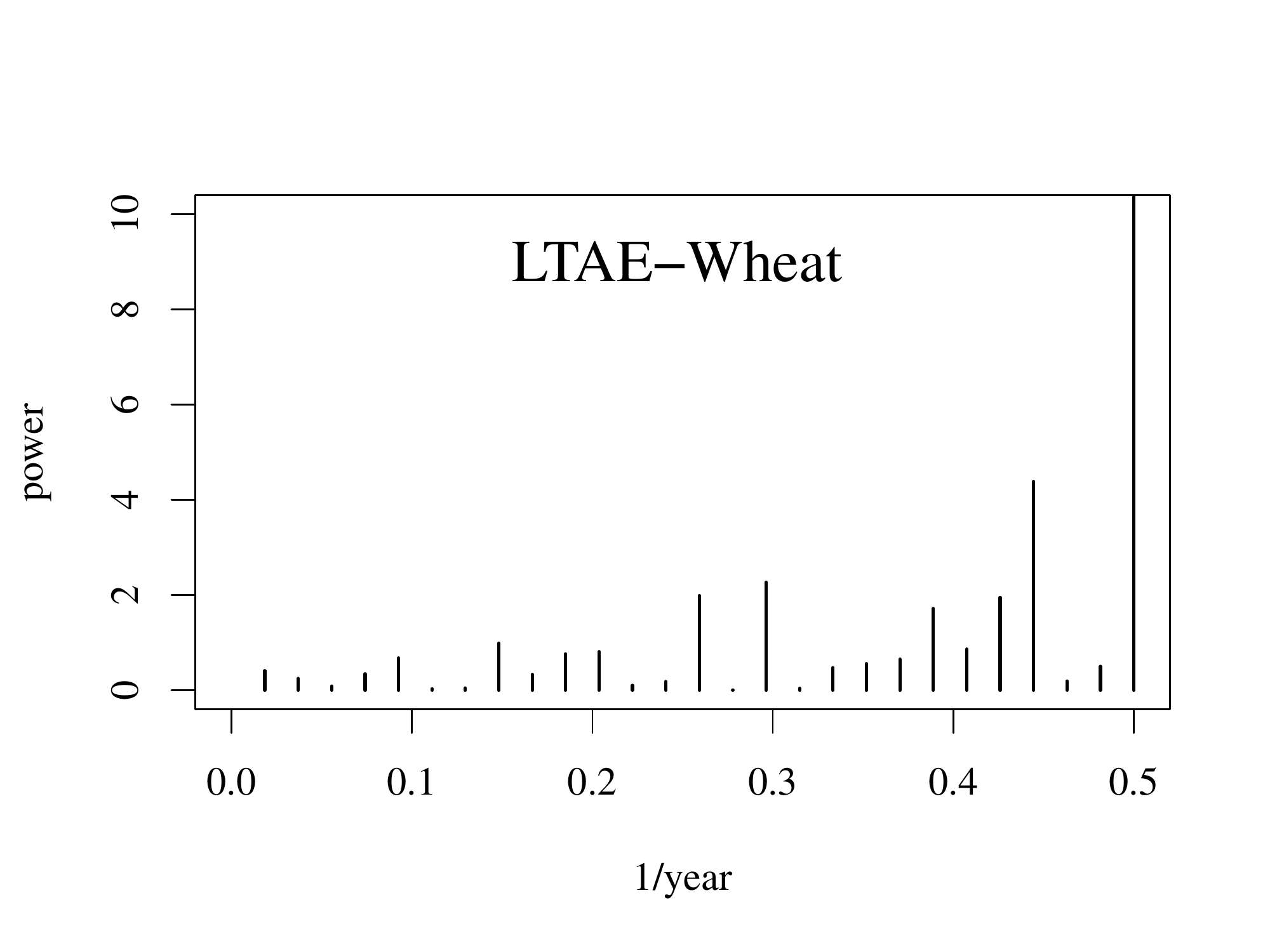}\tabularnewline
\end{tabular}
\end{turn}
\par\end{centering}
\caption{\label{fig:SSD-FFT}EMD-based noise identification}
\end{figure}

In figure \ref{fig:SSD-FFT} the detrended signals are shown together
with the signal reconstructed from EMD excluding the first component,
whose FFT is shown on the right.

This was not possible for SSN where first component incorporates the
main oscillation, how it is evidenced from FFT. SSN and WeMO preliminary
annual average seems in fact to be sufficient to filter out higher
frequencies in the first, being much less effective in the second,
so that the first component of EMD is still significantly high.

\begin{figure}[h]
\begin{centering}
\begin{turn}{90}
\begin{tabular}[t]{cccc}
\includegraphics[width=6cm]{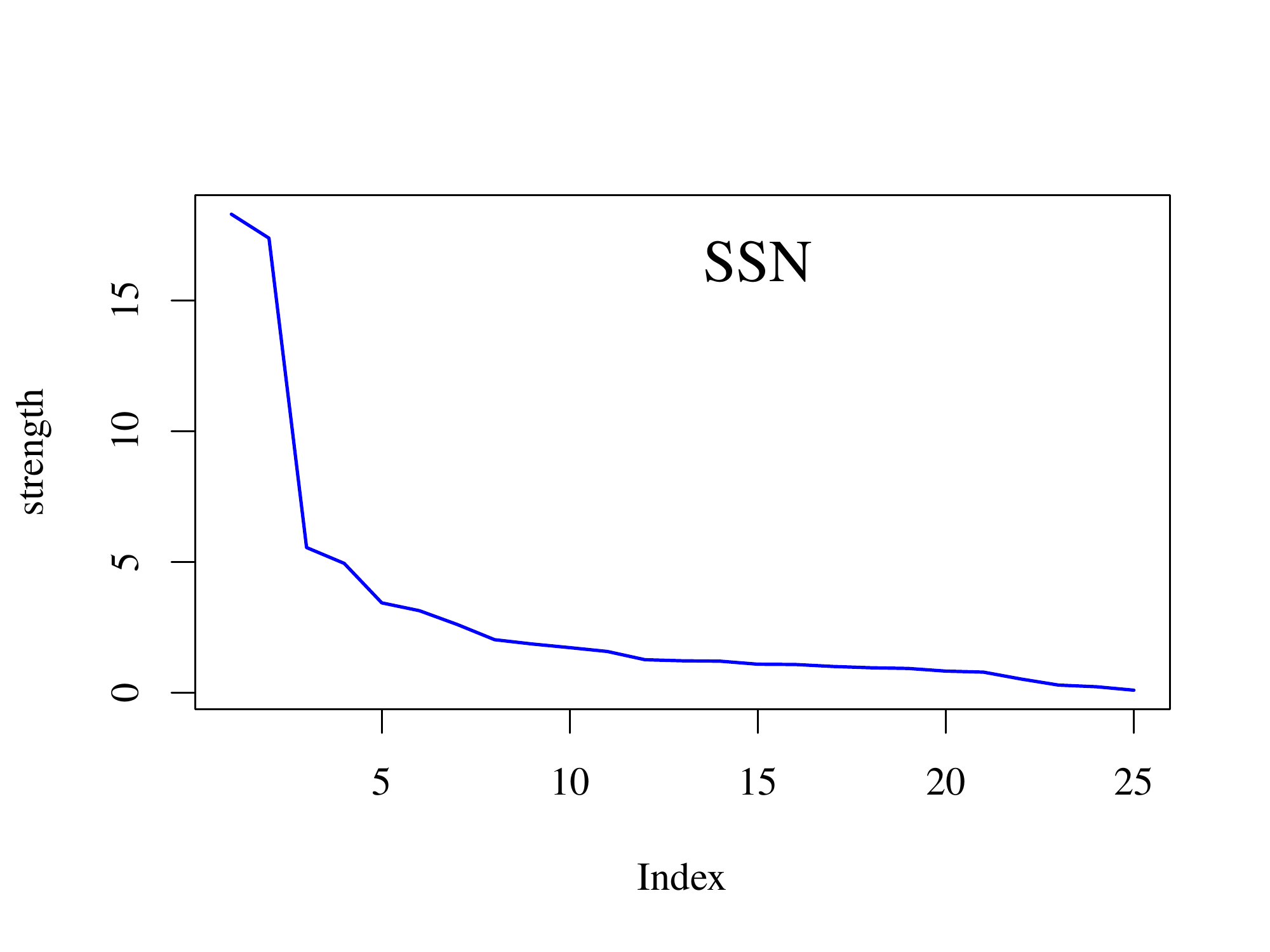} & \includegraphics[width=6cm]{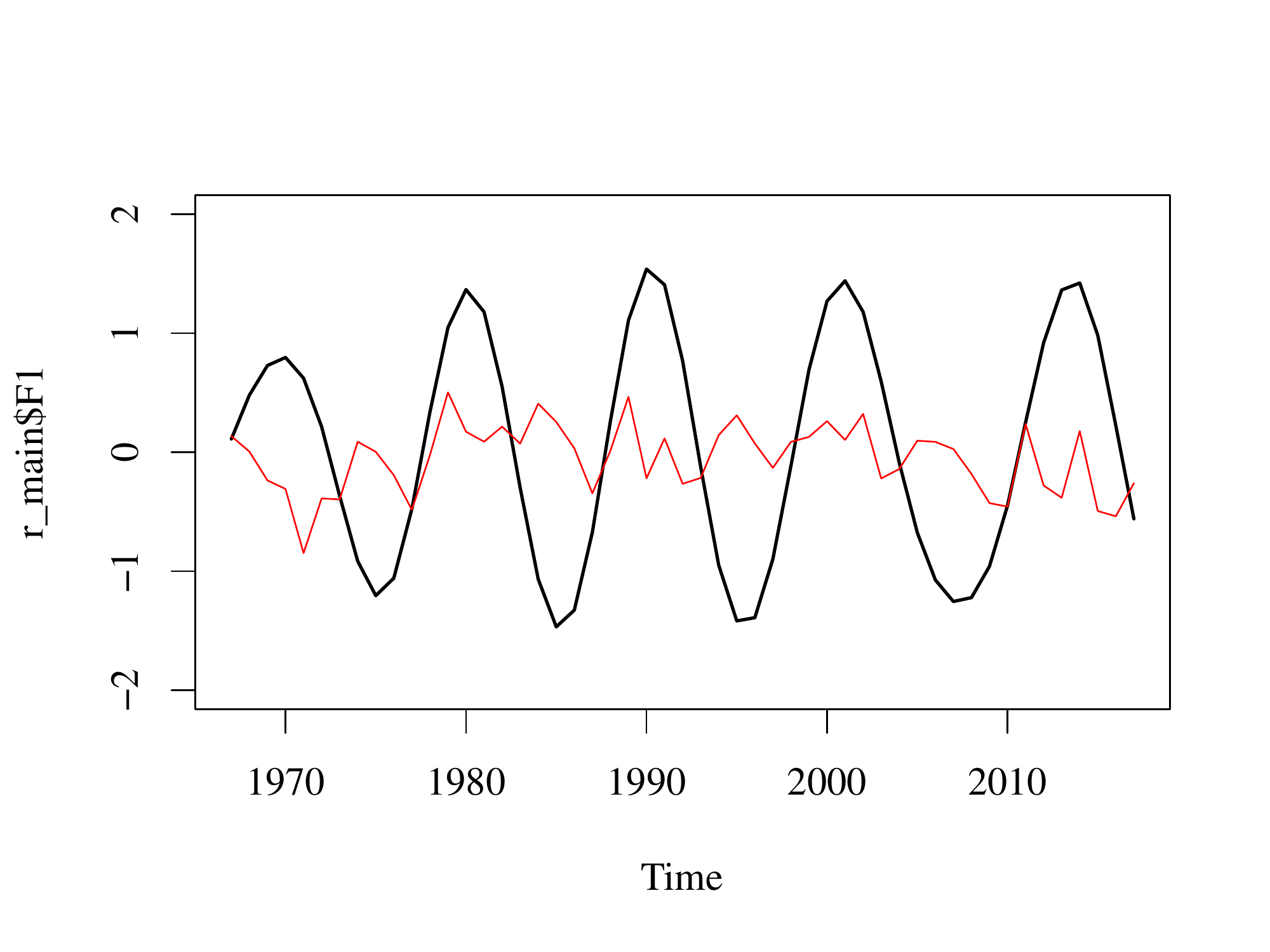} & \includegraphics[width=6cm]{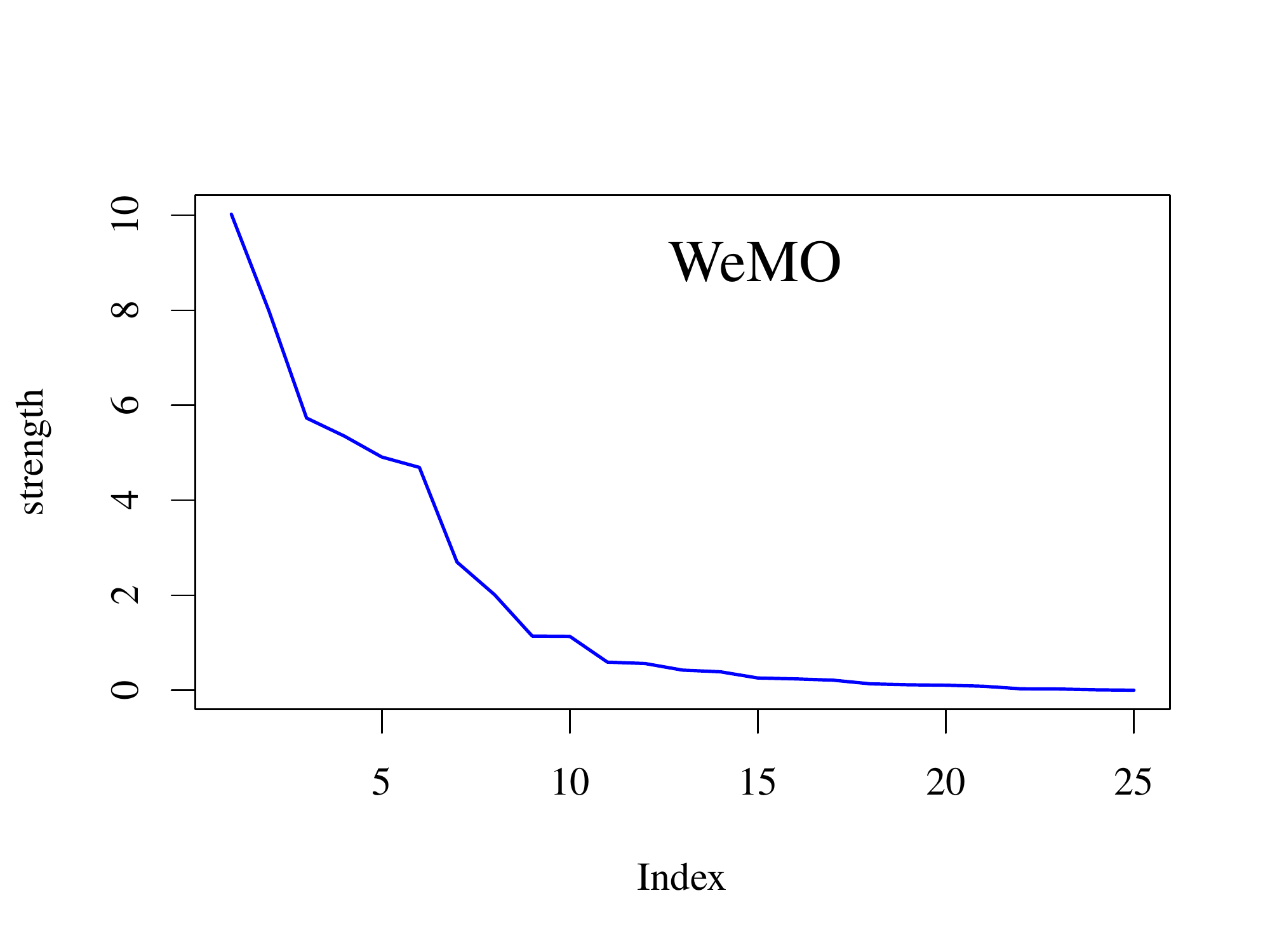} & \includegraphics[width=6cm]{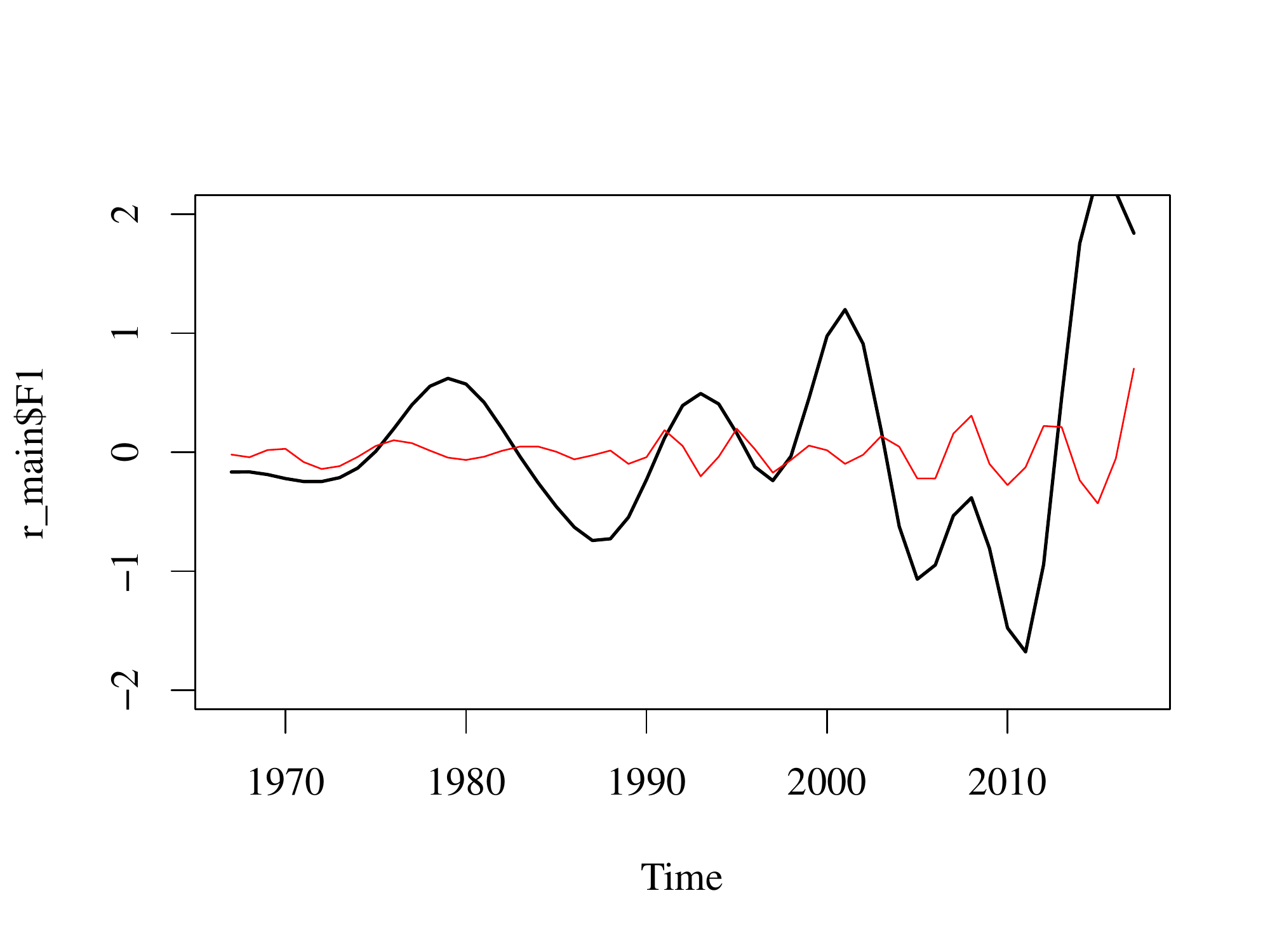}\tabularnewline
\includegraphics[width=6cm]{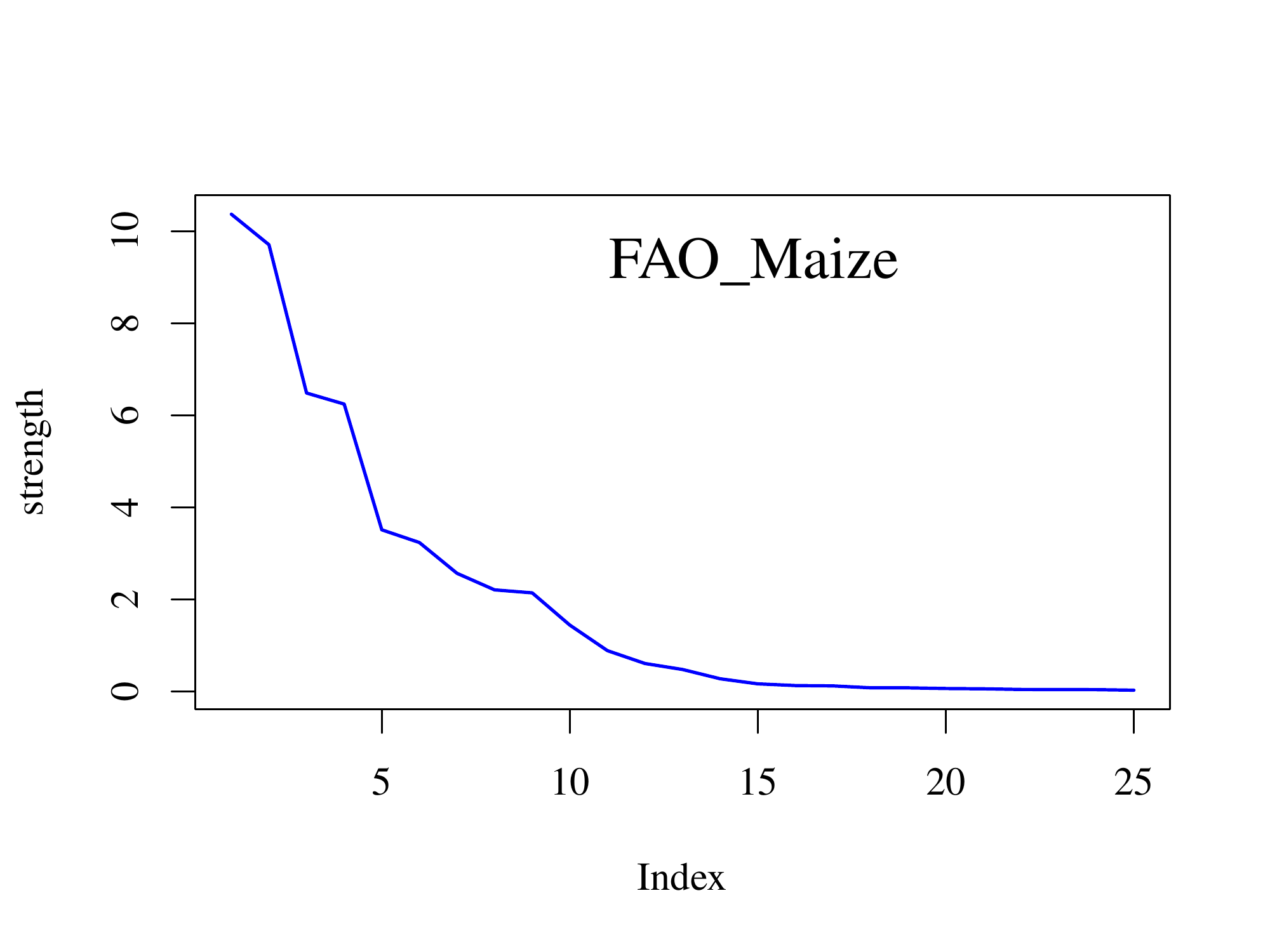} & \includegraphics[width=6cm]{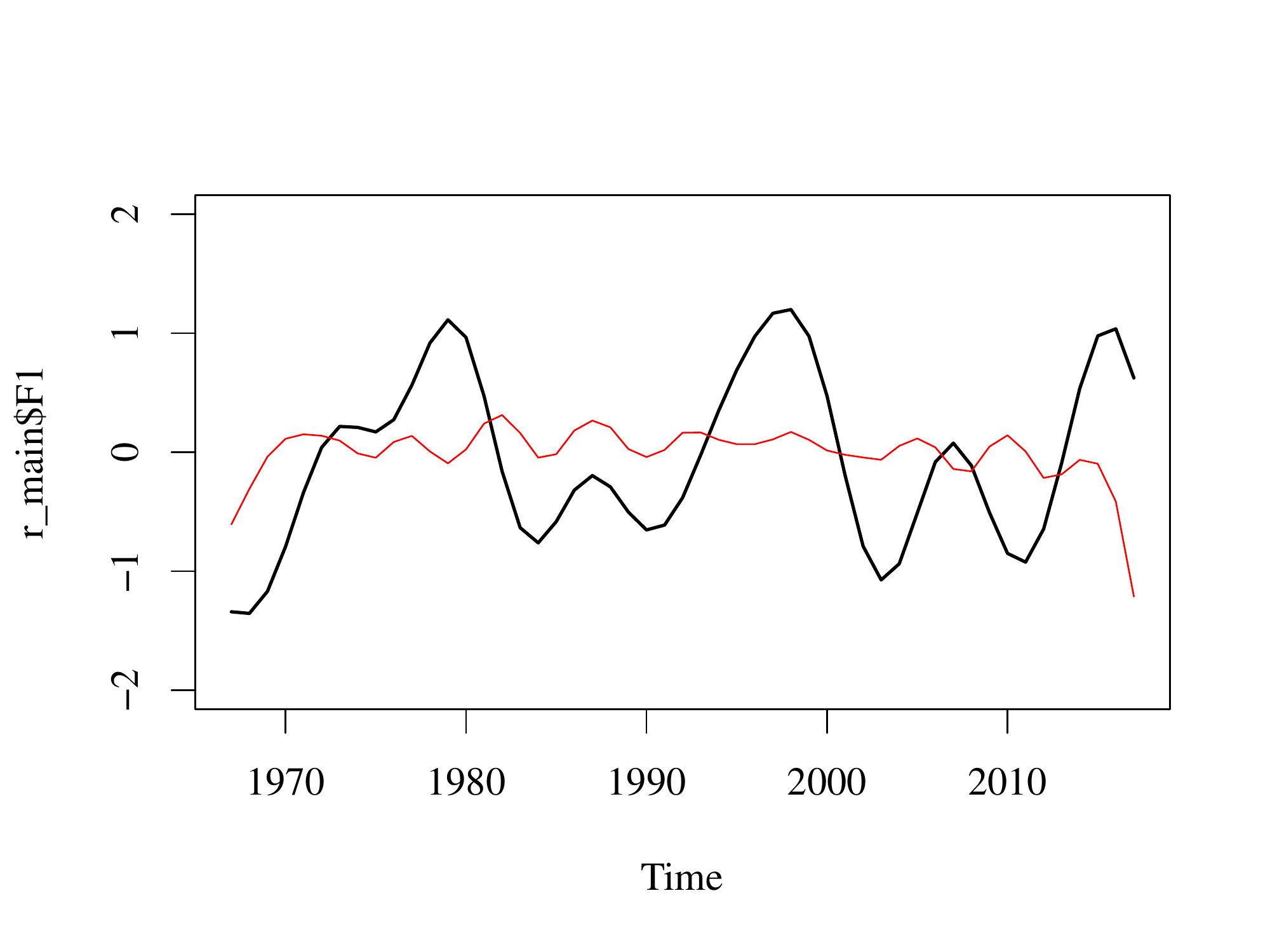} & \includegraphics[width=6cm]{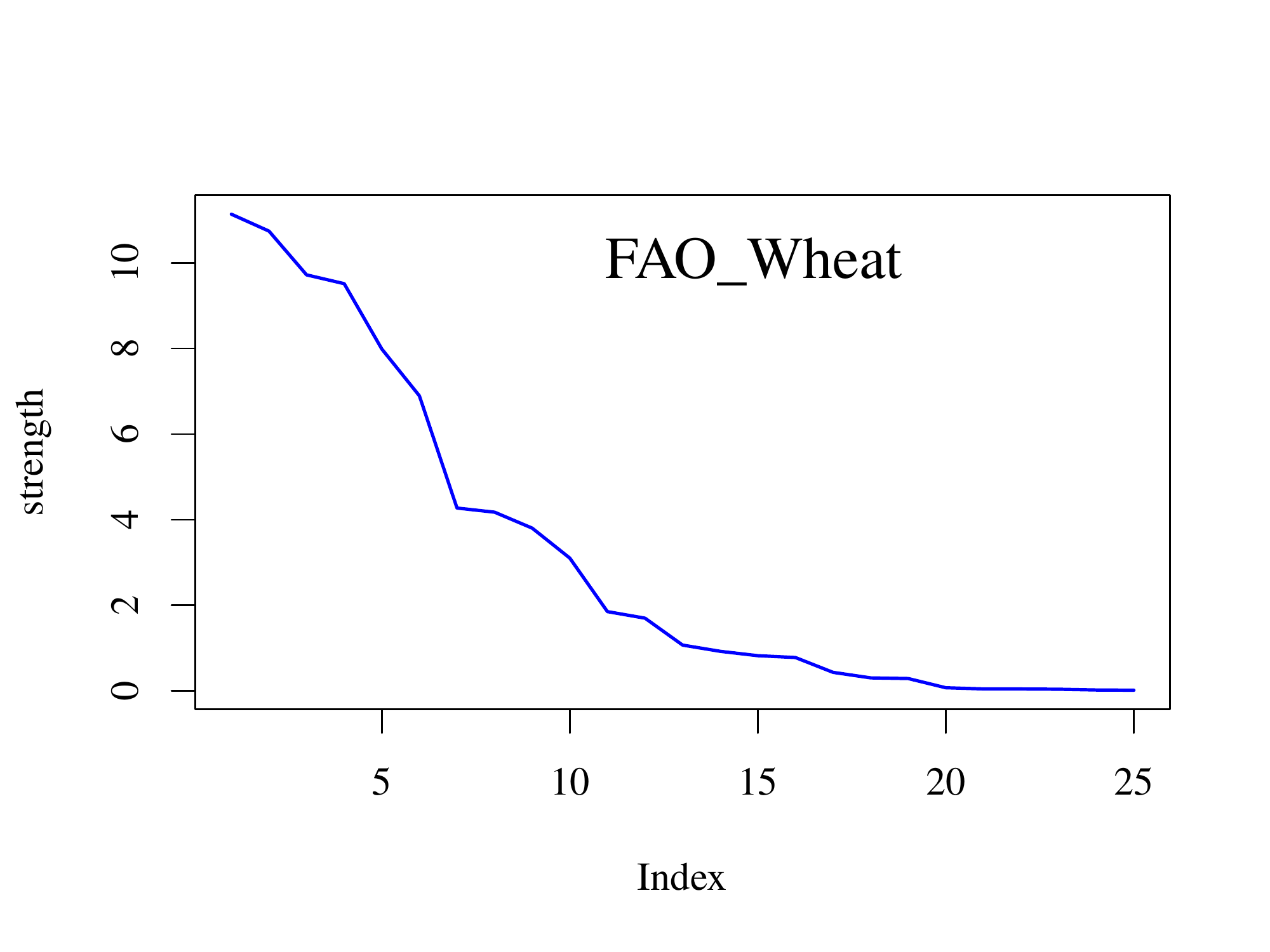} & \includegraphics[width=6cm]{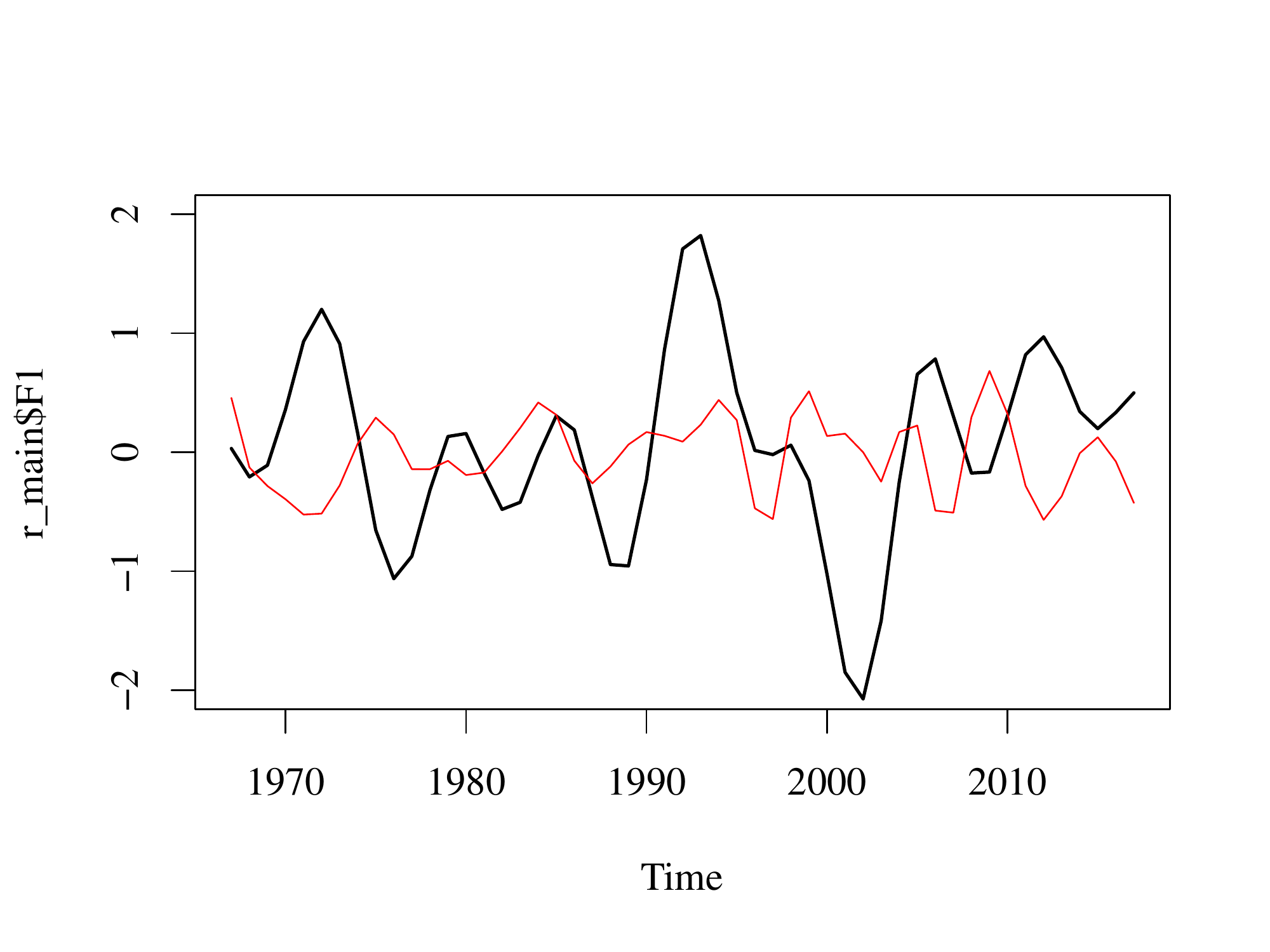}\tabularnewline
\includegraphics[width=6cm]{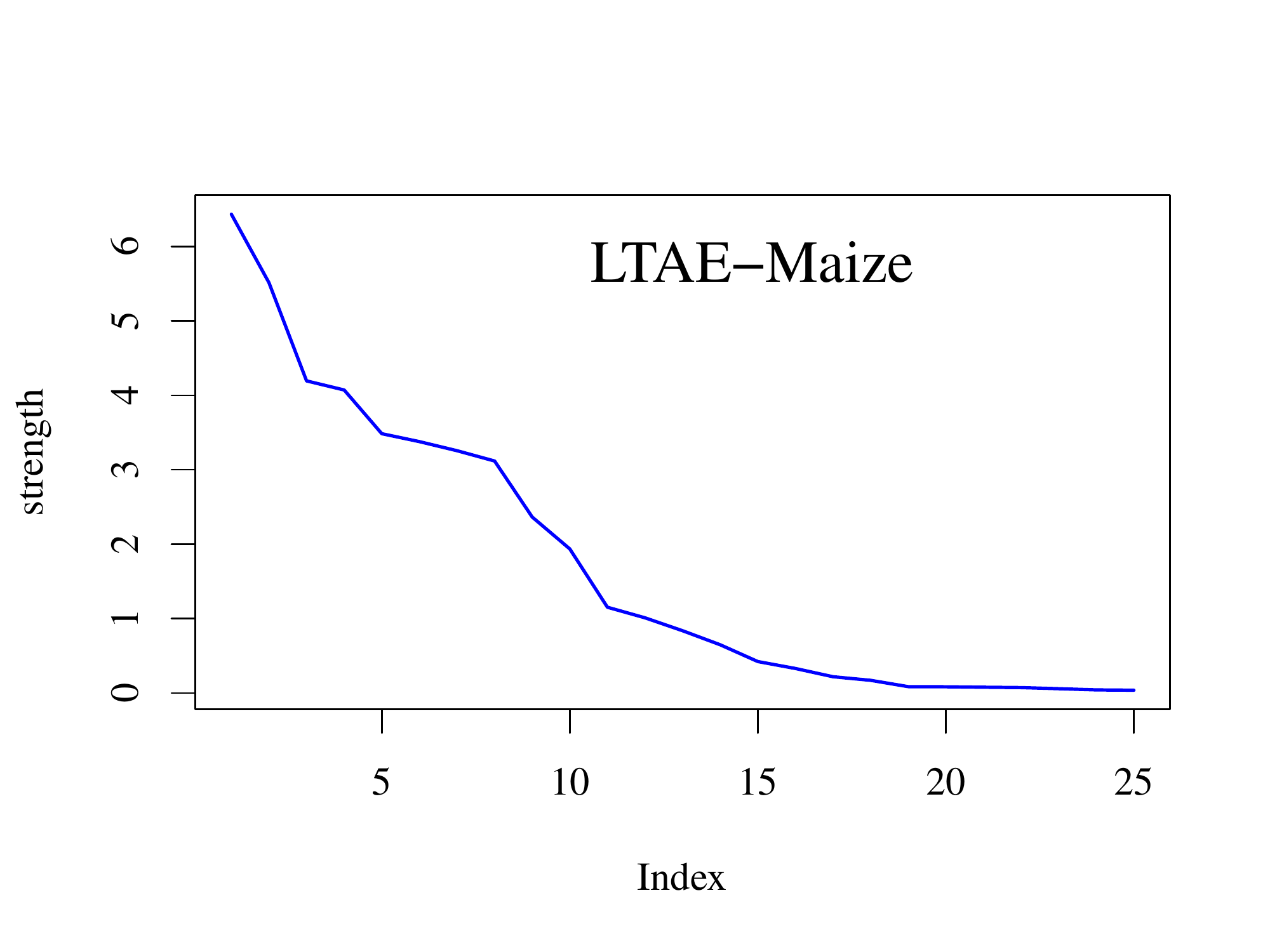} & \includegraphics[width=6cm]{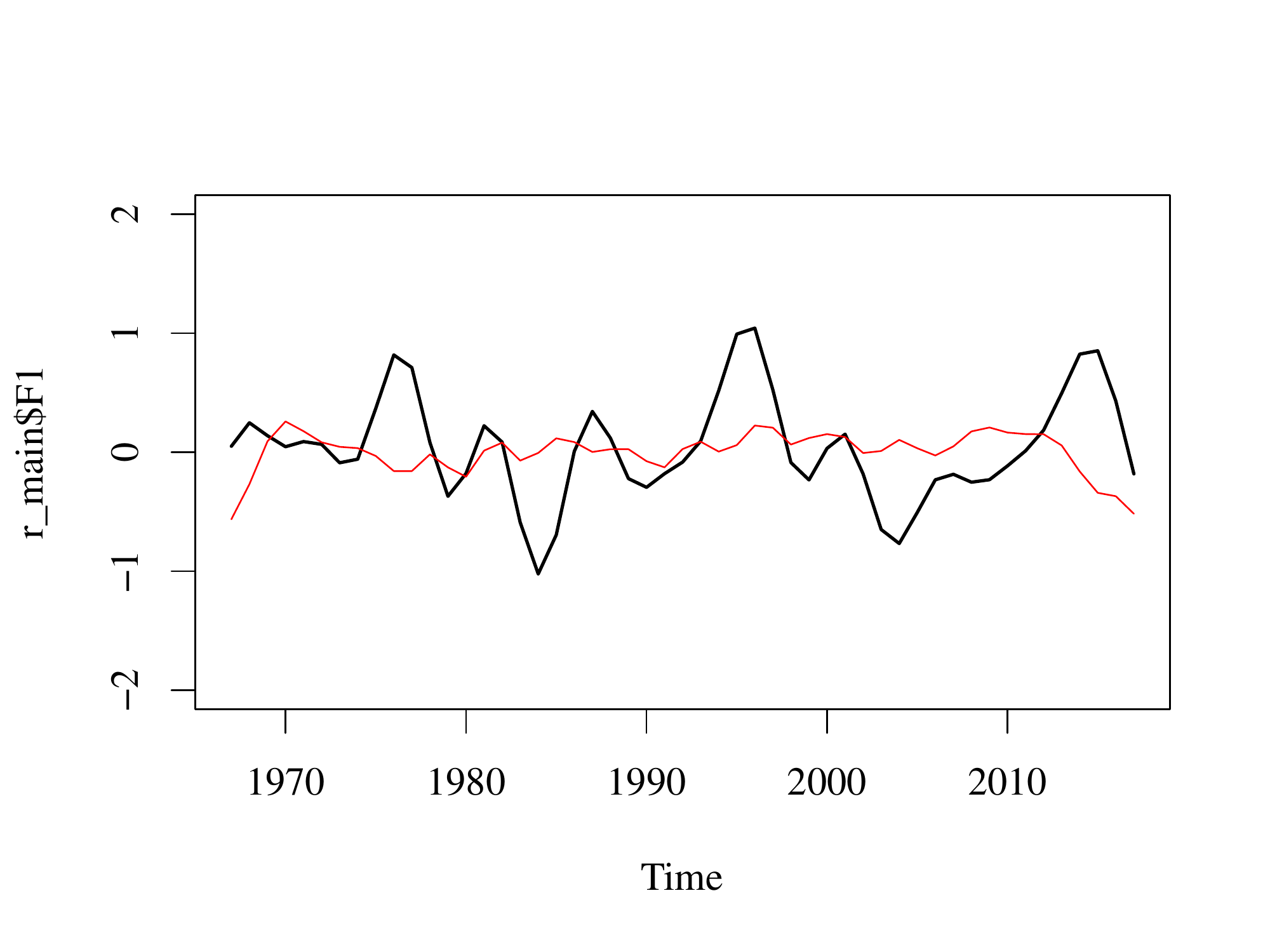} & \includegraphics[width=6cm]{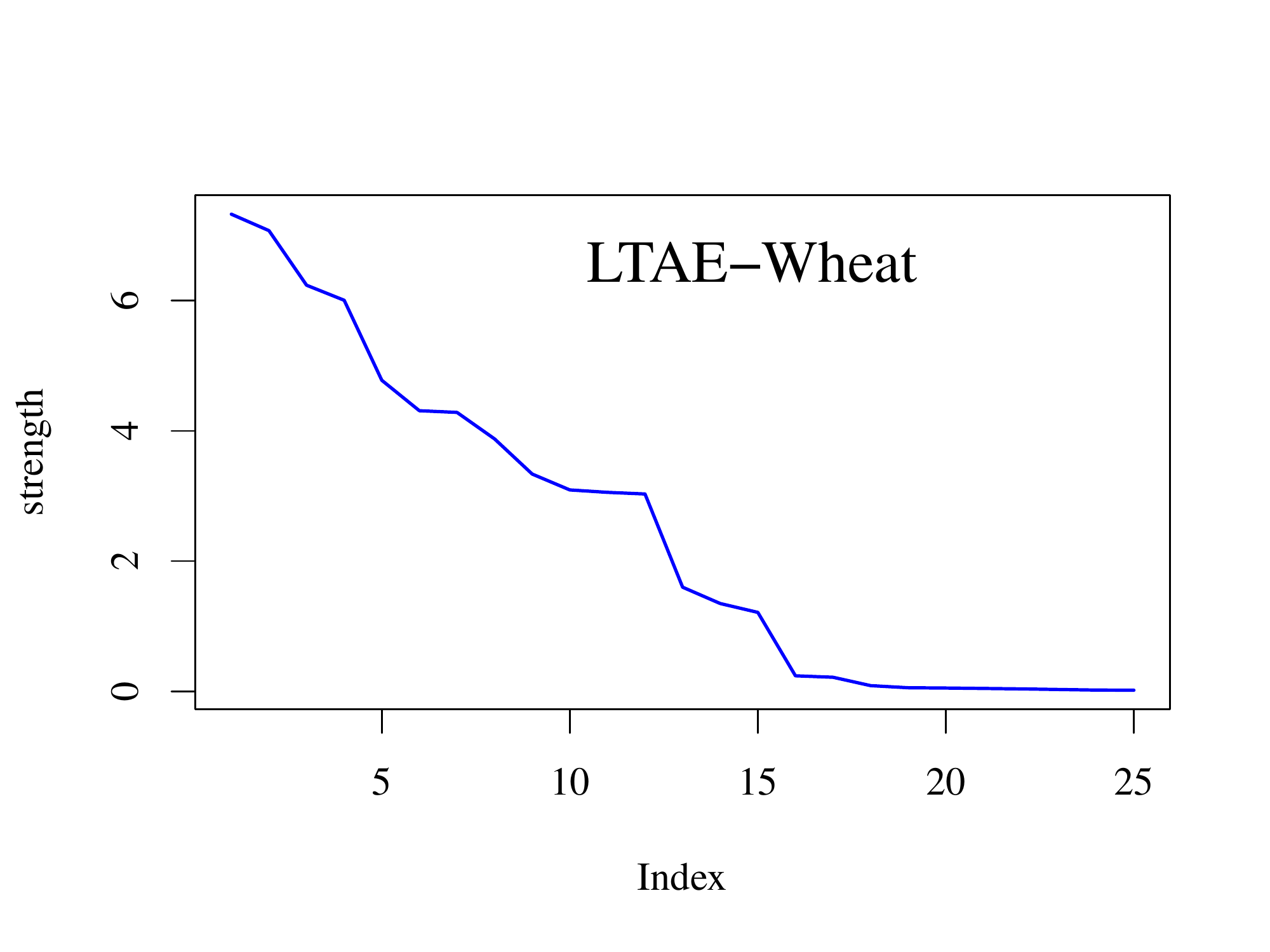} & \includegraphics[width=6cm]{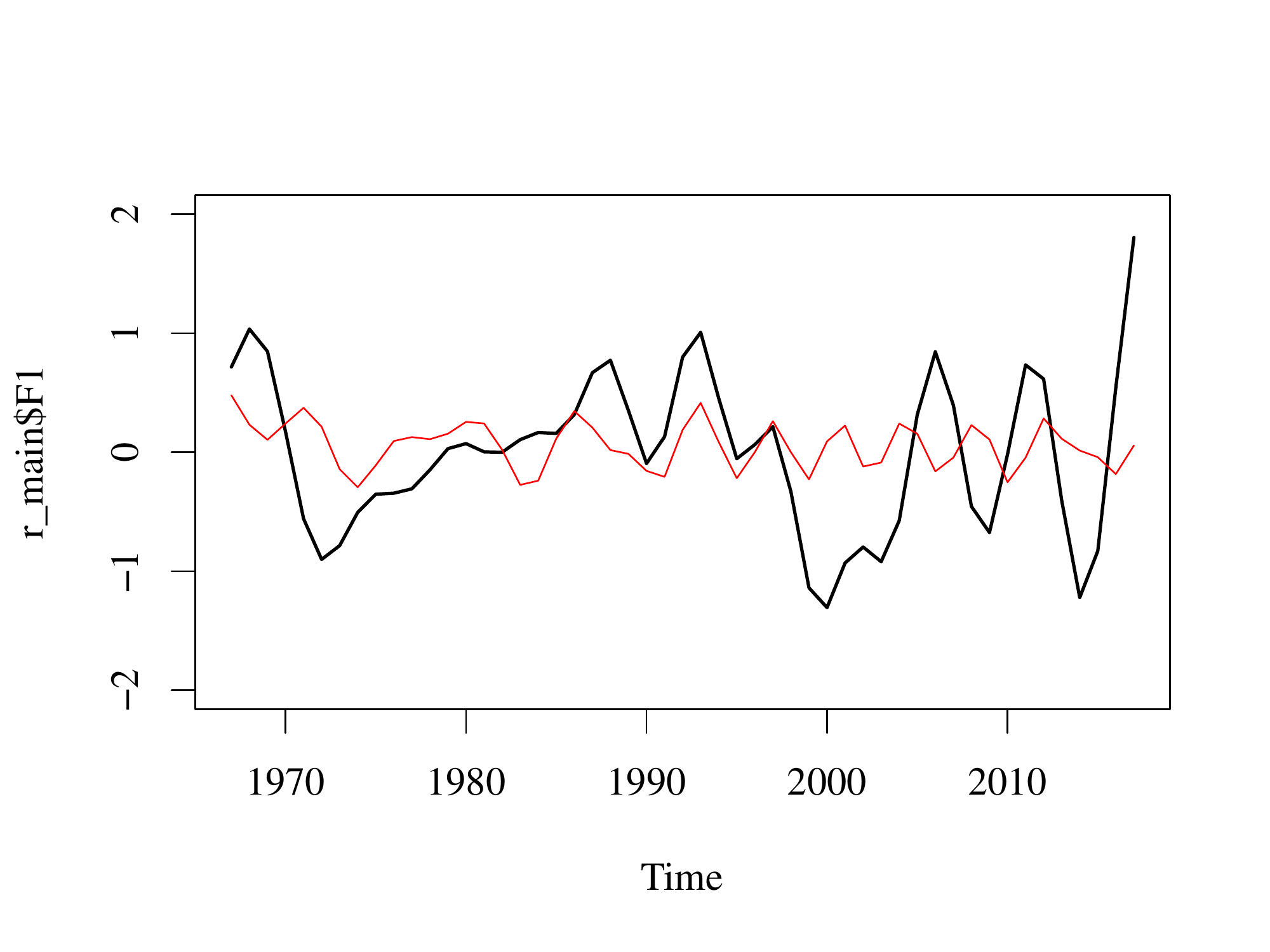}\tabularnewline
\end{tabular}
\end{turn}
\par\end{centering}
\caption{\label{fig:SSA}SSA-based noise identification}
\end{figure}

A similar procedure has been performed by SSA, to identify low amplitude
noise components with the aid of eigenvalues distribution (left plot
in figure \ref{fig:SSA}); the same figure reports denoized signals
together with the filtered noise.

\begin{figure}[h]
\begin{centering}
\begin{tabular}{cc}
\includegraphics[width=8cm]{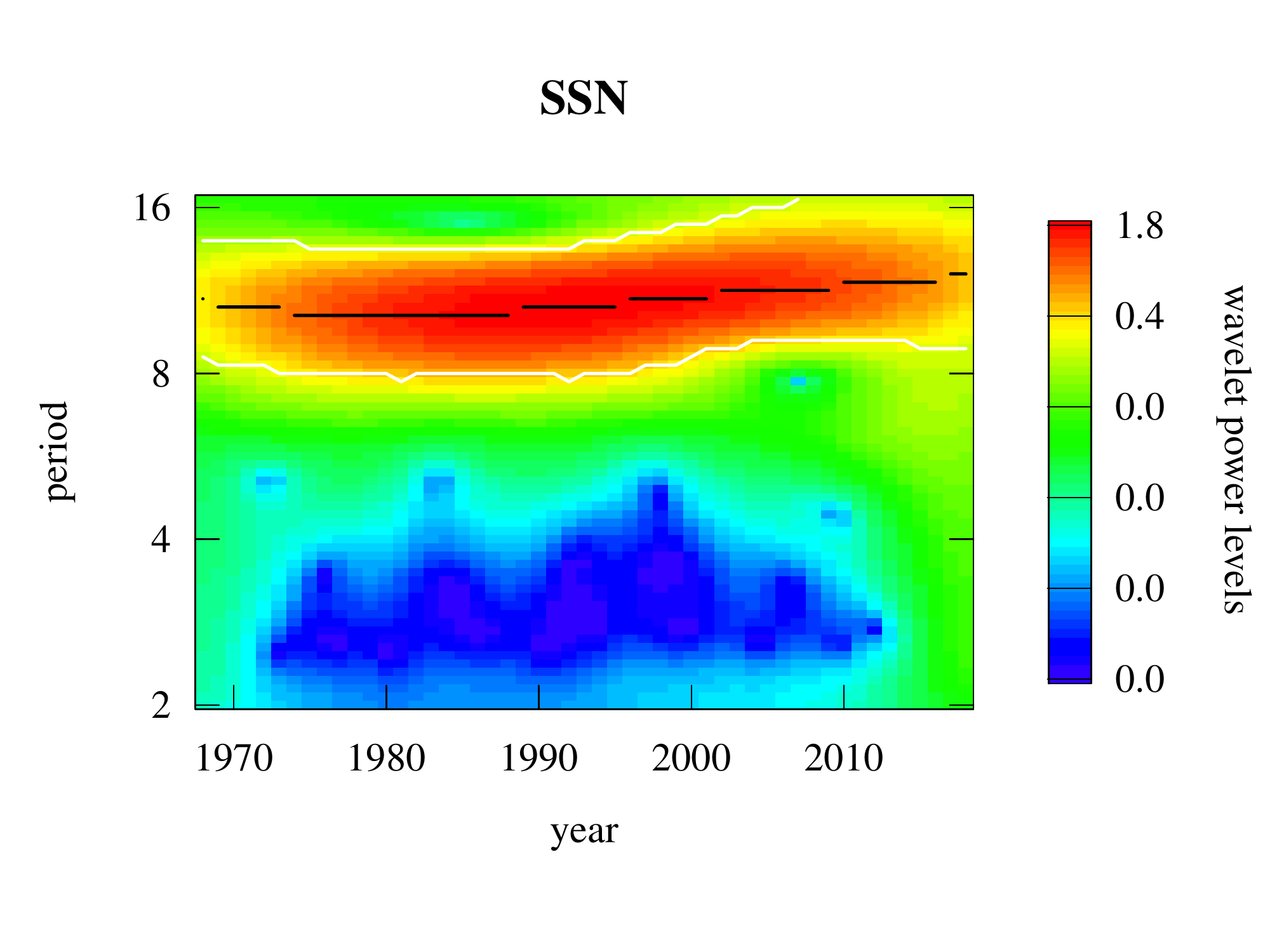} & \includegraphics[width=8cm]{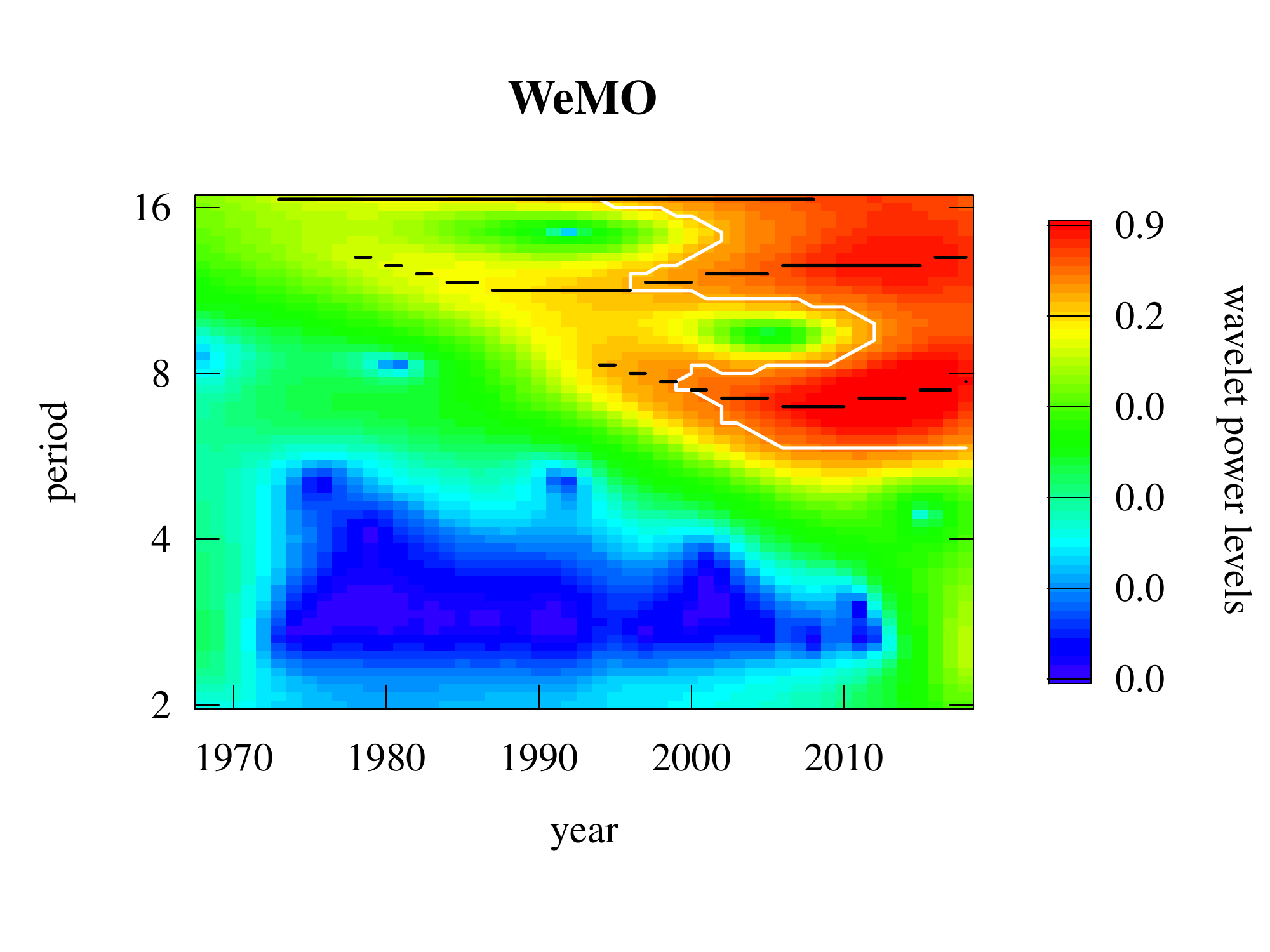}\tabularnewline
\includegraphics[width=8cm]{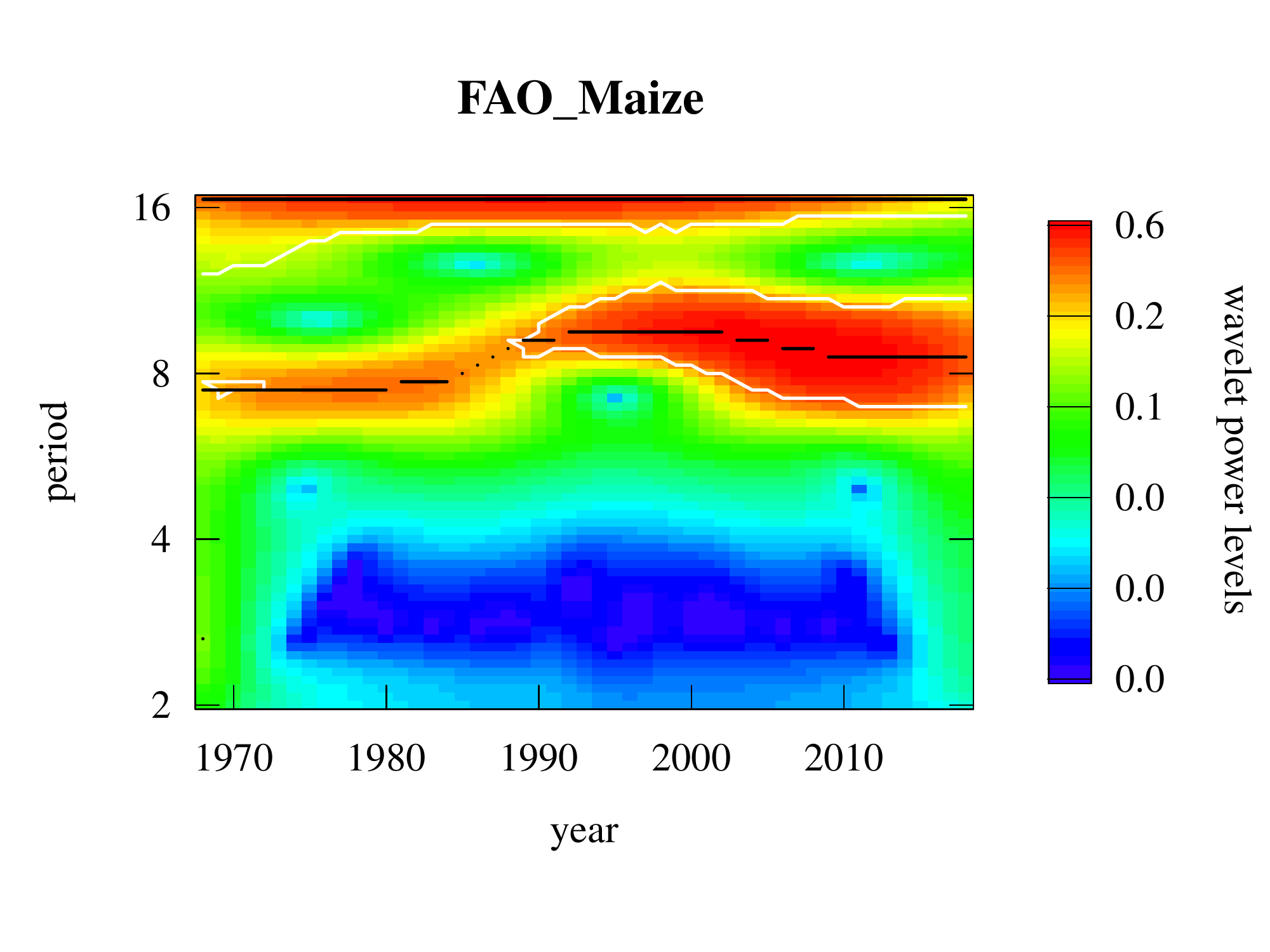} & \includegraphics[width=8cm]{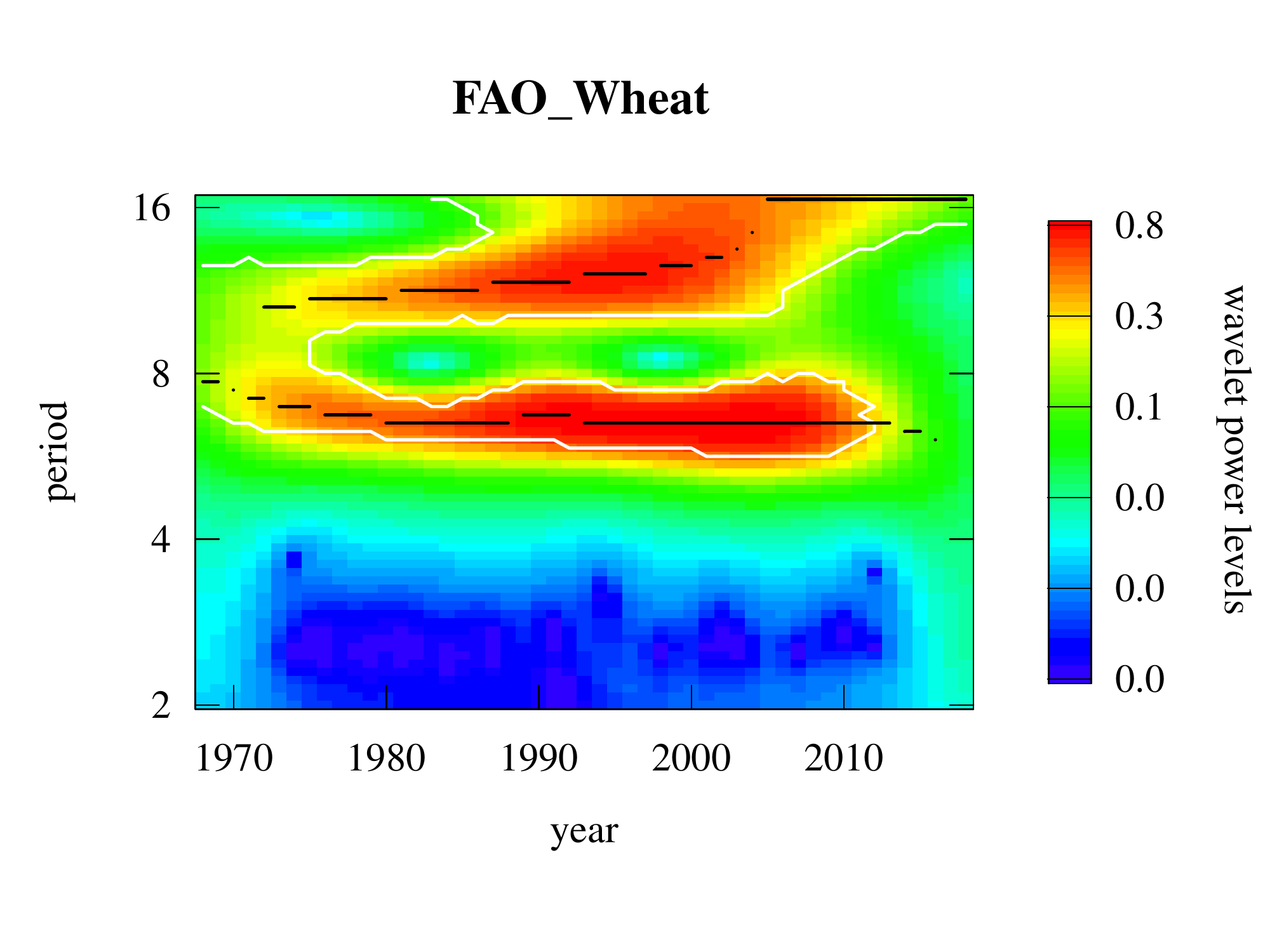}\tabularnewline
\includegraphics[width=8cm]{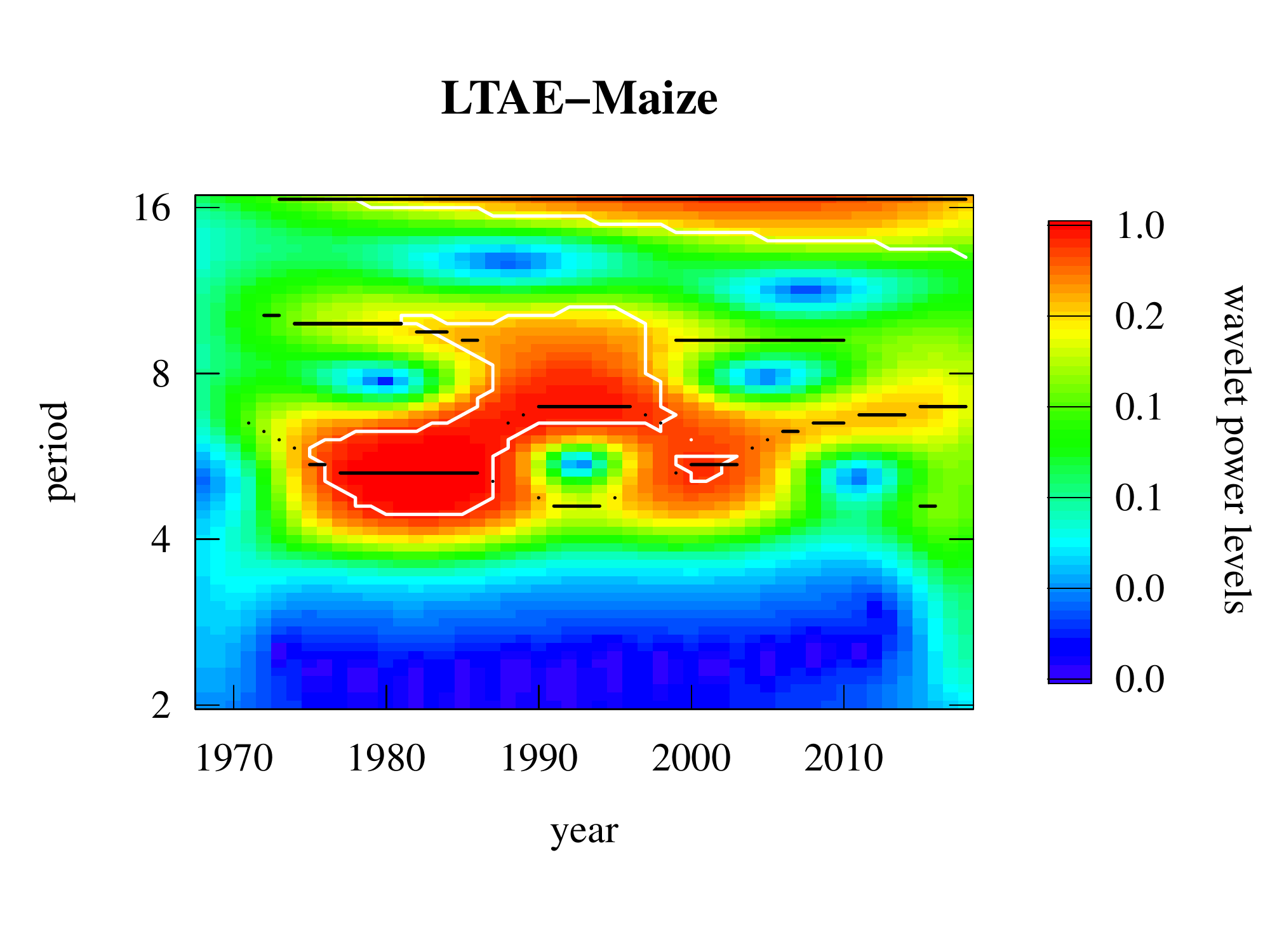} & \includegraphics[width=8cm]{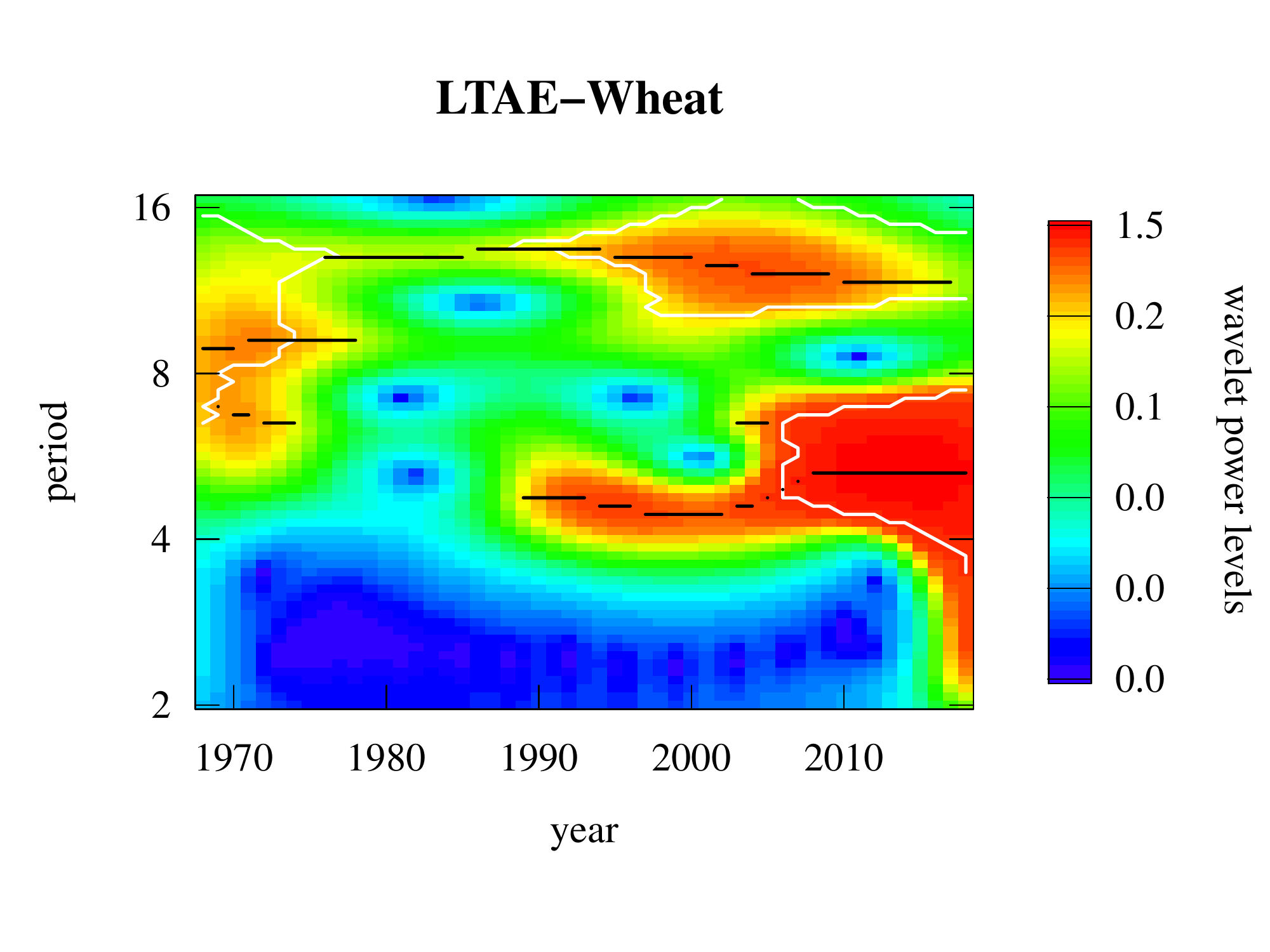}\tabularnewline
\end{tabular}
\par\end{centering}
\caption{\label{fig:WT}Wavelet transform of detrended normalized EMD-denoised
series}
\end{figure}

On detrended signals, the WA allows to locate frequencies in different
time domains making things more readable (figure \ref{fig:WT}).\textcolor{black}{{}
In the periodograms it is possible to notice an arc-shaped region,
which reflects the interval of validity of wavelet analysis: the greater
is the period, the narrow is the interval of data (year window). Black
and white lines represent 90\% and 95\% confidence limits of the power
values (see side bar).}

\textcolor{black}{From figure it is quite evident a shift in frequency
moving from the 11 years of 1970-2000 to higher values in the last
two decades. More evident is in WeMO the passage from a rather quiescent
zone (small amplitude fluctuations) of the first 3 decades to oscillations
within the range of 7-16 years for the last two decades. Such a behavior
can be slightly recognized in FAO-Maise, wheres FAO-Wheat slower oscillations
seems to follow more SSN, apart of a more rapid fluctuation around
7 years which seems to disappear in the last decade. Fluctuations
in LTAE-Maize with a period of 5-9 years evident in 1980-2000, whereas
in LTAE-Wheat strong fluctuations around 4-7 years are appearing in
the last two decades.}

Other suggestions are searched with the help of cross-wavelet analysis,
which put in evidence how SSN 11-year fluctuation and its shift can
be detected in almost every other signals. Also WeMO increase of fluctuation
with a large spectrum (4-16 years) seems to affect yield both at a
national and local scale.

More difficult is interpreting spectrogram related to comparison of
yield fluctuations, though those of maize seems to be more concentrated
in the past (central decades) while those of wheat seem to increase
in the last decade.
\begin{flushright}
\begin{figure}[ph]
\begin{centering}
\begin{tabular}{>{\centering}p{2cm}>{\centering}p{3cm}>{\centering}p{3cm}>{\centering}p{3cm}>{\centering}p{3cm}>{\centering}p{3cm}}
 & SSN & WeMO & FAO-Maize & FAO-Wheat & LTAE-Maize\tabularnewline
WeMO & \includegraphics[width=3.3cm]{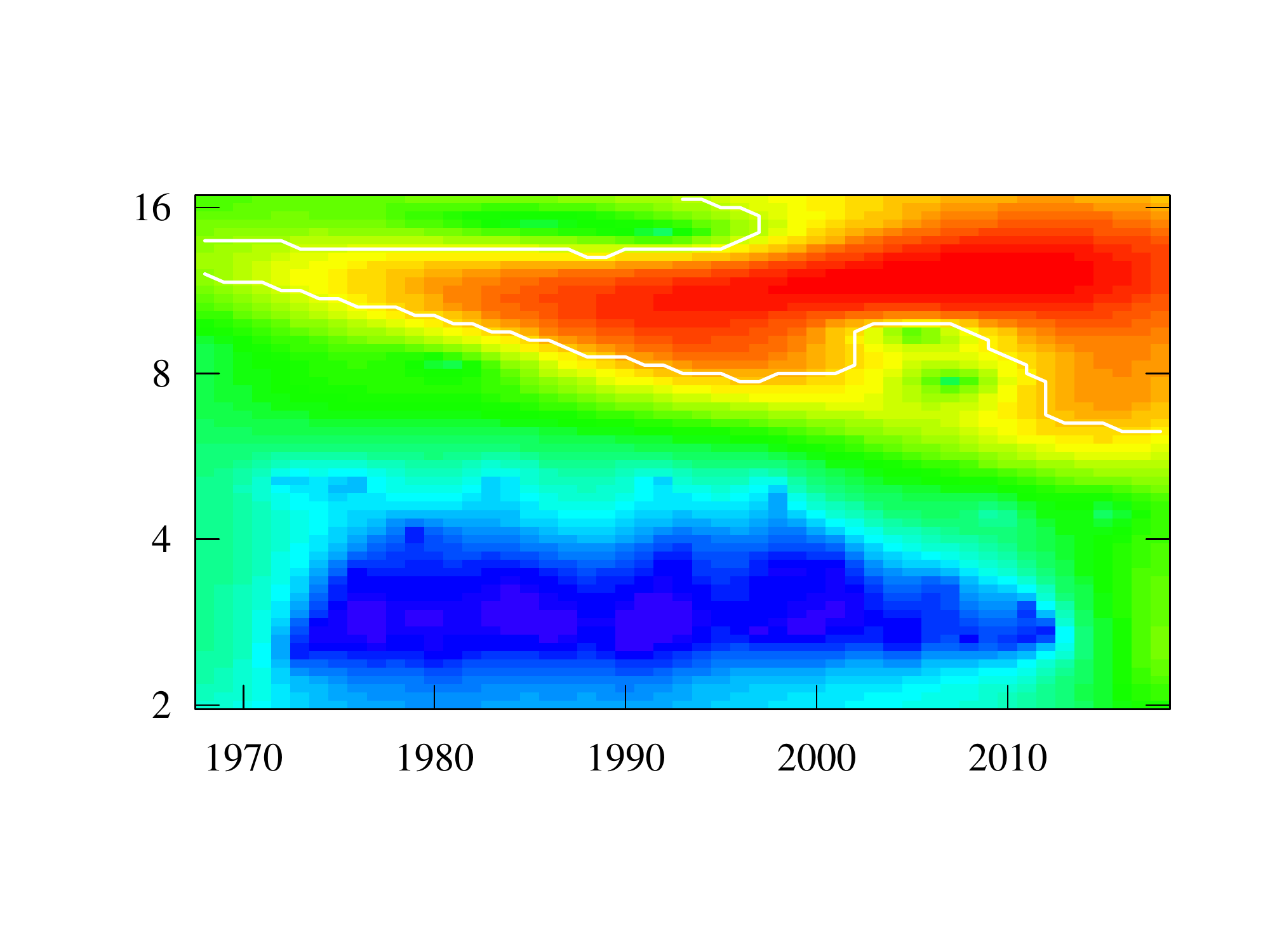} &  &  &  & \tabularnewline
FAO-Maize & \includegraphics[width=3.3cm]{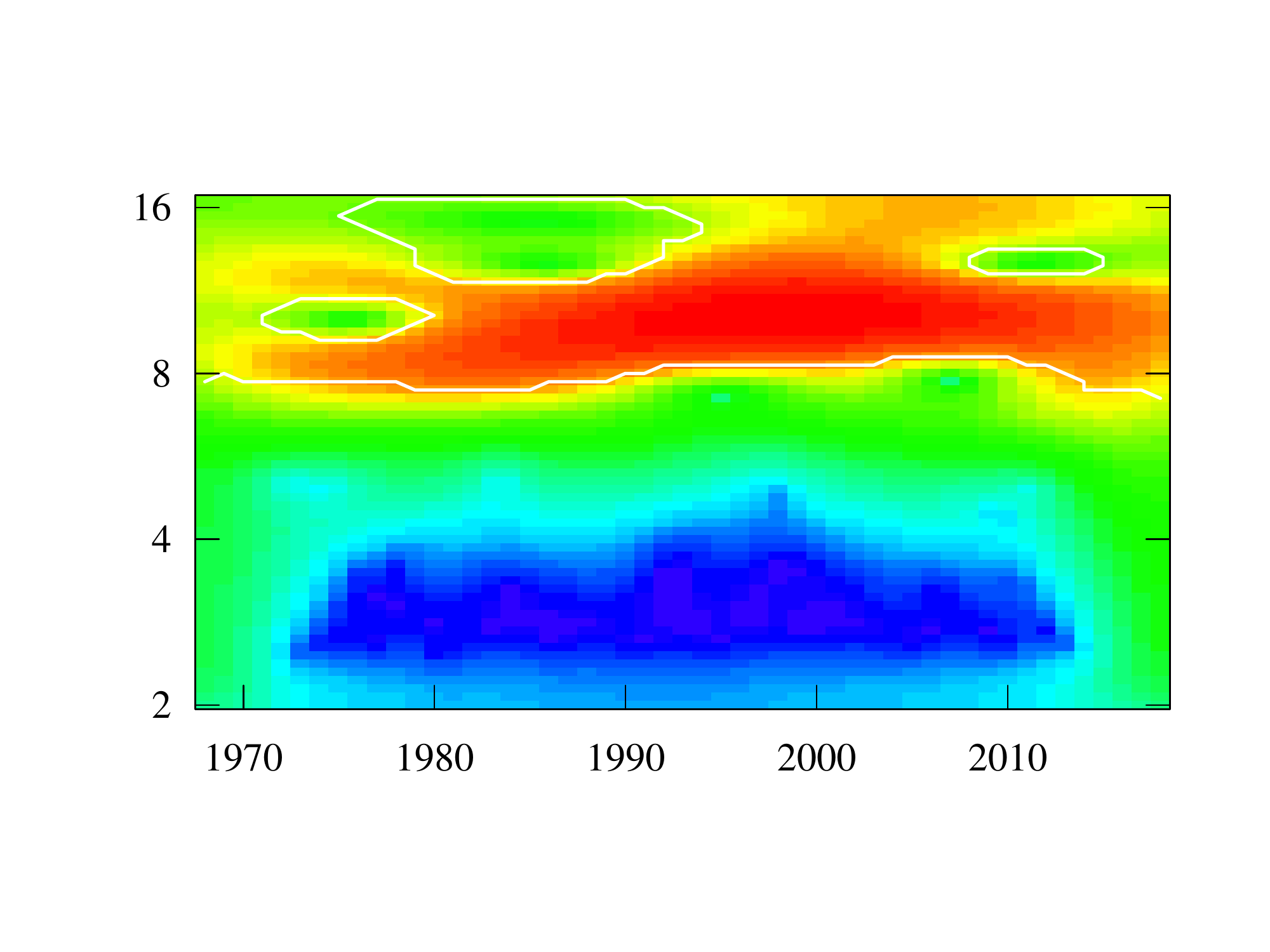} & \includegraphics[width=3.3cm]{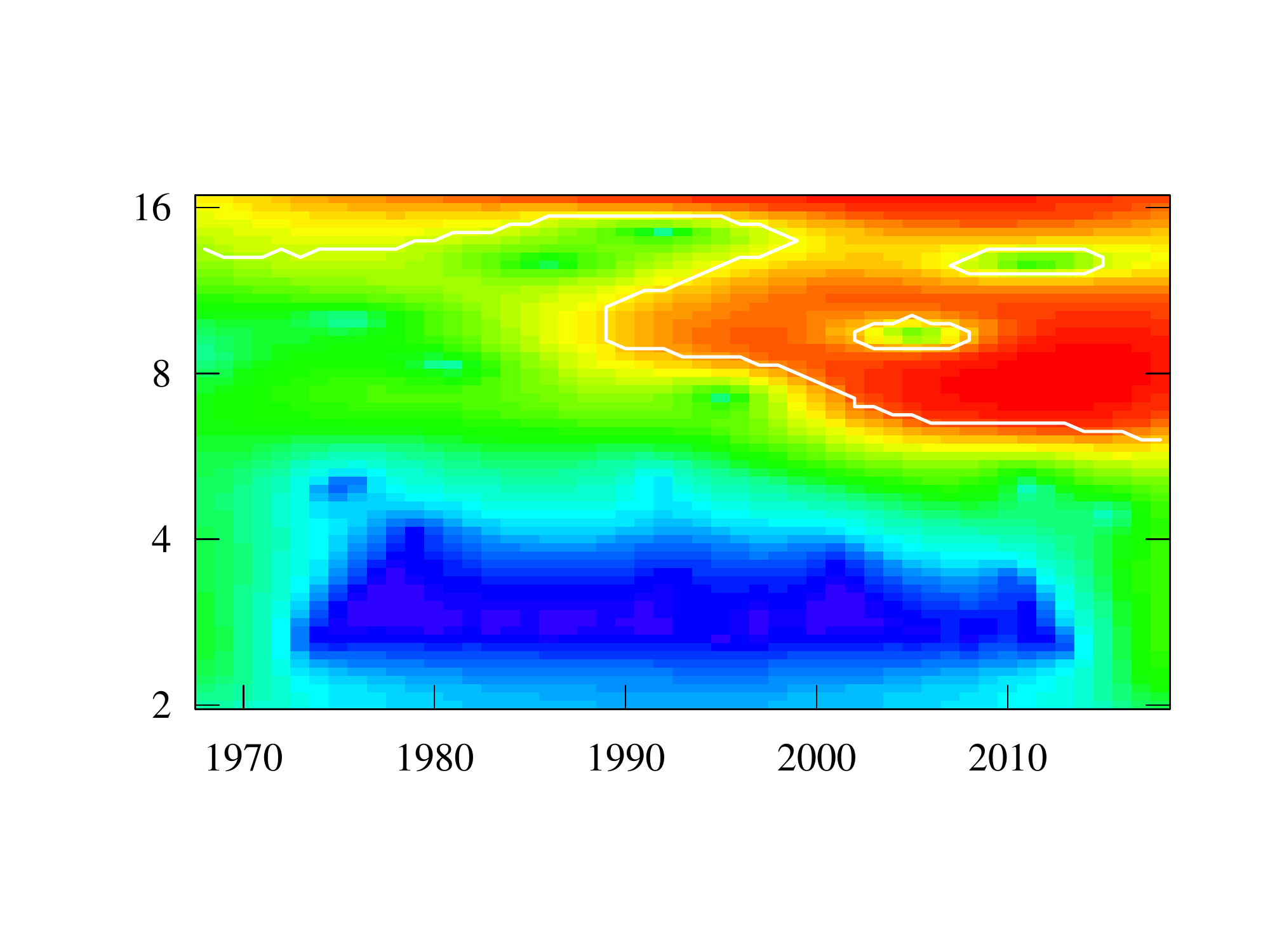} &  &  & \tabularnewline
FAO-Wheat & \includegraphics[width=3.3cm]{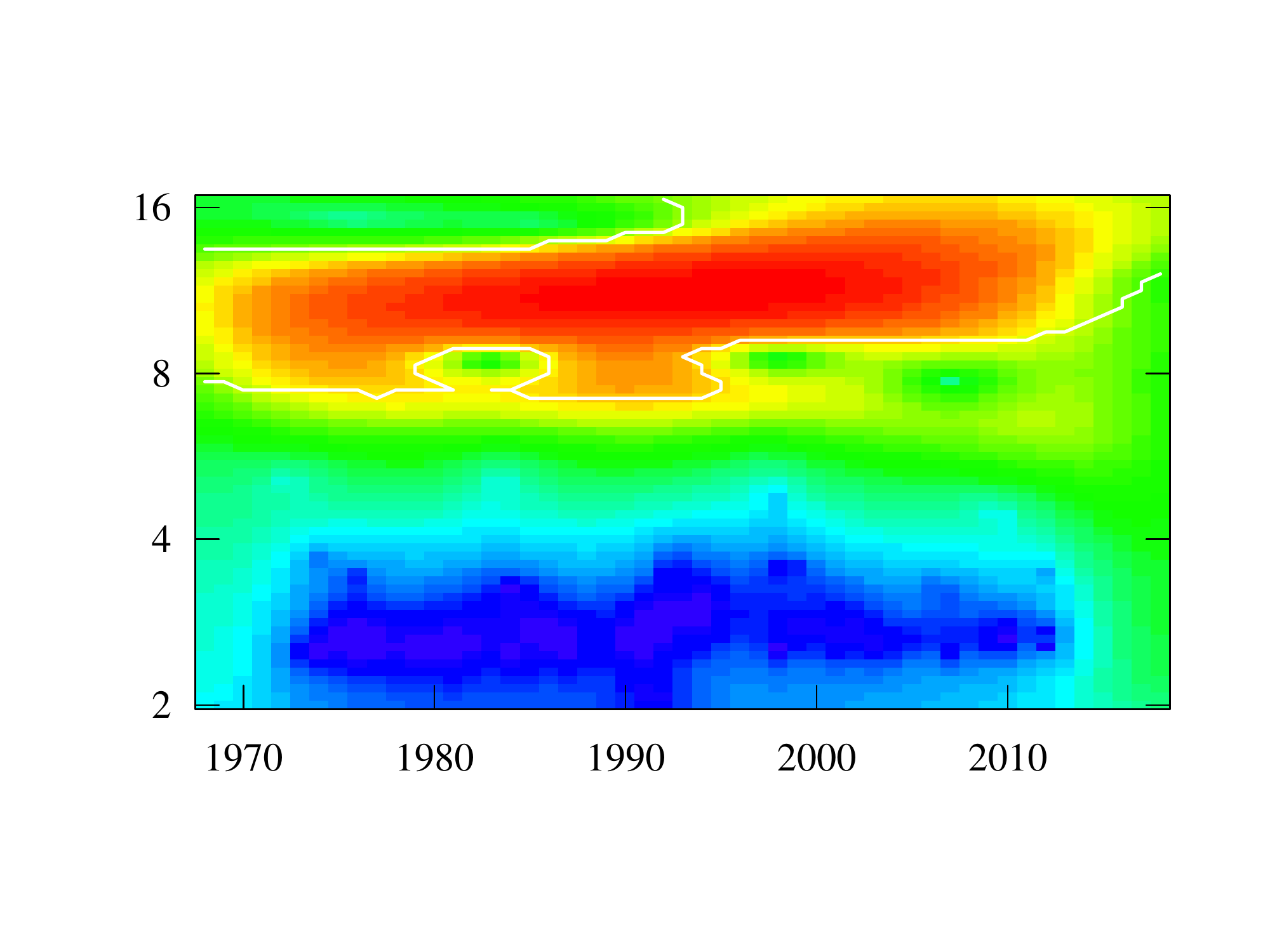} & \includegraphics[width=3.3cm]{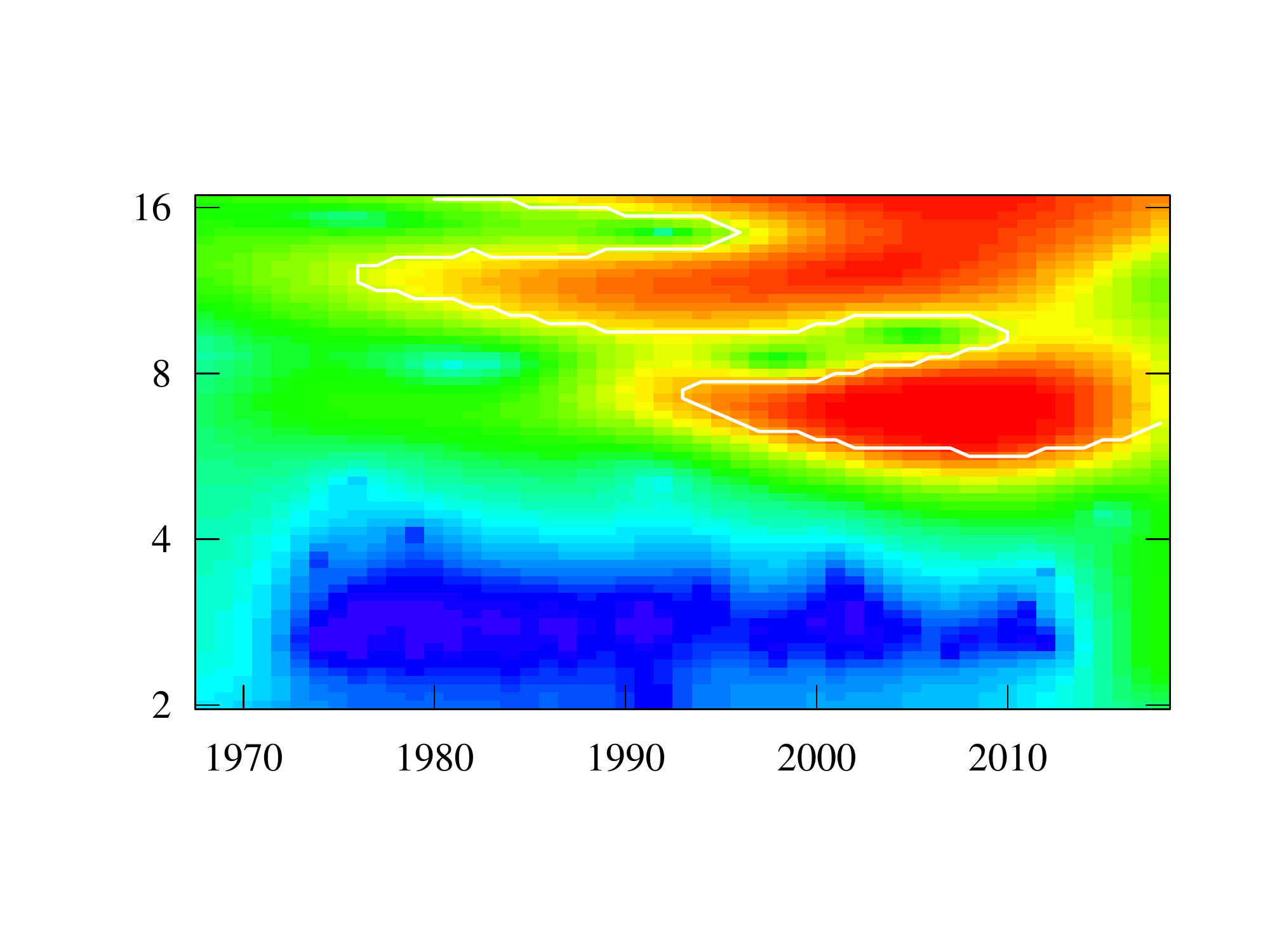} & \includegraphics[width=3.3cm]{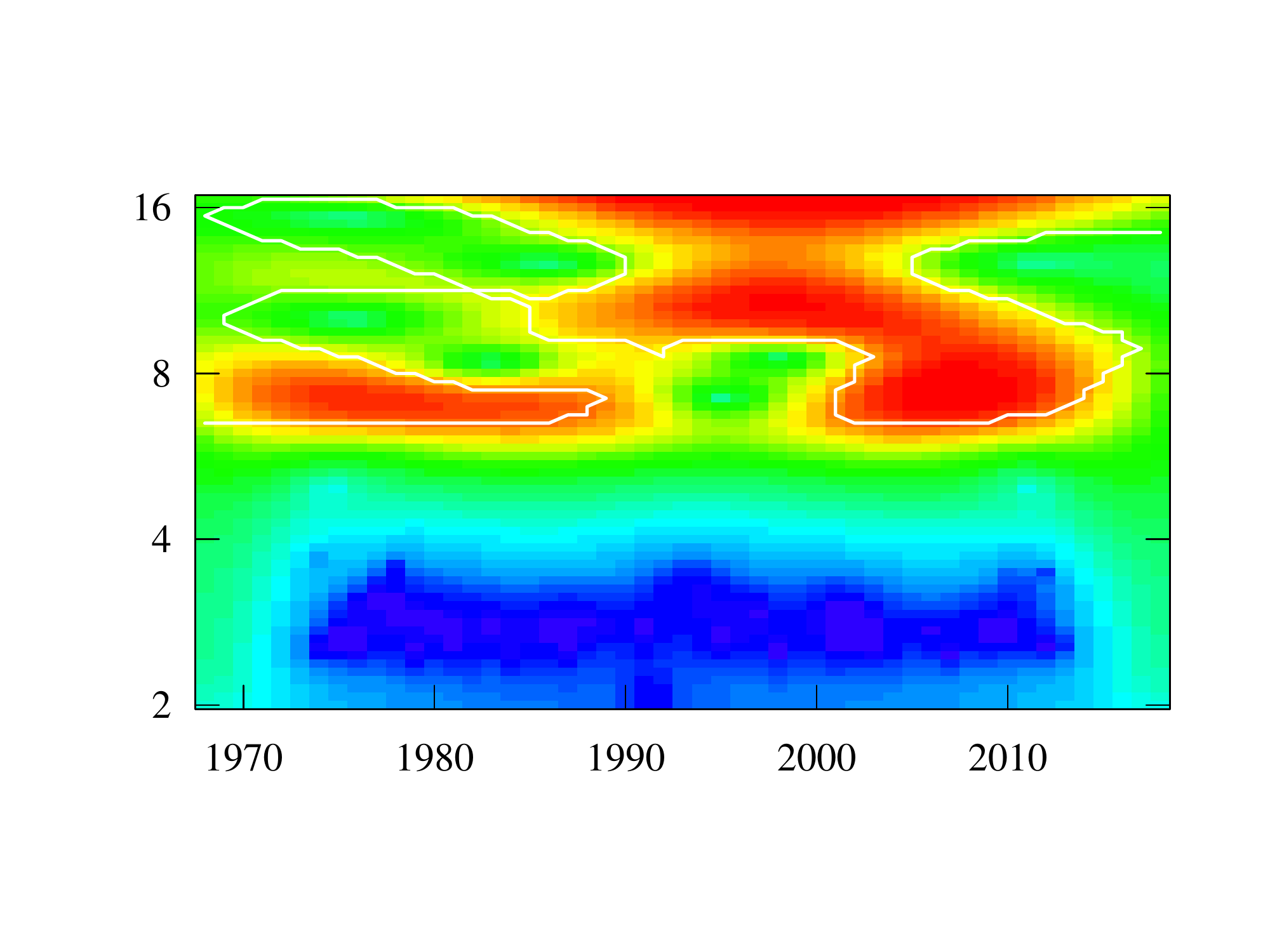} &  & \tabularnewline
LTAE-Maize & \includegraphics[width=3.3cm]{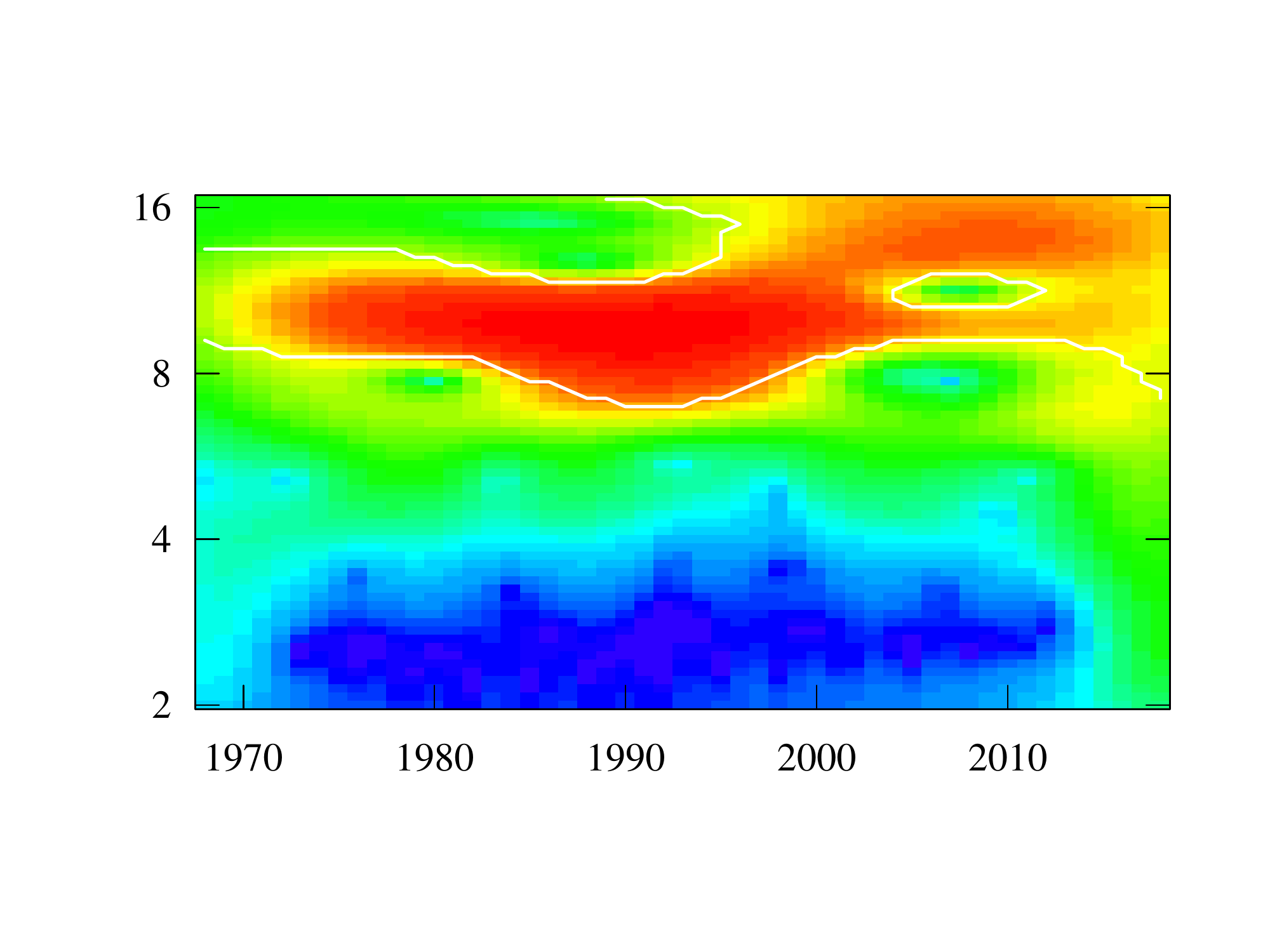} & \includegraphics[width=3.3cm]{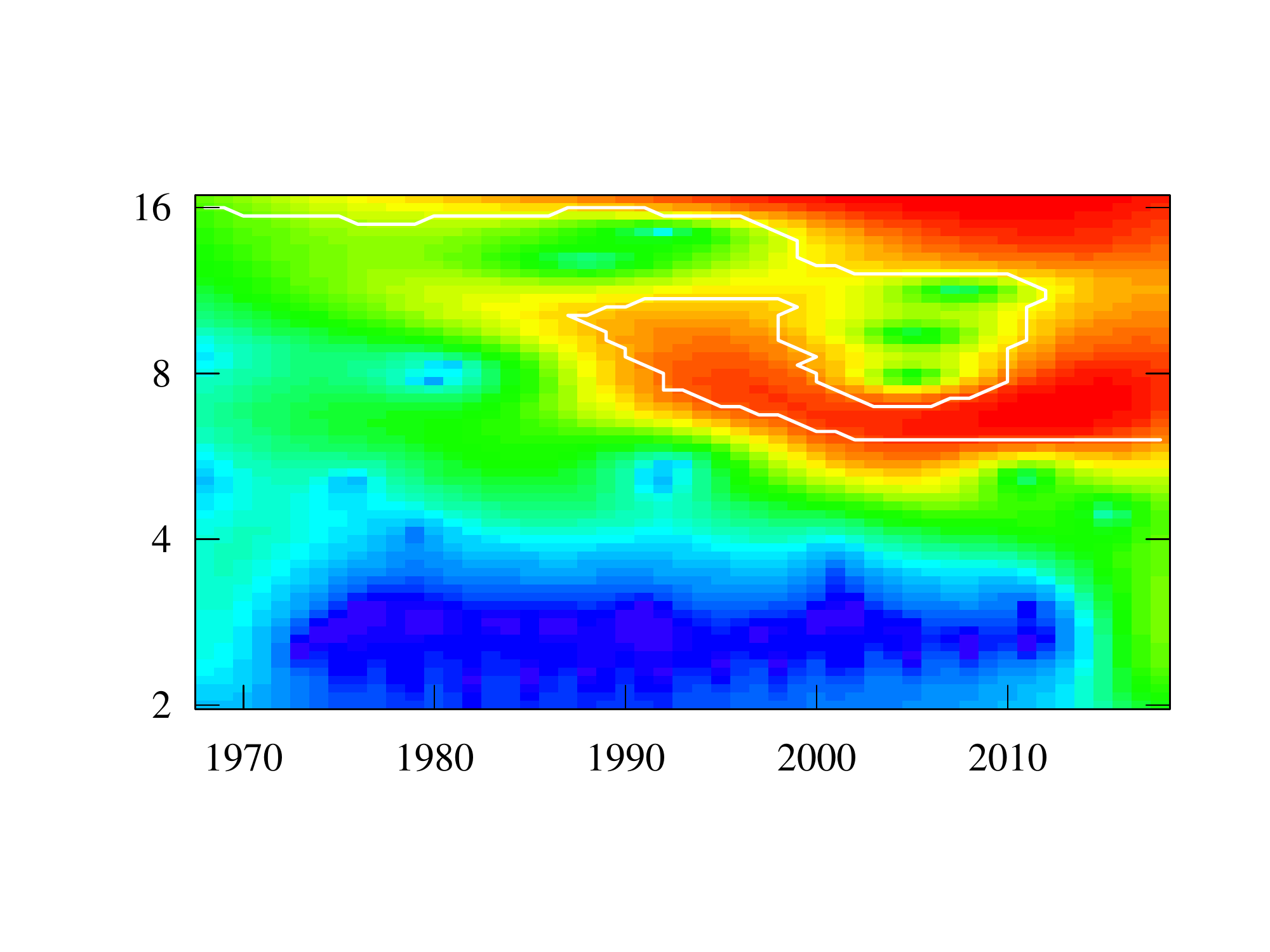} & \includegraphics[width=3.3cm]{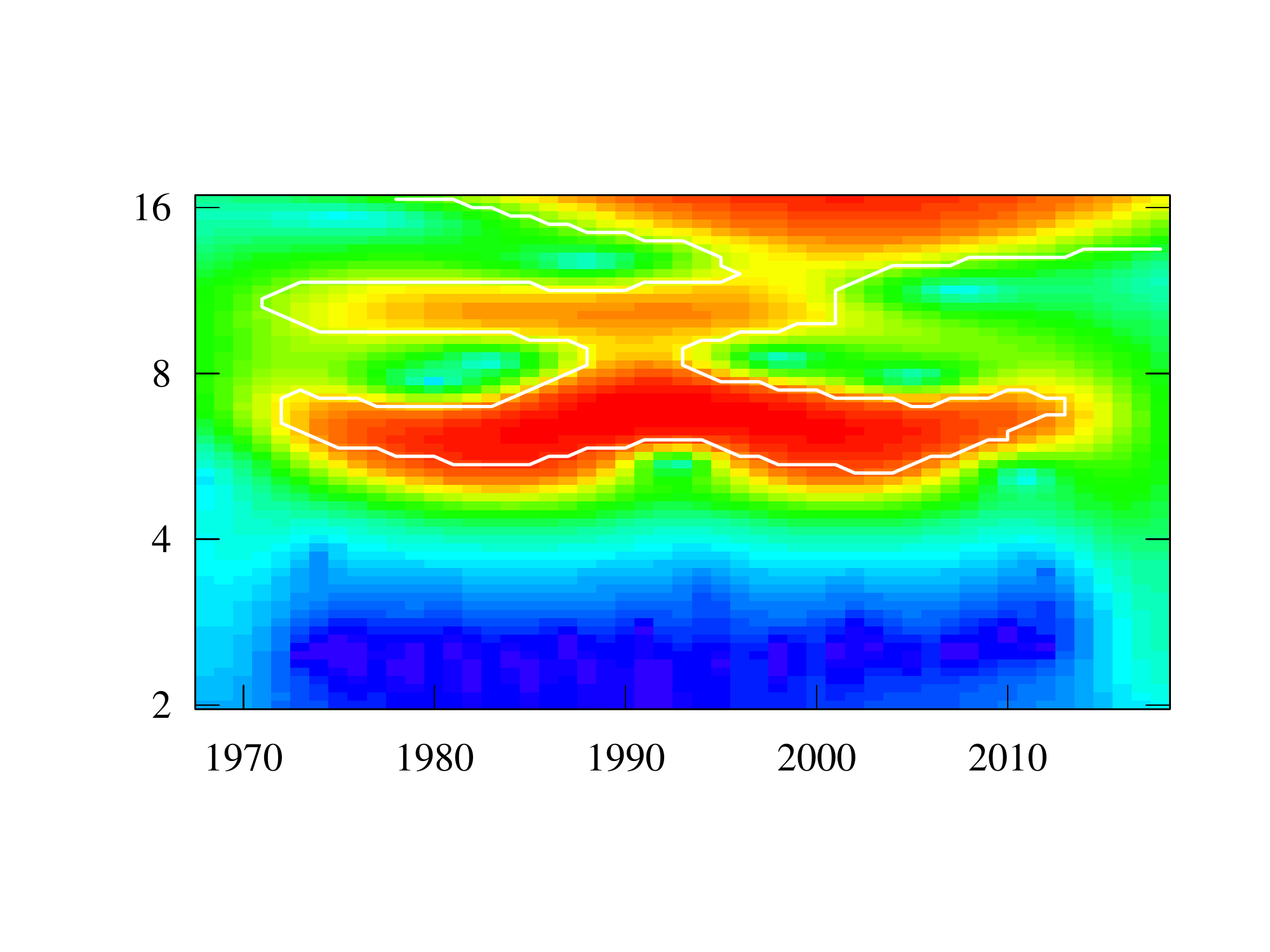} & \includegraphics[width=3.3cm]{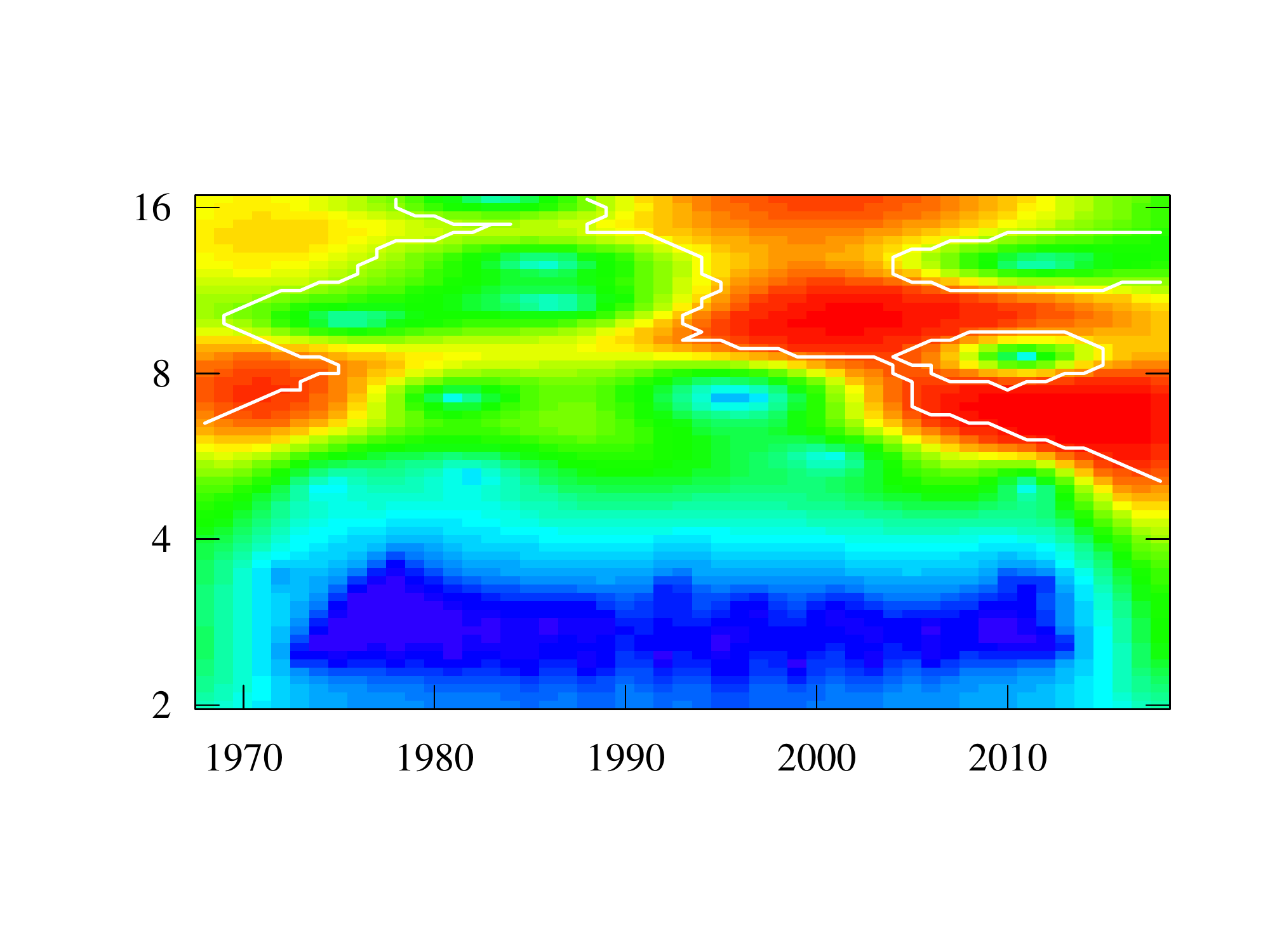} & \tabularnewline
LTAE-Wheat & \includegraphics[width=3.3cm]{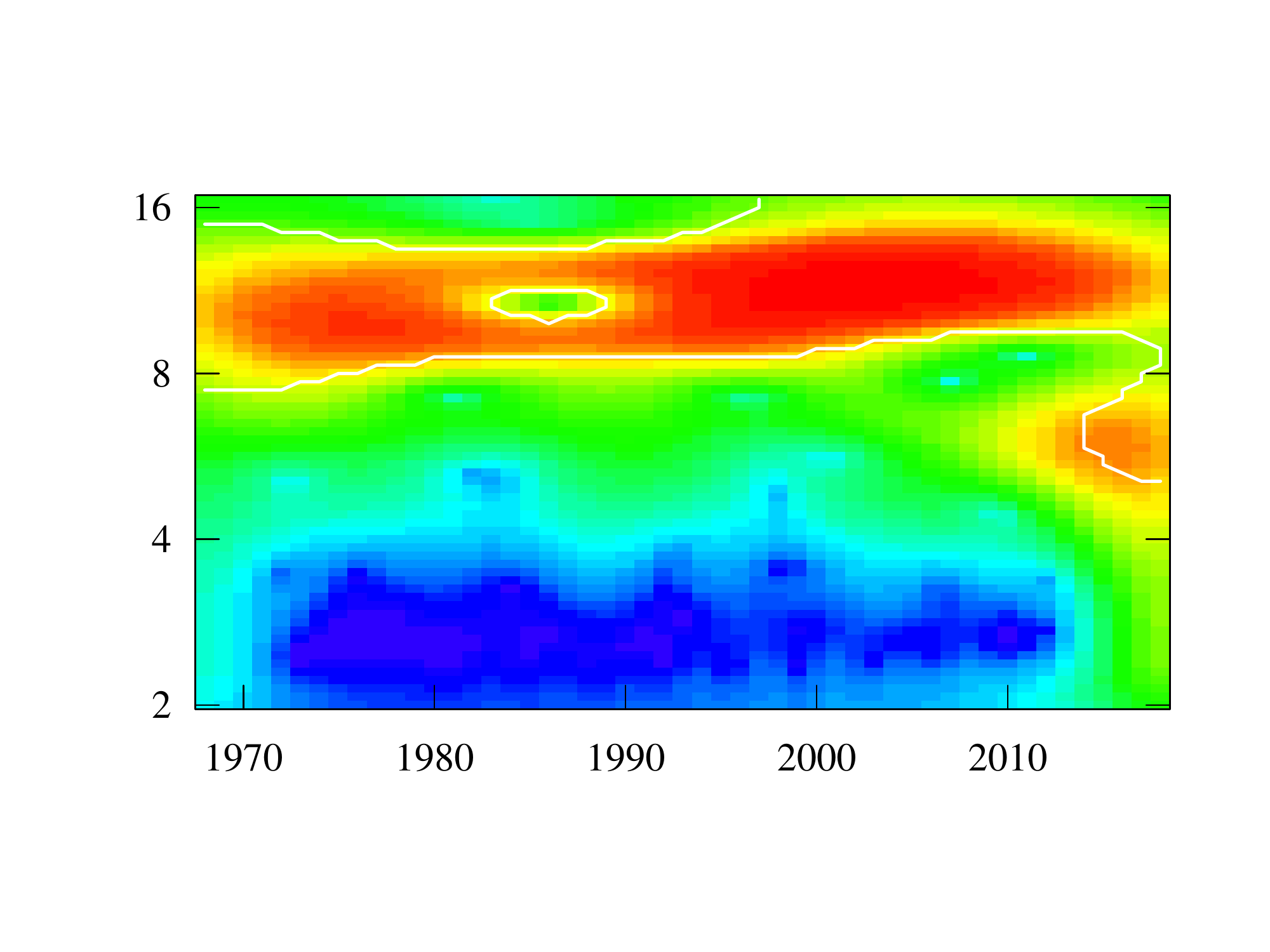} & \includegraphics[width=3.3cm]{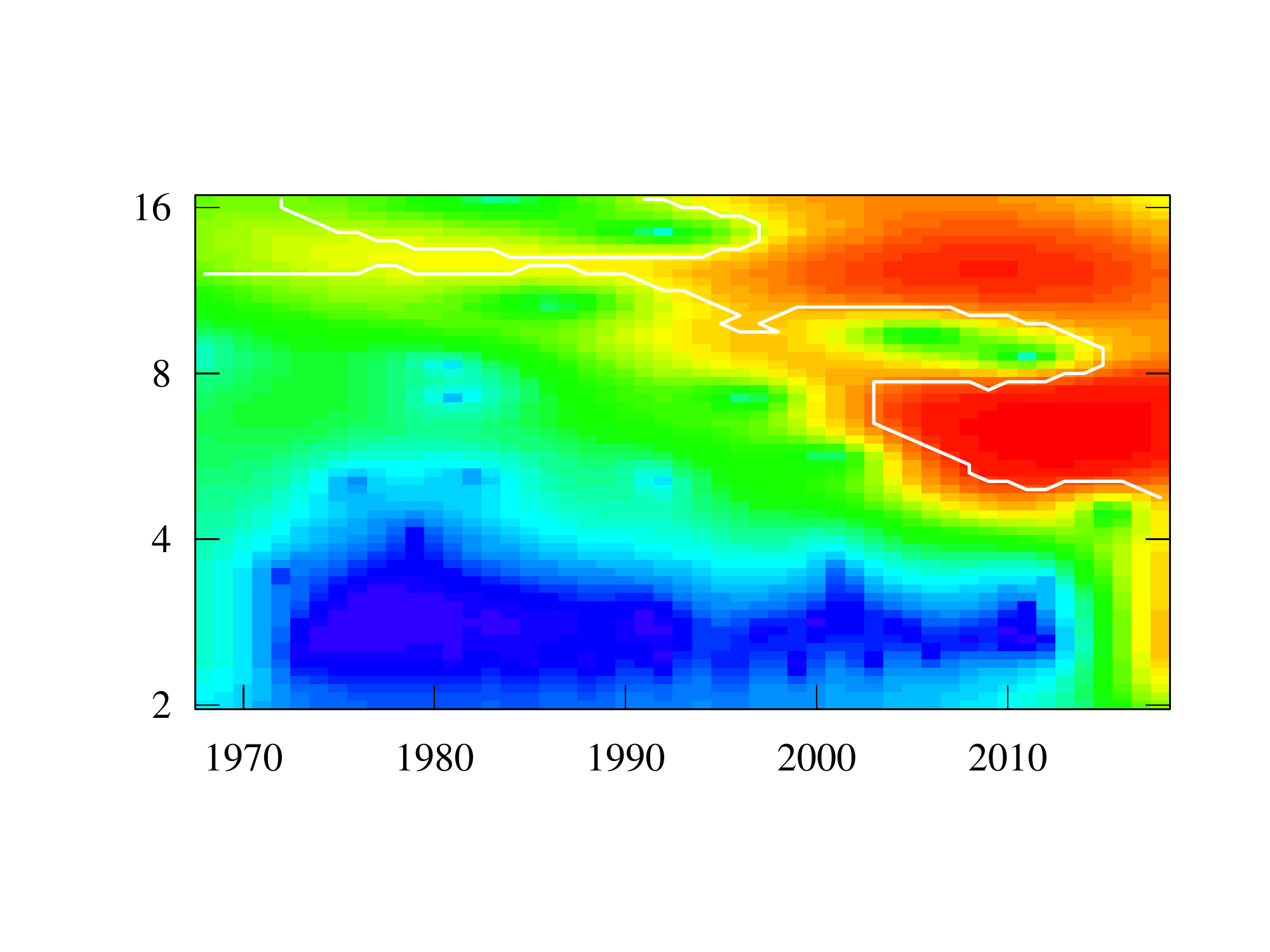} & \includegraphics[width=3.3cm]{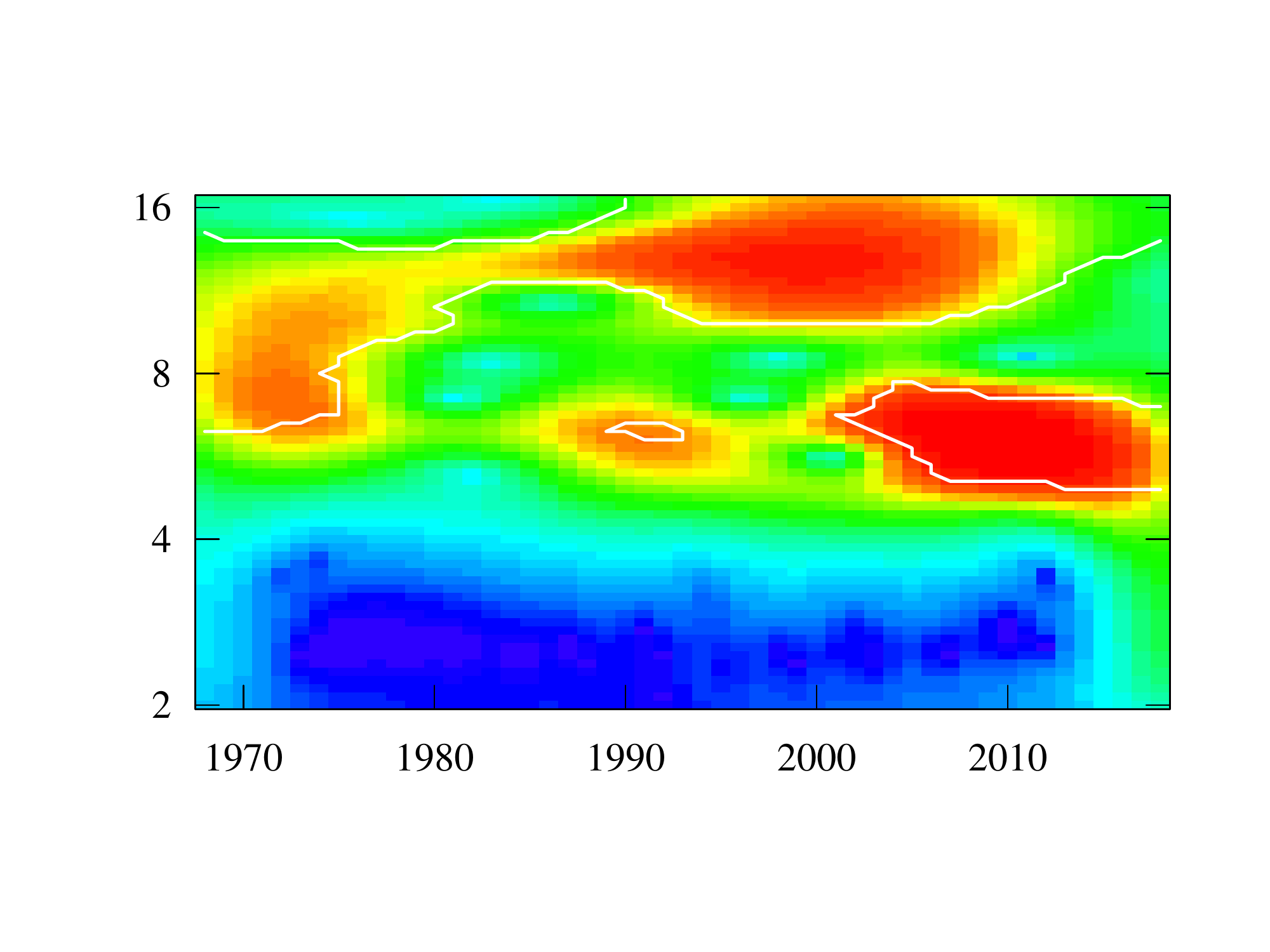} & \includegraphics[width=3.3cm]{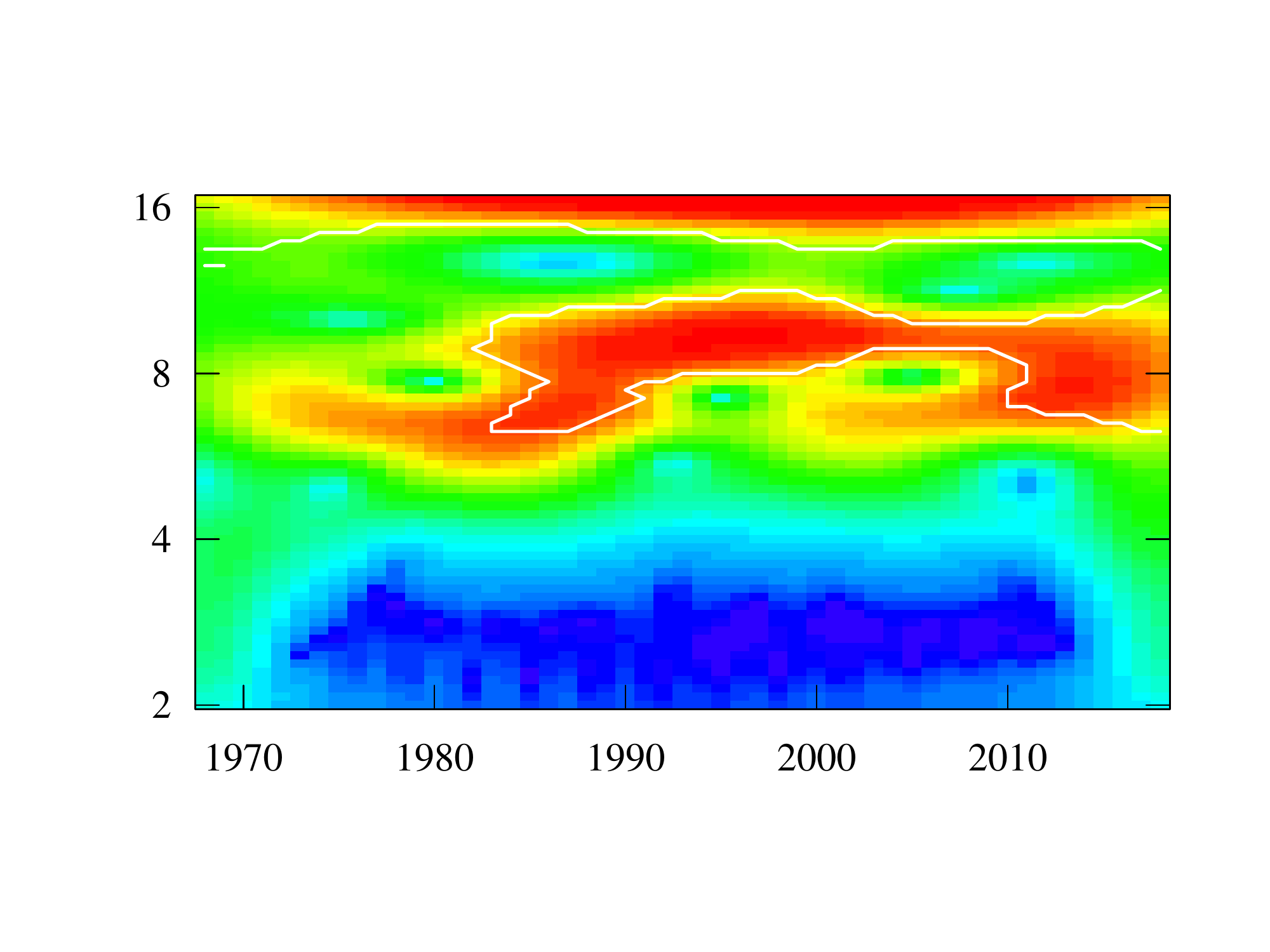} & \includegraphics[width=3.3cm]{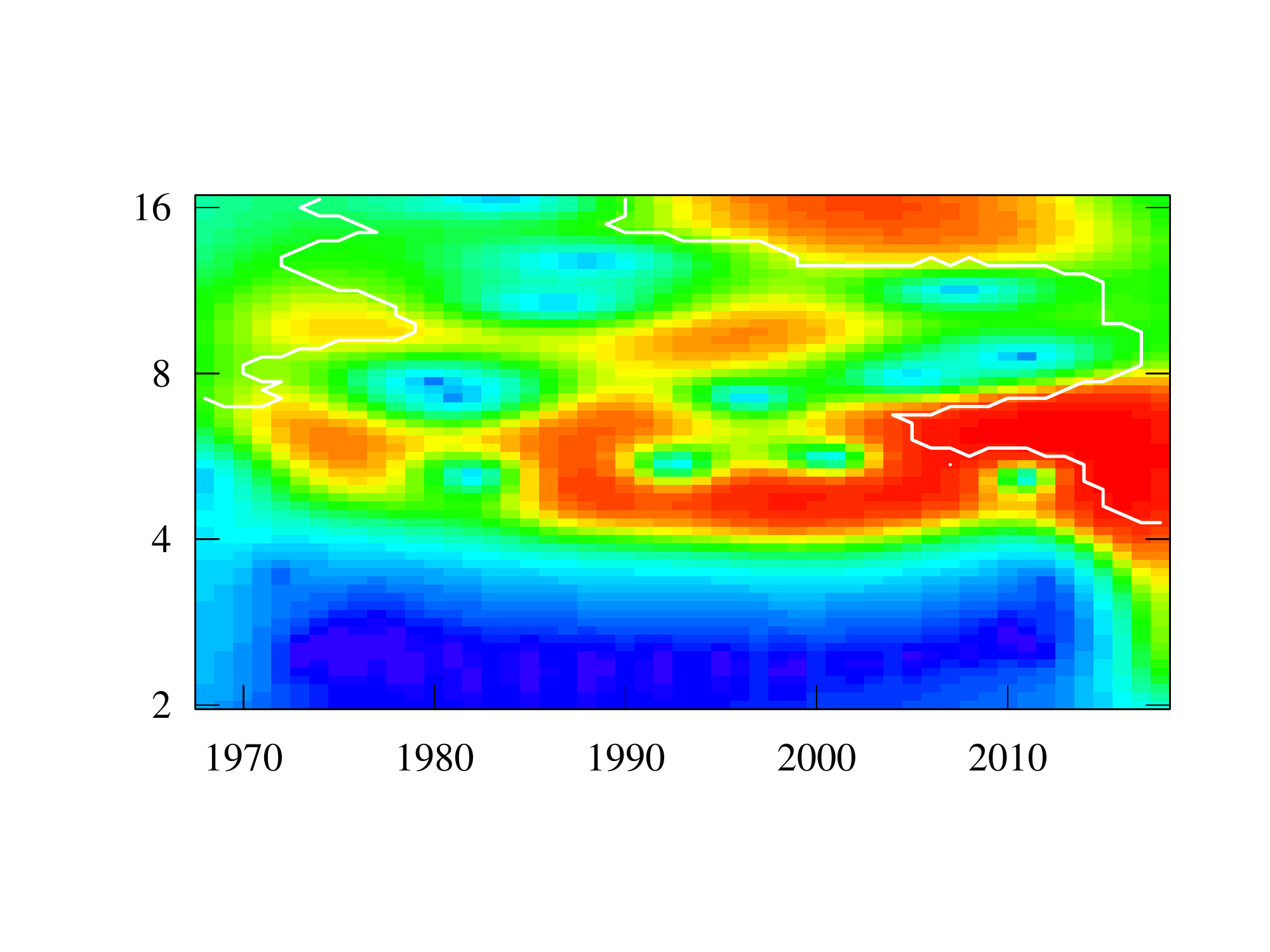}\tabularnewline
\end{tabular}
\par\end{centering}
\caption{\label{fig:WTC}Cross Wavelet Transform of detrended, normalised and
EMD-denoised reconstructed series. In the second column the lower
frequency have been dumped to mask the effect of SSN.}
\end{figure}
\par\end{flushright}

\section*{Conclusions}

\textcolor{black}{The aim of this study is to investigate how climate
change affects yields of major crops as maize and wheat, focusing
on trends and fluctuations}, useful to estimate production forecasts
and related risks.

\textcolor{black}{The analysis investigates possible relations of
local yields with national ones, a classical regional climatic index
as Western Mediterranean Oscillation Index, and }a global one given
by Sun Spot Number. 

\textcolor{black}{The present investigation shows how SSN oscillations
affects regional scale dynamics, recorded by indicators as WeMO, which
in turn reflects on the yield of Wheat and Maize.}

The analysis suffered the a small size of time series (51 years) though
non-parametric methods as EMD and SSA seemed efficient in identifying
and remove noise components.

\textcolor{black}{On the back of any derived conclusions there is
the analysis some historical series of climate indicators, whose causal
connections are not jet clearly perceivable to important stakeholders
as the farmer.}

\textcolor{black}{Another consideration regards the number of years
requested to obtain a view of a significant trend in climatic conditions
that influence crop yield. It seems that both at a global level and
locally, the annual anomalies around an average are so wide that a
clear trend, even if significant, needs a high number of years to
be perceived. During the lifespan of a farmer he would be hardly understand
if our planet is going toward a colder or a hotter end. Therefore
he has not any mean, based only on his experience, to decide if it\textquoteright s
better to change is production system in a way or another.}

\textcolor{black}{Such connections, having an empathized impact on
communication media, citizen, stakeholders and decision makers, are
based on indicators and variables related to measured variables, but
often far from common perception: this is the case of El-Nino Souther
Oscillation index, which long has been related to astronomical components
\citep{Attolini-1990}, both far from real world experience.}

\textcolor{black}{Understanding such fluctuations is of strategic
importance not only for economy of commodities, but also concerning
food security. The agro-environmental system is characterized by variables
reflecting the behavior of complex dynamics that cannot be ascribed
to specific causes. Recently many researches are addressing the climate
change issue, trying to identify and confirm the reasons at the base
of the many expected (not always tangible), effects of such environmental
drift on human life. Though most of the studies focus on food security,
an analysis of the real effects of climate change on crop production
is still lacking.}

\textcolor{black}{Finally it has to be put in evidence the importance
of Long Term Agronomic Experiments. While the trend of the average
Italian temperatures and annual rainfall can be obtained for a long
period \citep{Galli-1988}, yields are less reconstructable \citep{Cavazza-1966}.
Therefore they represent a source of data that, though influenced
by soil type and technical management, it embeds long-range seasonality.}

Finally, to avoid any manipulation of present results, authors want
to underline that the effects of human activities on climate change
is not discussed in this work.

\bibliographystyle{plain}
\bibliography{C:/Users/GV/Documents/BIB/BibTeX/MATH}

\subsection*{Acknowledgments}

Authors are thankful to ``World Data Center for the production, preservation
and dissemination of the international sunspot number'', the ``Climatic
Research Unit of East Anglia University'' for Western Mediterranean
Oscillation index records availability, FAO for data dissemination
efforts, and R, R-Studio developers and contributors for allowing
and maintain free access to their products.

\subsection*{Appendix 1 - Time Series Analysis}

Instead of additive model, multiplicative, $x_{i}=\mu_{i}\times y{}_{i}\times\varepsilon_{i}$
or mixed ones (e.g. $x_{i}=\mu_{i}\times\varepsilon_{i}+y_{i}$) can
be adopted emerging from a Time Series \textcolor{black}{Analysis},
where\textcolor{black}{{} observations are used to develop a mathematical
model which captures the underlying Data Generation Process (DGP),
which could be extremely useful for predictions. A common strategy
is }interpreting a time series ${x(t)}$ as a realization of a stochastic
process, a sample of the totality of realizations which is called
\textbf{ensemble}.\textcolor{black}{{} Such a sample can be used to
estimate the joint probability distribution of the random variable
involved, describing the probability structure of the time series
as a stochastic process.}

\textcolor{black}{Most of time series analysis relay on linearity
assumption, meaning that it can be modeled by a }\textbf{\textcolor{black}{linear
stochastic process}}\textcolor{black}{, where the current value of
the series is a linear function of past observations.}

\textsl{Linear processes include AutoRegressive (AR) model, Moving
Average (MA) and derived ones (ARMA, ARIMA, SARIMA, Box-Jenkins):}

\textsl{
\begin{eqnarray*}
AR) &  & x(t)=\mu+\varepsilon(t)+\sum_{j=1}^{p}\alpha_{j}\cdot x(t-j)\\
MA) &  & x(t)=\mu+\varepsilon(t)+\sum_{j=1}^{q}\beta_{j}\cdot\varepsilon(t-j)\\
ARMA) &  & x(t)=\mu+\varepsilon(t)+\sum_{j=1}^{p}\alpha_{j}\cdot x(t-j)+\sum_{j=1}^{q}\beta_{j}\cdot\varepsilon(t-j)\\
ARIMA_{1}) &  & x(t)=\varepsilon(t)+\sum_{j=1}^{p}\alpha_{j}\cdot x(t-j)-\sum_{j=1}^{q}\beta_{j}\cdot\varepsilon(t-j)
\end{eqnarray*}
}

where $p$ is the lag order for AR model, $\alpha_{j}$ representing
the weight/memory of past observations $x(t-j)$, $q$ is the order
of moving average, $\beta_{j}$ representing the weights over past
forecast (non observed) errors $\varepsilon(t-j)$. While ARMA is
just a simple combination of the two, ARIMA is based on differentiating
(in the reported equation the degree of differentiating d=1).

\textbf{\textcolor{black}{Linearity }}\textcolor{black}{may be tested
by a number of methods (\citep{Bisaglia-2014}, all of them lay on
surrogate data generation \citep{Theiler-1992},\citep{Lancaster-2018})
as Teraesvirta's and White's test (based on neural network), verifying
HP0: ``Linearity in mean\textquotedbl , Keenan's, Tsay's and likelihood's
tests, with HP0: ``The time series is generated from a Non-Linear
AR process'', McLeod-Li's test for Heteroschedasticity with HP0:
``The time series is generated from a Non-Linear ARIMA process''.
Such test are based on quasi-stationarity of time series.}

Table \ref{tab:LINEARITY-TEST} show the results of linearity test
performed with 6 different methods mentioned above on spline detrended
data.

\begin{table}[h]
\begin{tabular}{cccccccc}
\cline{3-8} 
 & test & SSN & WeMO & FAO-W & FAO-M & LTAE-W & LTAE-M\tabularnewline
\hline 
\multirow{2}{*}{linearity in 'mean'} & Teraesvirta & \textbf{0.618} & \textbf{0.202} & 0.069 & \textbf{0.125} & \textbf{0.741} & \textbf{0.282}\tabularnewline
\cline{2-8} 
 & White & \textbf{0.581} & \textbf{0.195} & \textbf{0.164} & \textbf{0.100} & \textbf{0.640} & \textbf{0.322}\tabularnewline
\hline 
\multirow{3}{*}{NL-AR} & Keenan & \textbf{0.228} & \textbf{0.883} & \textbf{0.099} & \textbf{0.932} & \textbf{0.703} & \textbf{0.117}\tabularnewline
\cline{2-8} 
 & Tsay & 0.022 & 0.041 & 0.044 & \textbf{0.386} & \textbf{0.732} & 0.003\tabularnewline
\cline{2-8} 
 & likelihood & $1e^{-7}$ & \textbf{0.430} & 0.061 & \textbf{0.257} & \textbf{0.334} & 0.003\tabularnewline
\hline 
NL-ARIMA & McLeod-Li & 0.014 & \textbf{1.000} & \textbf{0.557} & \textbf{0.979} & \textbf{0.585} & \textbf{0.791}\tabularnewline
\hline 
\end{tabular}\caption{\label{tab:LINEARITY-TEST}p-values of stability test applied on detrended
time series}
\end{table}

From table \ref{tab:LINEARITY-TEST} we can see that null hypothesis
(linearity in 'mean') can be accepted in the majority of cases, with
the exceptions for FAO-W emerging from Teraesvirta test. However there
are also strong clues on non-linearity, after Keenan's test in every
series, which are confirmed by every test on FAO-M and LTAE-W.

\subsection*{Appendix 2 - Stationarity and ACF}

Stationarity is a property making the time series useful for forecasts,
which could be meant in several ways. \textbf{Strict stationarity,
}wanting\textcolor{black}{{} $X={x_{i}}={x_{i+h}}$ for every $h$ is
hardly used; }\textbf{\textcolor{black}{first order stationarity}}\textcolor{black}{{}
requires a constant local average $E(X_{t})$, being $X_{t}$ is a
subset of original series around $t$; finally the }\textbf{\textcolor{black}{second
order}}\textcolor{black}{{} or }\textbf{\textcolor{black}{weak stationarity}}\textcolor{black}{{}
requires a constant variance $\sigma^{2}(X_{t})$ and an Auto Correlation
Function (ACF) not depending on time. The latter is the one commonly
adopted.}

\textcolor{black}{Being ACF: }\textbf{\textcolor{black}{
\begin{equation}
\rho_{k}=cov(X,X_{k})/\left(Var(X)\cdot Var(X_{k})\right)
\end{equation}
}}\textcolor{black}{where $X_{k}$ is the series shifted of a time
lag $k$, in stationary conditions it reduces to:}

\textbf{\textcolor{black}{
\begin{equation}
\rho_{k}=cov(X,X_{k})/\sigma^{2}
\end{equation}
}}

\textcolor{black}{ACF allows to reveal both the presence of a trend
and to spot periodical components. }

For the adopted time series, ACF can be seen in the figure BELOW

\begin{figure}[h]
\begin{centering}
\begin{tabular}{cc}
\includegraphics[width=8cm]{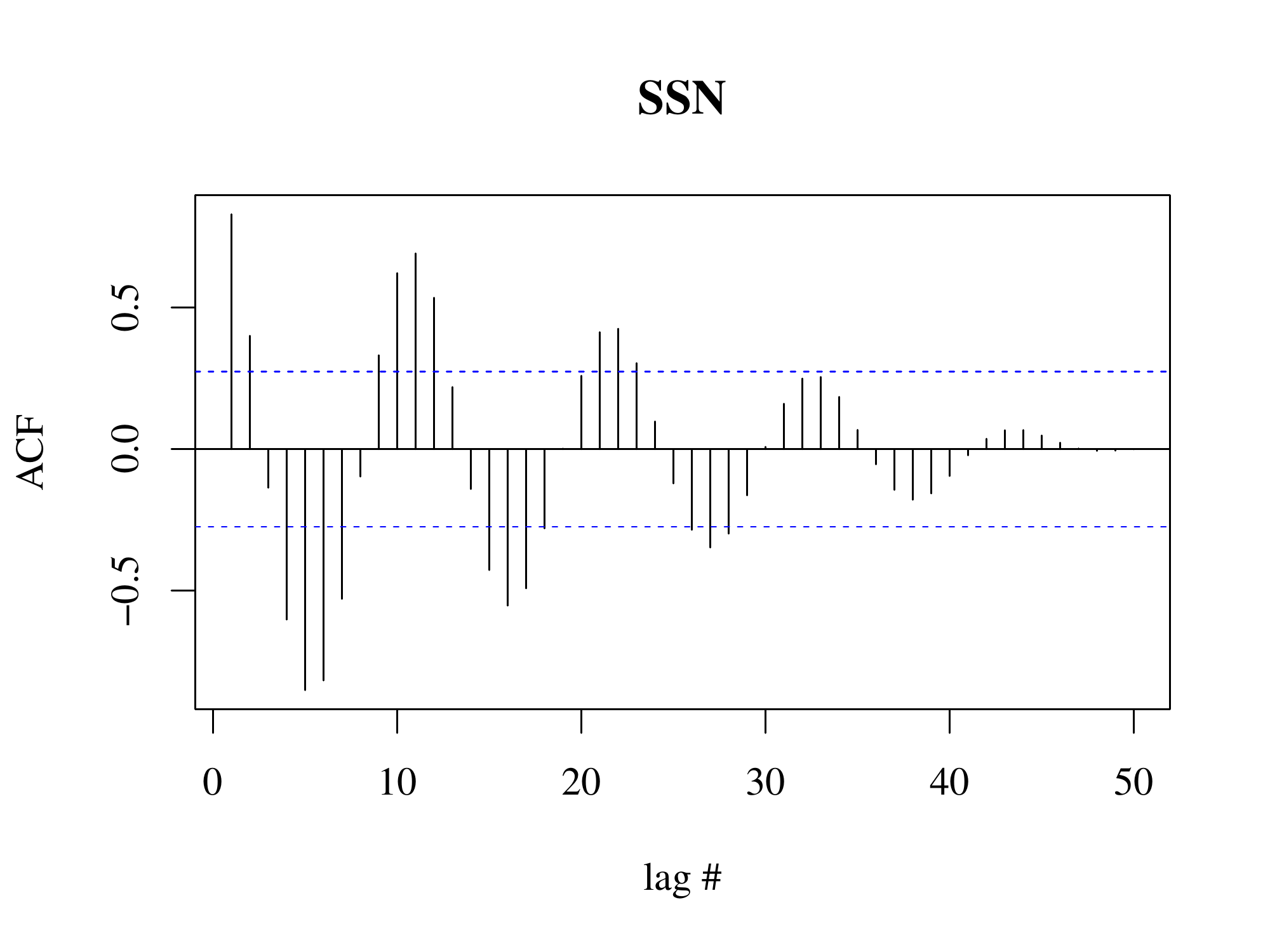} & \includegraphics[width=8cm]{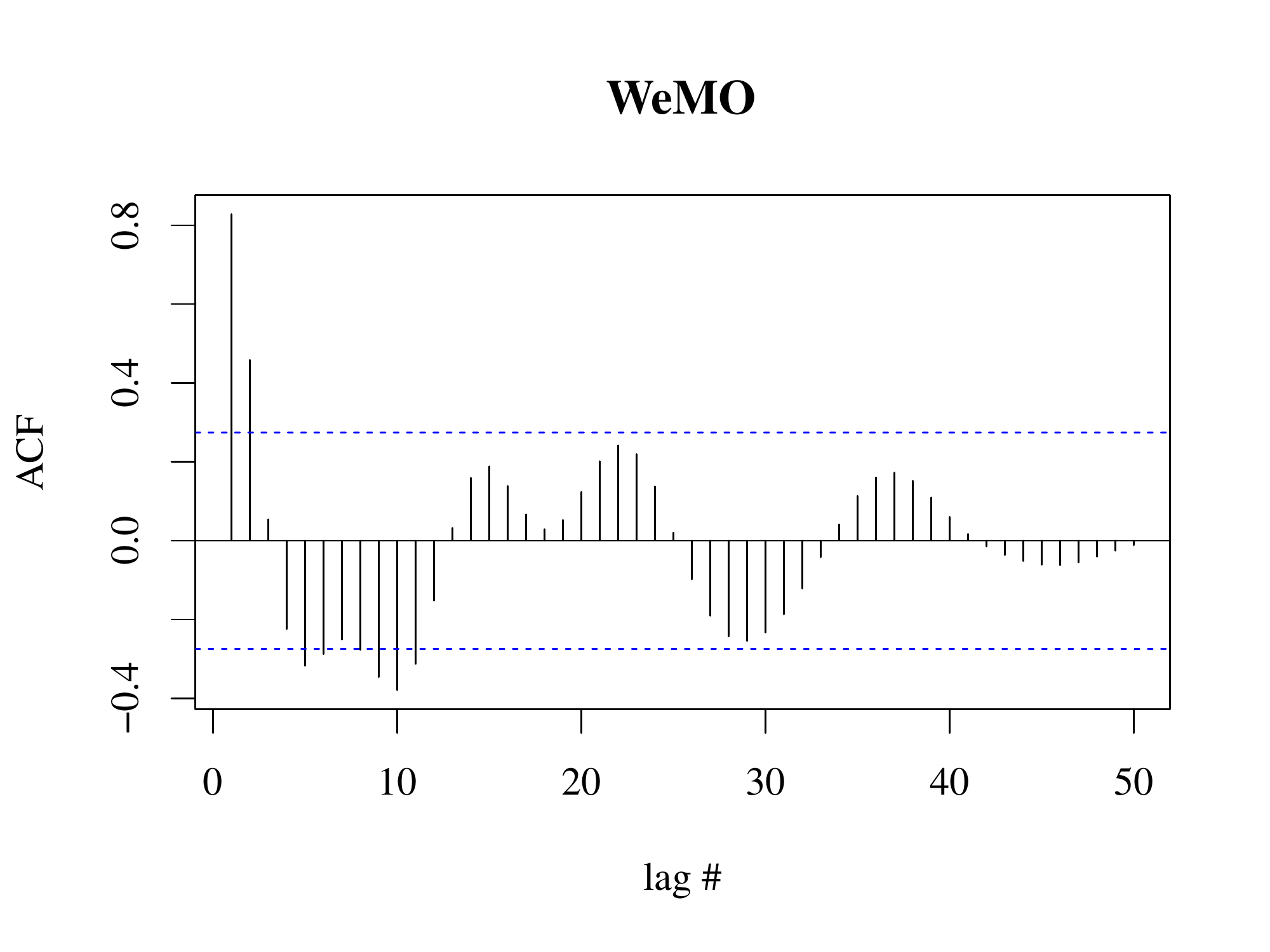}\tabularnewline
\includegraphics[width=8cm]{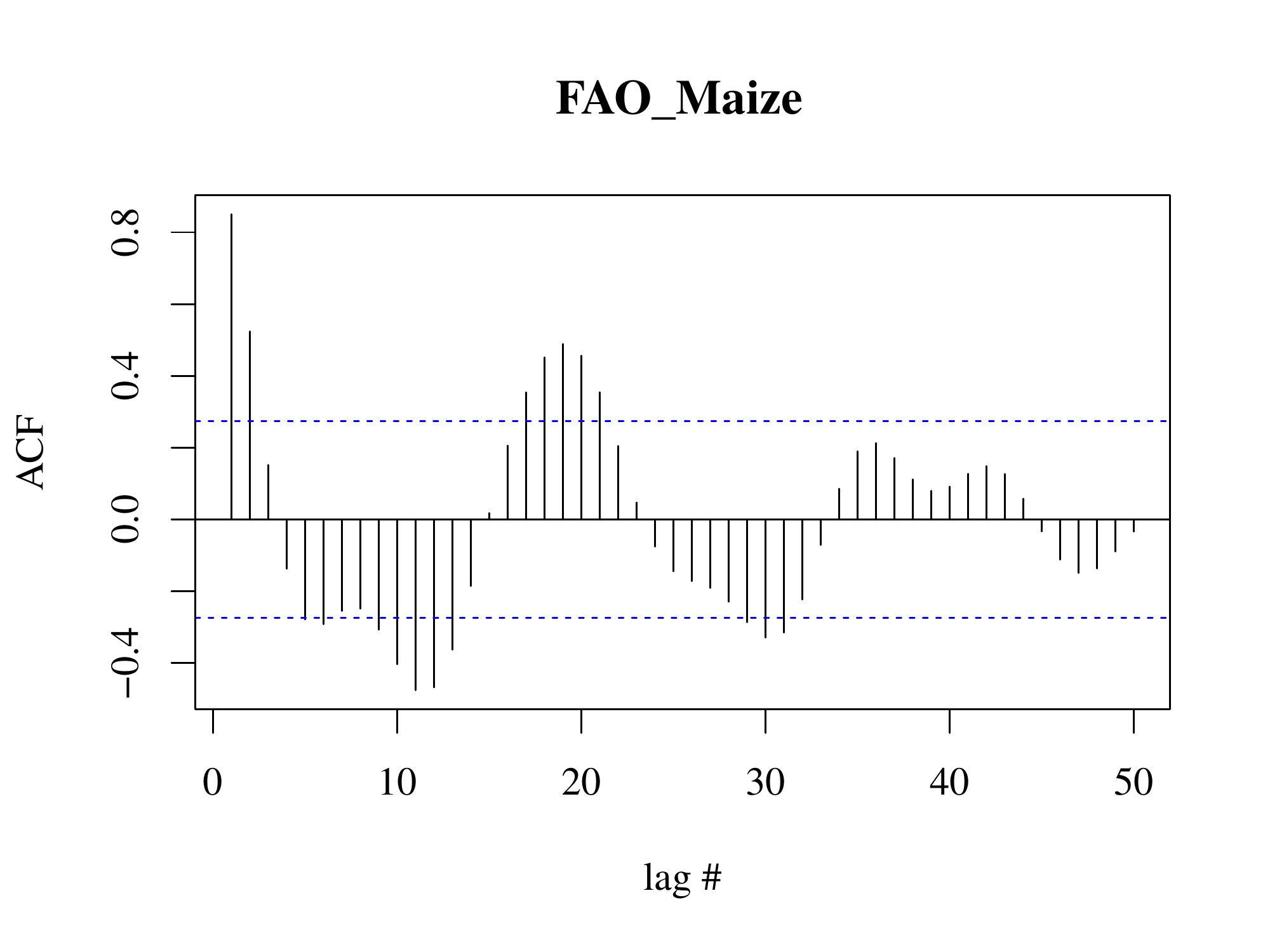} & \includegraphics[width=8cm]{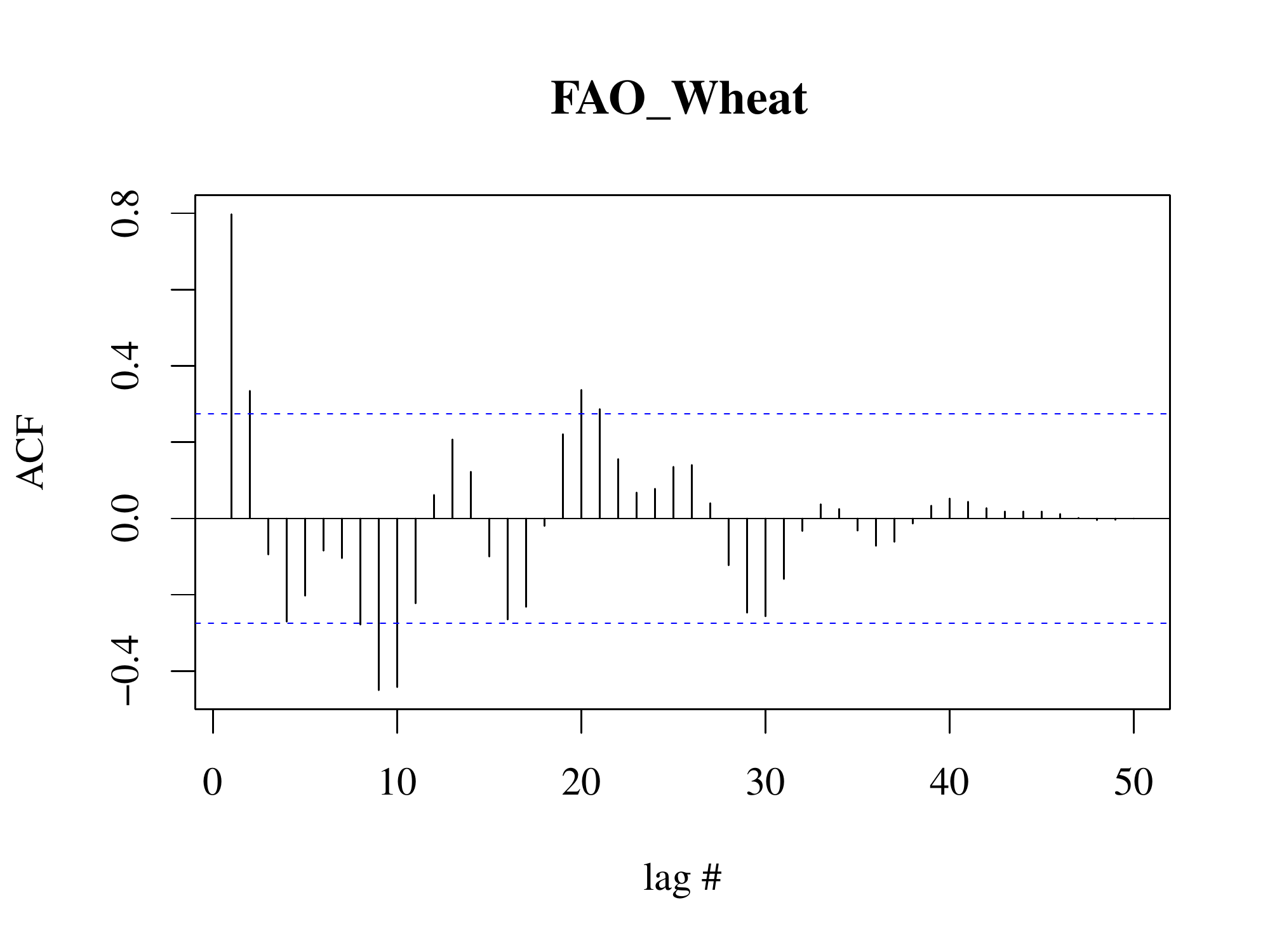}\tabularnewline
\includegraphics[width=8cm]{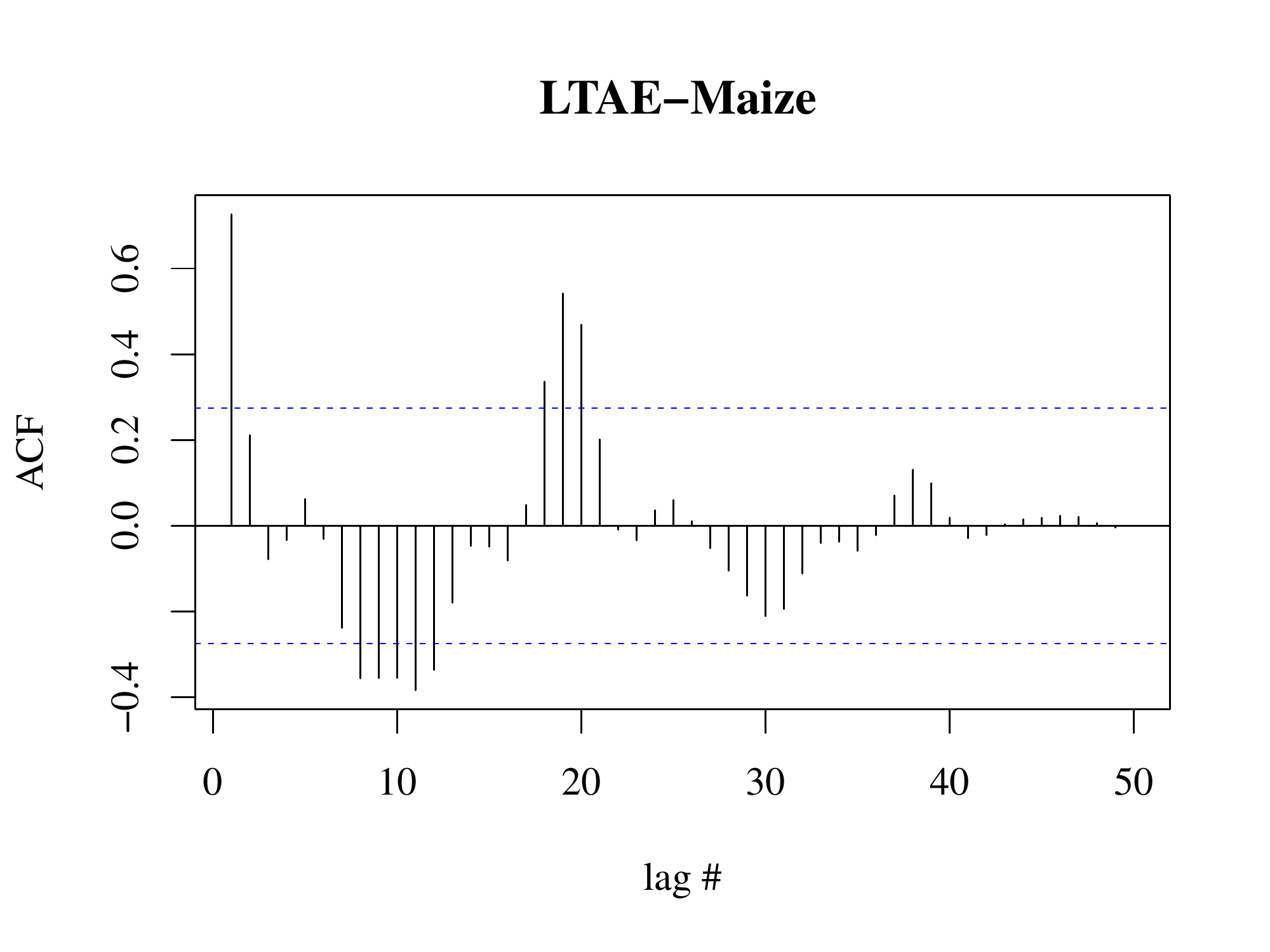} & \includegraphics[width=8cm]{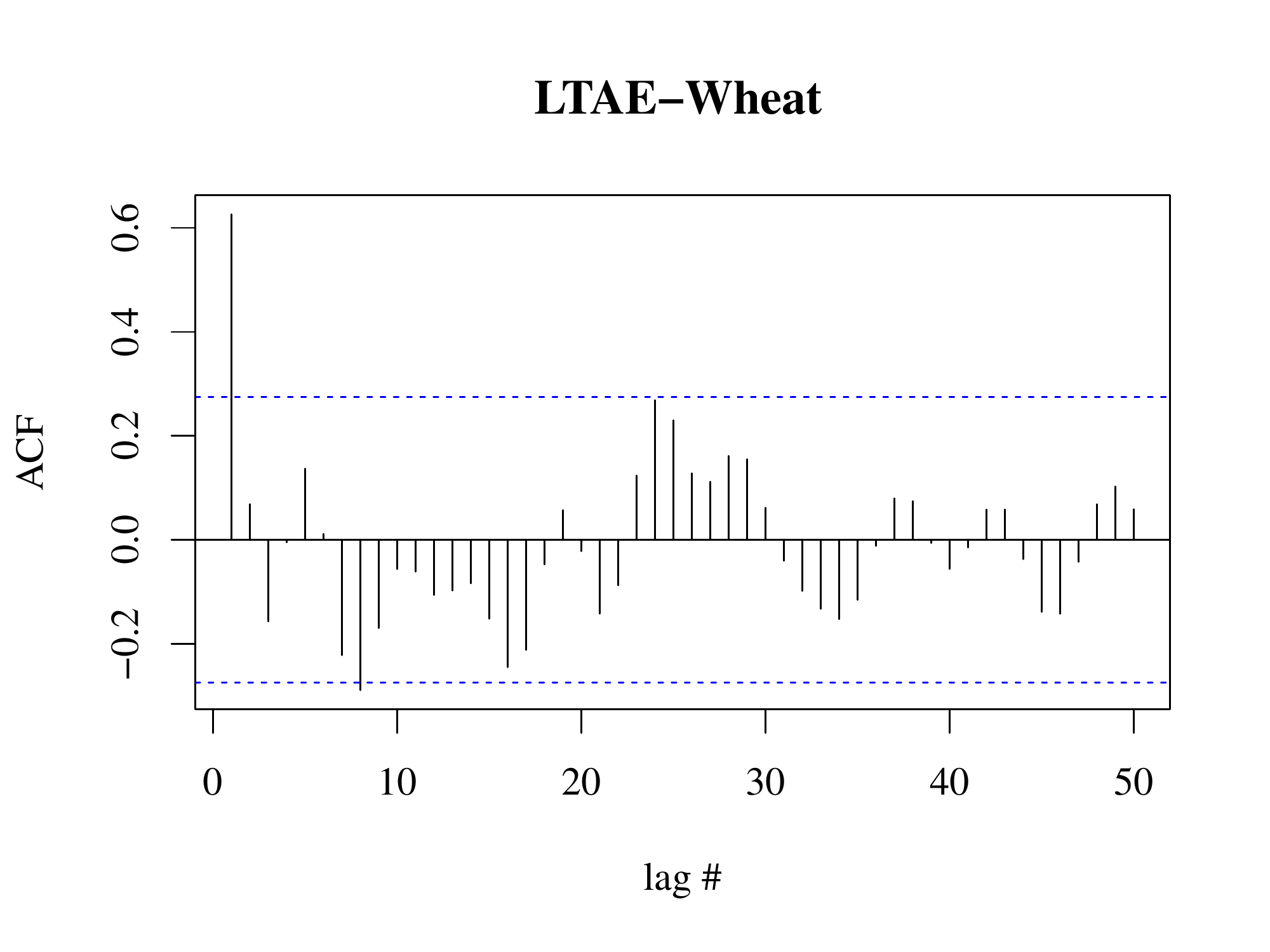}\tabularnewline
\end{tabular}
\par\end{centering}
\caption{Autocovariance functions for time series, showing a clear decrease
in SSN and WeMO in the last decades.}
\end{figure}

\end{document}